\documentclass[structabstract]{aa}  

\usepackage{graphicx}
\usepackage{natbib}
\usepackage{morefloats}

\bibpunct{(}{)}{;}{a}{}{,} 
\usepackage{txfonts}
\usepackage{epstopdf}
\begin{document}
 \title{Radio continuum observations of new radio halos and relics from the NVSS and WENSS surveys} 
  \subtitle{Relic orientations, cluster X-ray luminosity and redshift distributions}

   \author{R.~J. van Weeren\inst{1}
         \and M.~Br\"uggen \inst{3}
         \and H.~J.~A. R\"ottgering\inst{1}
         \and M.~Hoeft \inst{2}
        \and  S.~E.~Nuza \inst{4}
         \and H.~T.~Intema \inst{5}
          }

   \institute{Leiden Observatory, Leiden University,
              P.O. Box 9513, NL-2300 RA Leiden, The Netherlands\\
              \email{rvweeren@strw.leidenuniv.nl}
                \and Th\"uringer Landessternwarte Tautenburg, Sternwarte 5, 07778, Tautenburg, Germany
                 \and Jacobs University Bremen, P.O. Box 750561, 28725 Bremen, Germany
                 \and Astrophysikalisches Institut Potsdam, An der Sternwarte 16, 14482 Potsdam, Germany
                                  \and National Radio Astronomy Observatory, 520 Edgemont Road, Charlottesville, VA 22903-2475, USA
                 }


 
\abstract
   {Radio halos and relics are diffuse radio sources found in galaxy clusters showing significant substructure at X-ray wavelengths . These sources 
 provide important information about  non-thermal processes taking place in the intracluster medium (ICM). 
  Until now only a few dozen relics and halos are known, while 
   models predict that a much larger number of these sources exist. In this paper we 
   present the results of an extensive observing 
   campaign to search for new diffuse radio sources in galaxy clusters.
}
   {The aim of the observations is to create a large sample of diffuse radio sources in galaxy clusters that help to understand 
the formation of radio relics and halos and can be used to probe the physical conditions of the ICM. 
   }
   {We carried out radio continuum observations with the 
Westerbork Synthese Radio Telescope (WSRT),  Giant Metrewave Radio Telescope (GMRT) and Very Large Array (VLA) 
of clusters with diffuse radio emission visible in NVSS and WENSS survey images. Optical images were taken with the William Herschel and Isaac Newton Telescope (WHT, INT). 
}
    {We discovered 6 new radio relics, including a probable double relic system, and 2 radio halos. In addition, we confirm the presence of diffuse radio emission in four galaxy clusters. By constructing a sample of 35 radio relics we find that relics are mostly found along the major axis of the X-ray emission from the ICM, while their orientation is perpendicular to this axis. We also compared the X-ray luminosity and redshift distributions of clusters with relics to an X-ray selected sample from the NORAS and REFLEX surveys. We find tentative evidence for an increase of the cluster's relic fraction with X-ray luminosity and redshift. The major and minor axis ratio distribution of the ICM for clusters with relics is broader than that of the NORAS-REFLEX sample.
   }
{The location and orientation of radio relics with respect to the ICM elongation is consistent with the scenario that relics trace merger shock waves.
}
   \maketitle

\section{Introduction}
Radio halos and relics are found in massive merging galaxy clusters. These radio sources indicate the presence of magnetic fields and in-situ particle acceleration within the ICM \citep[e.g.,][]{1977ApJ...212....1J, 2004IJMPD..13.1549G}. Galaxy clusters form through mergers with other clusters and galaxy groups as well as through continuous accretion of gas from the intergalactic medium. Since giant radio halos and relics are found in merging clusters \citep[e.g.,][]{2001ApJ...553L..15B, 2004ApJ...605..695G, 2007A&A...467...37B, 2009A&A...507..661B, 2010ApJ...721L..82C}, it has been proposed that a small fraction of the energy released during a cluster merger event is channeled into the (re)acceleration of particles.

One scenario for the origin of radio relics is that they trace merger shock waves within the ICM in which particles are accelerated by the diffusive shock acceleration (DSA)  mechanism  \citep{1977DoSSR.234R1306K, 1977ICRC...11..132A, 1978MNRAS.182..147B, 1978MNRAS.182..443B, 1978ApJ...221L..29B, 1983RPPh...46..973D, 1987PhR...154....1B, 1991SSRv...58..259J, 2001RPPh...64..429M}. 
In the presence of a magnetic field these particles emit synchrotron radiation at radio wavelengths \citep[e.g.,][]{1998A&A...332..395E, 2000ApJ...542..608M}. 
The efficiency with which collisionless shocks can accelerate particles is unknown and may not be enough to produce the observed radio brightness of relics. A closely linked scenario is that of  shock re-acceleration of pre-accelerated electrons in the ICM, which is a more efficient mechanism for weak shocks \citep[e.g.,][]{2005ApJ...627..733M, 2008A&A...486..347G, 2011ApJ...734...18K}.

An alternative scenario has been proposed by \cite{2010arXiv1011.0729K} based on a secondary cosmic ray electron model, where the time evolution of magnetic fields and the cosmic ray distribution are taken into account to explain both halos and giant relics. Detailed spectral maps, at multiple frequencies, and measurements of the magnetic field via the polarization properties can test this model. For relics, strong magnetic fields (${B > B_{\rm{cmb}} = 3.25 \times (1+z)^2}$) are predicted and the spectral index\footnote{$F_{\nu} \propto \nu^{\alpha}$, where $\alpha$ is the spectral index} at the outer edges of relics should be flat, with $\alpha \simeq -1$.

Unlike relics, radio halos are found in the center of merging galaxy clusters and follow the X-ray emission from the ICM. Radio halos have been explained by turbulence injected by recent merger events. The injected turbulence is thought to be capable of re-accelerating relativistic particles \citep[e.g., ][]{2001MNRAS.320..365B, 2001ApJ...557..560P, 2005MNRAS.357.1313C}. Alternatively, the energetic electrons are secondary products which originate from hadronic collisions between relativistic protons and thermal ions \citep[e.g.,][]{1980ApJ...239L..93D, 1999APh....12..169B, 2000A&A...362..151D, 2010ApJ...722..737K, 2011A&A...527A..99E}.  Recent observations put tension on the secondary models \citep[e.g, ][]{2010MNRAS.401...47D, 2010MNRAS.407.1565D, 2011ApJ...728...53J, 2011MNRAS.412....2B, 2011A&A...530A..24B}. The existence of ultra-steep spectrum radio halos is also claimed not to be in agreement with these secondary models \citep{2008Natur.455..944B}, but they can be explained by the turbulent re-acceleration model. However, currently only a few of these ultra-steep spectrum radio halos are known so more observations, such as presented in this paper, are needed to increase this number and provide better measurements of the radio spectra. 

In the last decade a number of successful searches have been carried out to find new diffuse radio sources in galaxy clusters \citep[e.g.,][]{1999NewA....4..141G, 2000NewA....5..335G, 2001A&A...376..803G, 2001ApJ...548..639K, 2003A&A...400..465B, 2007A&A...463..937V, 2008A&A...484..327V,2009A&A...507.1257G, 2009A&A...508...75V, 2009ApJ...697.1341R}. However, our understanding of the formation of these sources is still limited. Models for the formation of relics and halos can be tested through statistical studies of correlations between the non-thermal radio emission and global properties of the clusters, such as mass, temperature, and dynamical state \citep{2000ApJ...544..686L,2006MNRAS.368..544F,2006MNRAS.369.1577C, 2007MNRAS.378.1565C, 2008A&A...480..687C, 2010A&A...509A..68C}. 

We recently discovered two large radio relics  
in the NVSS and WENSS surveys \citep{2010Sci...330..347V,2011A&A...528A..38V}. Interestingly, these relics remained unnoticed for about 15 years. This suggests that more diffuse radio sources could be discovered by inspection of the NVSS \citep{1998AJ....115.1693C} and WENSS \citep{1997A&AS..124..259R} survey images. We therefore carried out a visual inspection of the NVSS and WENSS images around known clusters detected by ROSAT \citep{1999A&A...349..389V, 2000IAUC.7432R...1V, 1998MNRAS.301..881E, 2000ApJS..129..435B, 2007ApJ...662..224K, 2002ApJ...580..774E, 2004A&A...425..367B}. 

Candidate clusters hosting diffuse radio emission were observed with the WSRT, GMRT and/or VLA. Clusters with existing published observations were not re-observed. In this paper we present the radio images and global properties of the clusters. In addition, we investigate the position and orientation of radio relics with respect the ICM, and compare the X-ray luminosity and redshift distributions of clusters with relics to an X-ray selected sample. In a follow-up paper we will present the polarization and detailed spectral properties of the radio emission in these clusters. The layout of this paper is as follows. In Sect.~\ref{sec:obs-reduction} we give an overview of the observations and the data reduction. In Sect.~\ref{sec:results} we present the radio images and give an short overview of the cluster's properties. We end with discussions and conclusions in Sects.~\ref{sec:discussion} and \ref{sec:conclusion}.

Throughout this paper we assume a $\Lambda$CDM 
cosmology with $H_{0} = 71$ km s$^{-1}$ Mpc$^{-1}$, 
$\Omega_{m} = 0.27$, and $\Omega_{\Lambda} = 0.73$. 
All images are in the J2000 coordinate system.

\begin{table*}
\begin{center}
\caption{Observations}
\begin{tabular}{lllllll}
\hline
\hline
Cluster &observation date & frequency & bandwidth   & integration time & map resolution & rms noise\\
              &                             & MHz          & MHz           &  hr                         &   arcsec              &  $\mu$Jy~beam$^{-1}$\\
\hline
Abell 1612  &  GMRT, 13 May, 2009 & 325  & 32 & 4.0& $11.6\arcsec~\times9.5\arcsec$ & 236\\
       & GMRT, 22 Nov, 2009 & 610, 241 & 32, 6 & 2.5, 2.5 &  $7.7\arcsec~\times4.7\arcsec$, $21\arcsec \times 12\arcsec$& 64, 777\\
Abell 523    & VLA Aug, 25, 2005;&  1425 & 37.5 & 1.1$^{a1}$, 3.7$^{a2}$ & $21\arcsec~\times20\arcsec^{d}$& 40\\
                      & Nov 28, 2005; Dec 28, 2009                                                                       & \\ %
Abell 697$^{e}$  &   WSRT, 24 Aug, 2009 & 1382, 1714& 160, 160& 6.0, 6.0&$34\arcsec~\times21\arcsec^{b}$ & 24, 32\\
Abell 3365$^{e}$  & WSRT, 22 Feb, 2009 & 1382& 160& 12.0&$108\arcsec~\times13\arcsec$&  29\\

Abell 3365  & VLA, 30 Sep, 2009   & 1425& $2 \times 50$ & 0.7$^{g1}$, 2.5$^{g2}$ & $43\arcsec~\times~35\arcsec$, $13.5\arcsec~\times~ 9.2\arcsec$ & 84, 27 \\
                      &  30 May, 2009    &  & \\
Abell 746$^{e}$    & WSRT, 19 Sep, 2009 & 1382& 160 & 6.0&$23\arcsec~\times18\arcsec$ & 28\\
Abell 2034$^{f}$ & WSRT, 26 Jul, 2009 & 1382& 160 & 6.0&$30\arcsec~\times16\arcsec$ & 24\\
Abell 2061$^{f}$ & WSRT, 23 Jul, 2009& 1382, 1714& 160, 160 & 6.0, 6.0&$32\arcsec~\times16\arcsec^{b}$ & 22, 26\\
RXC J1053.7+5452$^{e}$ & WSRT, 14 Mar, 2009 & 1382& 160 & 6.0 &$24\arcsec~\times18\arcsec$ & 30\\ %
CIZA J0649.3+1801  & GMRT, 22 Nov, 2009 & 610, 241& 32, 6  & 3.0, 3.0& $25\arcsec~\times25\arcsec^{c}$, $17\arcsec \times 14\arcsec$& 515, 1800\\
RX J0107.8+5408$^{e}$ & WSRT, 29 Aug, 2009& 1382 & 160 & 6.0& $21\arcsec~\times17\arcsec$ & 29\\
\hline
\end{tabular}
\label{tab:observations}
\end{center}
$^{a1}$ D array\\
$^{a2}$ C array\\
$^{b}$ the WSRT 1714~MHz image was convolved to the same resolution as the 1382~MHz image\\
$^{c}$ convolved to a beam of  $25\arcsec~\times25\arcsec$\\
$^{d}$ D and C array data were combined\\
$^{e}$ The minimum baseline length is 36~m. \\
$^{f}$ The minimum baseline length is 54~m. \\
$^{g1}$ DnC array\\
$^{g2}$ CnB array\\
\end{table*}
\section{Observations \& Data Reduction}
\label{sec:obs-reduction}

\subsection{Radio observations}
Most of candidate clusters with diffuse radio emission were observed with the WSRT. 
GMRT or VLA observations were taken of a few clusters which were missed by the 
WSRT observations. A summary of the observations is given in Table~\ref{tab:observations}. 
The WSRT observations were taken in frequency switching mode, alternating every 5 minutes 
between the 21 and 18 cm bands. In this paper we will only use the 21 cm data, except for the clusters Abell~697 and Abell 2061. The other data will be presented 
in a future paper which will focus on the polarization properties and rotation measure synthesis.
The total integration time for the WSRT observations was 6~hr per cluster, except for Abell~3365 (see Table~\ref{tab:observations}). 
VLA observations of Abell 523 were taken in D-array.  
We also included archival observations from project AB1180 (L-band D and C-array). 
Abell~3365 L-band VLA observations were taken in DnC and CnB array (project AR690).

GMRT 325~MHz observations (with the hardware correlator) were taken of Abell~1612 on May 13, 2009, recording both RR and 
LL polarizations with 32~MHz bandwidth. We observed Abell~1612 and CIZA~J0649.3+1801 at 610/241~MHz with the GMRT. 
The 610/241~MHz observations were taken in dual-frequency mode, recording 
RR polarization at 610~MHz and LL polarization at 241~MHz. Total recorded bandwidth was 32~MHz at 
610~MHz and 6~MHz at 241~MHz. 
The GMRT software backend \citep[GSB; ][]{2010ExA....28...25R} was used. 

For the data reduction we used the NRAO Astronomical Image 
Processing System (AIPS) package. Standard gain calibration and RFI removal were performed. Bandpass calibration 
was carried out for observations done in spectral line mode. 
For the 241~MHz observations RFI was fitted and subtracted from the data using the technique described by \cite{2009ApJ...696..885A} which was implemented in  
Obit \citep{2008PASP..120..439C}. The fluxes of the primary calibrators were 
set according to the \cite{perleyandtaylor} extension 
to the \cite{1977A&A....61...99B} scale. 
Several rounds of phase self-calibration were performed before doing two final rounds  
of amplitude and phase self-calibration.

For GMRT data in the imaging step we used the polyhedron method \citep{1989ASPC....6..259P, 1992A&A...261..353C} to 
minimize the effects of non-coplanar baselines. 
Images were made using ``Briggs'' weighting \citep[with robust set to 0.5,][]{briggs_phd} and cleaned 
down to $2$ times the rms noise level ($2\sigma_{\mathrm{rms}}$) using clean boxes. 
Finally, we corrected the image for the primary beam attenuation. For more details on the data reduction the reader is referred to \cite{2011A&A...527A.114V}.

\subsection{Optical WHT \& INT images}
For the clusters Abell~1612, Abell~523, Abell~3365, Abell~2034, CIZA~J0649.3+1801 and RX J0107.8+5408 we made use of optical V, R and I band images
 taken with the WHT (PFIP camera) and INT (WFC camera) telescopes between 15--19 
April (WHT) and 1--8 October (INT), 2009 \citep[for more details see][]{2011A&A...527A.114V}.

\section{Results}
\label{sec:results}
In this section we present the radio continuum images of the clusters. We briefly discuss the results of the radio observations for each cluster, a summary is given in Table~\ref{tab:sample}. To compute the integrated fluxes for the diffuse radio sources we subtracted the flux contribution from the discrete sources. We alphabetically labeled these discrete sources for each cluster where applicable. The fluxes of the discrete sources are reported in Table~\ref{tab:compact} and they are measured from images made with uniform weighting. We included the uncertainties in the subtraction of the discrete sources in the uncertainty for the integrated flux measurements of the diffuse cluster emission in Table~\ref{tab:sample}. 
 We assume that the uncertainties for the discrete sources are uncorrelated. More details are given in the subsections of the individual clusters. 

We also display  overlays onto ROSAT images and galaxy iso-density contours \citep[mostly derived from SDSS~DR7 photometric redshifts,][]{2009ApJS..182..543A}. The iso-densities were computed by counting the number of galaxies per unit surface area within a certain photometric redshift range (for SDSS) or V--R and R--I color range (for INT images). For the INT images we created a catalog of objects using Sextractor \citep{1996A&AS..117..393B}. We then removed all point-like objects (i.e., stars) from the catalogs. 
To exclude galaxies not belonging to the cluster we selected only galaxies with R$-$I and V$-$R colors within $0.15$ magnitude from the average color 
of a massive elliptical cD galaxy in the cluster. The range of $0.15$ in the colors was taken to maximize the contrast of the 
cluster with respect to the fore and background galaxies in the field, but not being too restrictive so that a sufficient number of candidate 
cluster members was selected.

\begin{table*}

\begin{center}
\caption{Cluster properties}
\begin{tabular}{lllllll}
\hline
\hline
Relic & $z$ &$S_{1382\mbox{ } \rm{MHz}}$ &  $P_{1.4\mbox{ }\rm{GHz}}$ & $L_{\rm{X,~0.1-2.4~keV}}$$^{d}$ & LLS$^{a}$  & classification$^{b}$\\
 &  &  mJy &  $10^{24}$ W~Hz$^{-1}$&$10^{44}$~erg~s$^{-1}$ & kpc\\
\hline
\object{Abell 1612}  & 0.179    & $62.8 \pm 2.6^{e}$  & 7.9 &2.41& 780 &R\\ 
\object{Abell 746} (center, periphery)    & 0.232& $18\pm4$, $24.5 \pm 2.0$  &3.8,  6.8  &3.68&  850, 1100 & H, R\\
\object{Abell 523}    &  0.10   & $61 \pm 7^{f}$  &1.7 &0.89& 1350 & H or R\\
\object{Abell 697}  & 0.282 & $5.2 \pm 0.5$, ${4.0\pm 0.5^{g}}$ &2.2 & 19.42 &  750 & H\\  
\object{Abell 2061} & 0.0784 &$27.6 \pm 1.0$, ${21.2 \pm 2.1^{g}}$ & 0.45 &3.95 & 675&R\\
\object{Abell 3365} & 0.0926 & $42.4 \pm 3.0^{j}$, $5.4 \pm 0.5^{k}$  &1.1$^{j}$, 0.14$^{k}$& 0.859 & 560$^{j}$, 235$^{k}$& DR\\
& & $42.6 \pm 2.6^{f,j}$, $5.3 \pm 0.5^{f,k}$  \\

\object{CIZA J0649.3+1801}  & 0.064 & $321 \pm 46^{i}$ &1.2$^{h}$ &2.38 &800 &R \\
\object{RX J0107.8+5408}$^{c}$ &0.1066&$ 55 \pm 5$&1.8&5.42&1100&H \\  %
\object{Abell 2034} (center, periphery) &0.113  & $7.3 \pm 2.0 $, $24 \pm 2$ &0.28, 0.89,  &3.56&600, 220  & H or R, R\\ 
\object{RXC J1053.7+5452} & 0.0704  &$15\pm2$&0.2 &0.44 & 600 &R \\
\hline
\hline
\end{tabular}
\label{tab:sample}
\end{center}
$^{a}$ largest linear size\\
$^{b}$ R = radio relic, H = radio halo, DR = double radio relic\\
$^{c}$ also known as \object{CIZA J0107.7+5408} or \object{ZwCl 0104.9+5350}\\
$^{d}$ Abell~3365 and Abell~1612: \cite{2004A&A...425..367B}, CIZA~J0649.3+1801 and RX~J0107.8+5408: \cite{2002ApJ...580..774E}, RXC~J1053.7+5452: \cite{2004A&A...423..449P},  Abell~697: \citep{2004A&A...423..449P}, Abell 2034~and Abell~2061: \cite{1998MNRAS.301..881E}, Abell~523: \cite{2000ApJS..129..435B}, Abell~746: this work\\
$^{e}$ NVSS flux\\
$^{f}$ VLA flux at 1425~MHz\\
$^{g}$ WSRT flux at 1714~MHz\\
$^{h}$ using a typical relic spectral index of $-1.3$\\
$^{i}$ GMRT flux at 610 MHz\\
$^{j}$ eastern relic\\
$^{k}$ western relic\\
\end{table*}

\begin{table*}
\begin{center}
\caption{Fluxes of compact sources embedded in the diffuse emission}
\tabcolsep=0.13cm
{\tiny\begin{tabular}{lllllllllll}
\hline
\hline
&A& B & C & D & E & F & H&  I & J & Figure\\
\hline
{A746}                  &   $1.52 \pm 0.09$& $0.36 \pm 0.06$& $0.24\pm0.06$& $0.37\pm0.06$ & $0.45\pm 0.07$ & $0.93 \pm 0.06$& $0.51\pm0.07$& $0.33\pm0.06$& $0.34\pm0.06$&\ref{fig:a746wsrt21cm} \\

{A523}                     &$10.9 \pm 0.2$& $5.7 \pm 0.1$ & $2.9 \pm 0.1$ & &&&&&& \ref{fig:a523_vla} \\

{A697$^a$}                     & $0.50 \pm 0.07$ & $0.29 \pm 0.06$  &  $0.22\pm0.06$    & $0.25\pm0.08$  & $0.16\pm0.06$& $0.19\pm0.05$ &&&&\ref{fig:a697_halo} \\
{A697$^b$}                     & $0.38 \pm 0.08$ & $0.24 \pm 0.07$   &  $0.17 \pm 0.05$ & ${0.12\pm0.05^{d}}$ &${0.15\pm0.06^{d}}$& ${0.07 \pm 0.05^{d}}$&&&&\ref{fig:a697_halo} \\
  
{A2061$^{a}$}                  & $0.53 \pm 0.08$ & $0.41 \pm 0.07$ & $0.36 \pm 0.07$ & &&&&&& \ref{fig:a2061_optical} \\
{A2061$^{b}$}                  & ${0.31\pm 0.08^{d}}$ &$0.33 \pm 0.08$ & ${0.20\pm0.08^{d}}$ & &&&&&& \ref{fig:a2061_optical} \\

{A3365$^{c}$}                  &   $1.5 \pm 0.1$  &  $0.53 \pm 0.06$ & $0.62\pm 0.06$ &&&&&&& \ref{fig:a3365_vla}\\
{CIZA J0649.3+1801}  & $7.7 \pm 0.8$& $1.2 \pm 0.1$& $2.5\pm 0.3$  &&&&&&&\ref{fig:rx18_xray}\\
{RX J0107.8+5408}     &  $2.6\pm 0.2$ &  $1.1 \pm 0.1$ &  $1.0\pm0.1$ & $0.7\pm0.1$&&&&&& \ref{fig:cizah_xray}    \\  
{A2034}                 & $0.93\pm0.06$ & $0.59\pm0.06$ &$0.34\pm0.05$ &&&&&&& \ref{fig:a2034_xray}\\
{RXC J1053.7+5452}  & $1.2\pm0.1$ & $1.4\pm 0.1$&  &   &&&&&&\ref{fig:rx1054_xray}\\
\hline
\hline
\end{tabular}}
\label{tab:compact}
\end{center}
Note: reported fluxes are in mJy\\
$^{a}$ 1382~MHz\\
$^{b}$ 1714~MHz\\
$^{c}$ from VLA CnB array image\\
$^{d}$ by measuring the flux at the 1382 MHz source position\\
\end{table*}

\subsection{Abell 1612}
\begin{figure*}
\begin{center}
\includegraphics[angle =90, trim =0cm 0cm 0cm 0cm,width=0.48\textwidth]{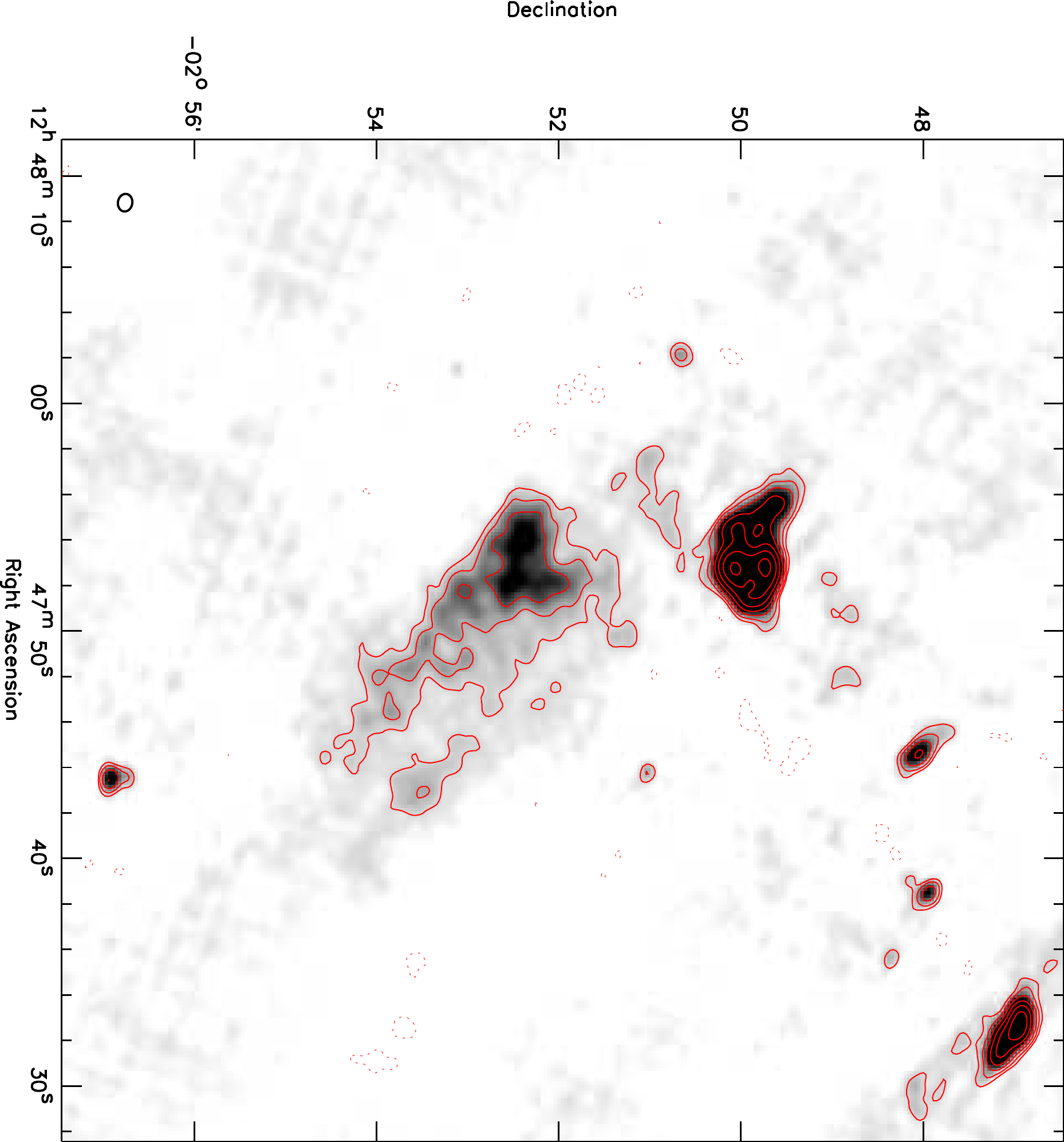}
\includegraphics[angle =90, trim =0cm 0cm 0cm 0cm,width=0.48\textwidth]{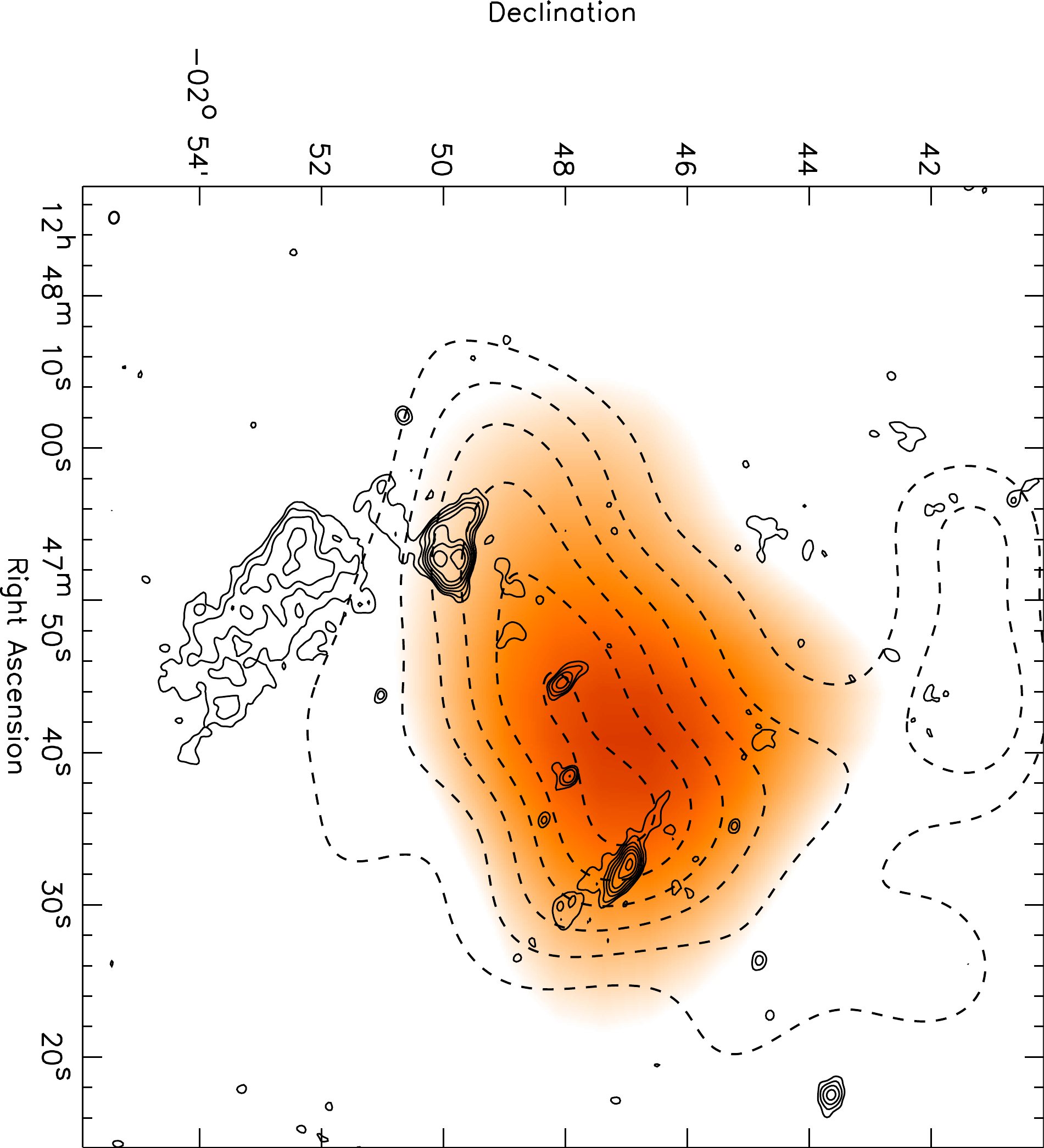}
\end{center}
\caption{Left: GMRT 325~MHz image of Abell 1612. Radio contours are drawn at levels of ${[1, 2, 4, \ldots]} \times 4\sigma_{\mathrm{rms}}$. Dashed contours 
are drawn at $-3\sigma_{\mathrm{rms}}$. Right: A1612 X-ray emission from ROSAT, tracing the thermal ICM, is shown by the color image. The original image from the ROSAT All Sky Survey was convolved with a 270\arcsec FWHM Gaussian. Solid contours are from the GMRT 325~MHz image and drawn at levels of ${[1, 2, 4,  \ldots]} \times 3\sigma_{\mathrm{rms}}$.  Dashed contours show the galaxy iso-density distribution derived from the SDSS survey. Contours are drawn at ${[1.0,1.4, 1.8, \ldots]}  \times 1.1$ galaxies arcmin$^{-2}$ selecting only galaxies with  $0.16 < z_{\rm{phot}} <  0.20$.}
\label{fig:a1612_gmrt325}
\end{figure*}

\begin{figure*}
\begin{center}
\includegraphics[angle =90, trim =0cm 0cm 0cm 0cm,width=0.49\textwidth]{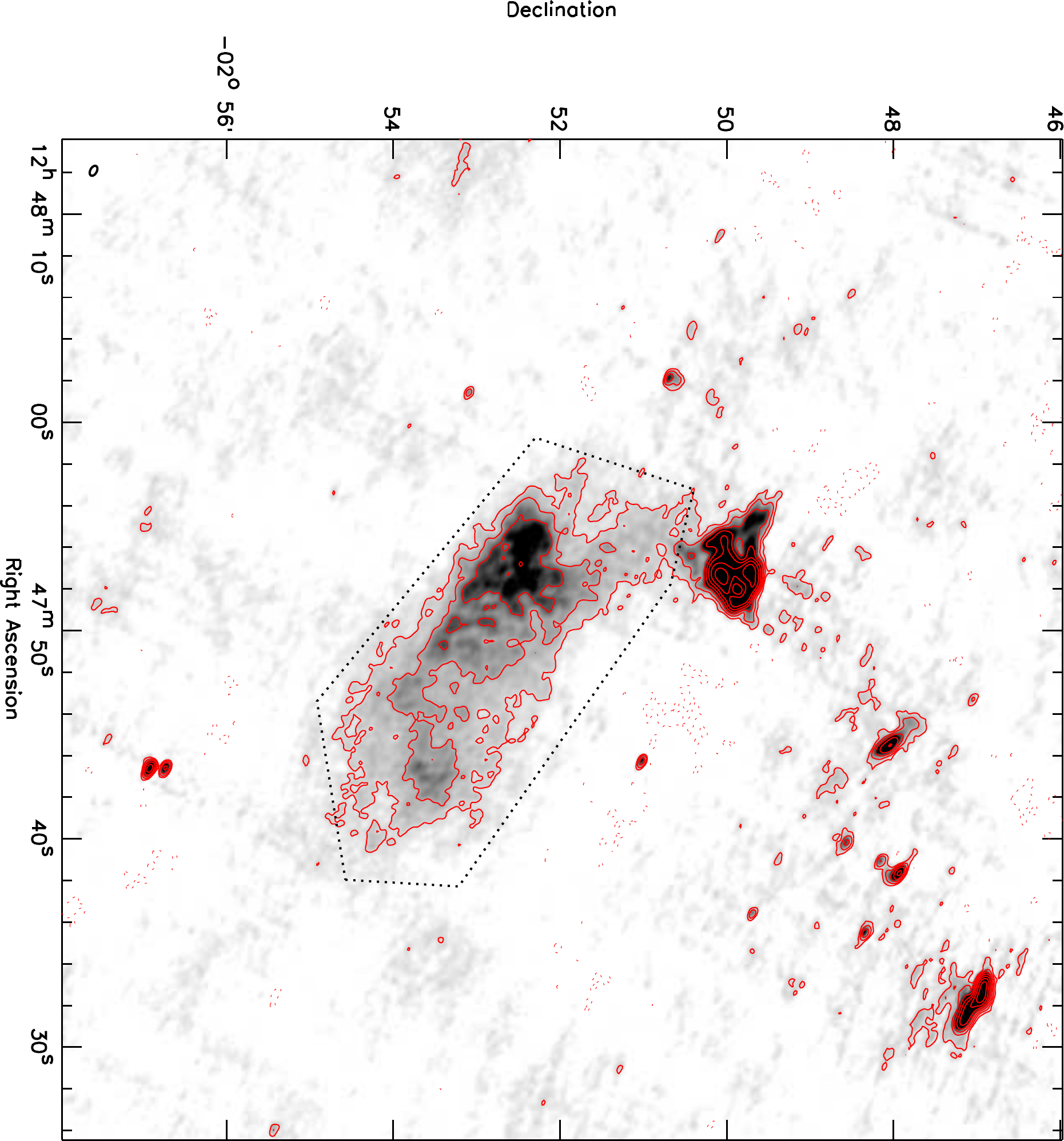}
\includegraphics[angle =90, trim =0cm 0cm 0cm 0cm,width=0.49\textwidth]{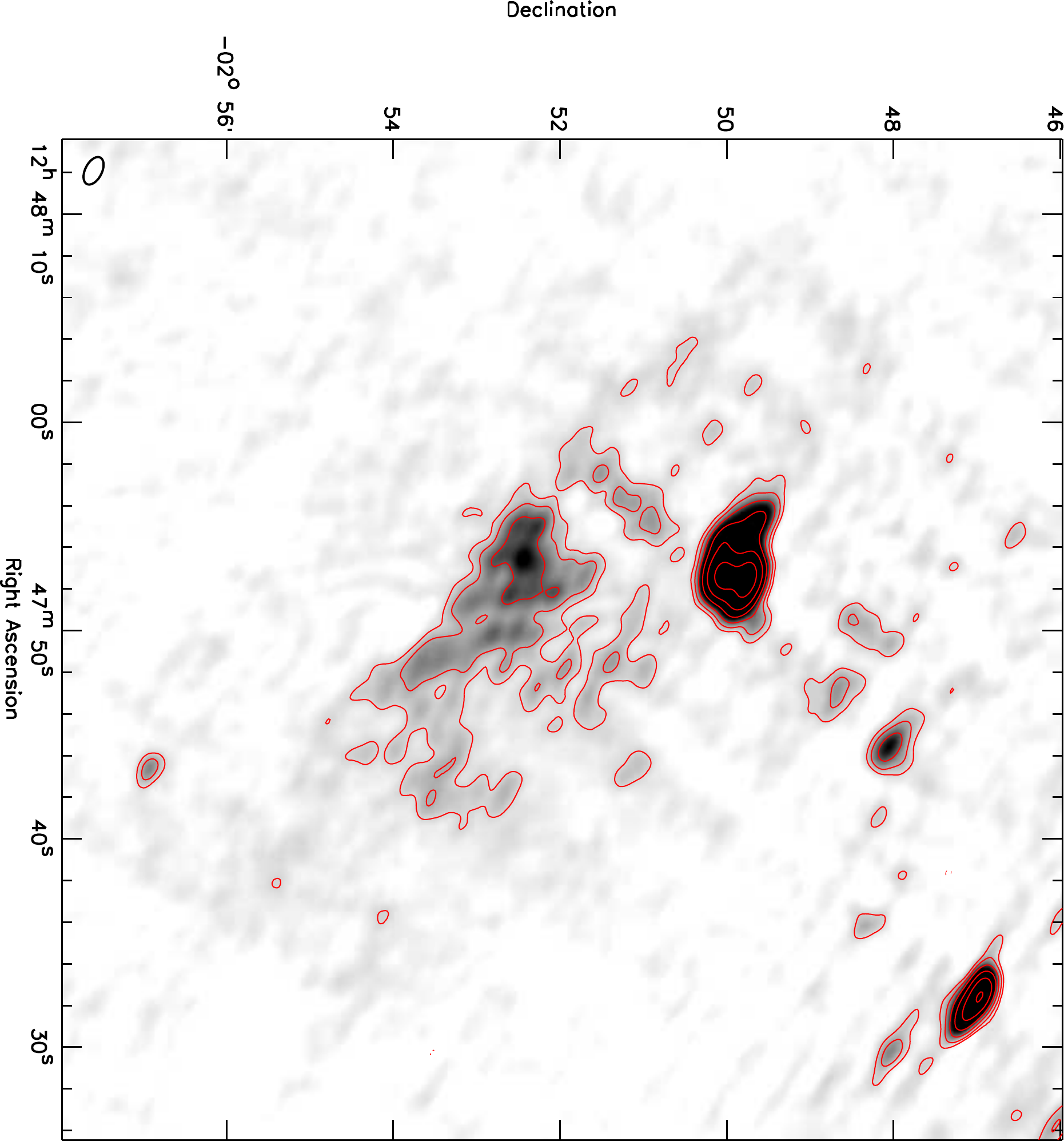}
\end{center}
\caption{GMRT 610 (left) and 241~MHz (right) image of A1612. Contour levels are drawn as in Fig.~\ref{fig:a1612_gmrt325}. Black dotted lines in the 610~MHz image indicate the integration area for the flux measurements.}
\label{fig:a1612_gmrt610}
\end{figure*}

\begin{figure*}
\begin{center}
\includegraphics[angle =90, trim =0cm 0cm 0cm 0cm,width=0.4\textwidth]{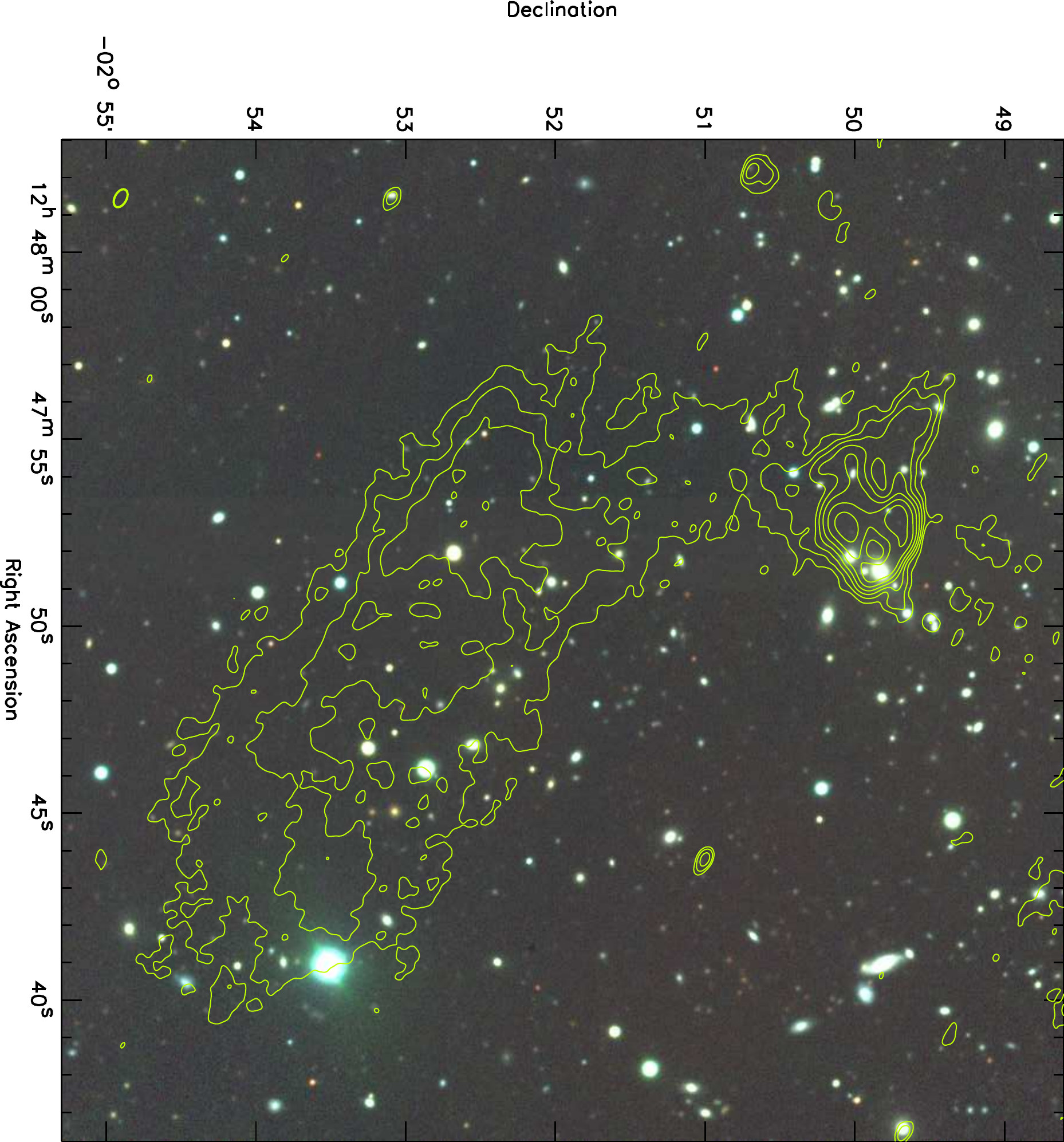}
\includegraphics[angle =90, trim =0cm 0cm 0cm 0cm,width=0.51\textwidth]{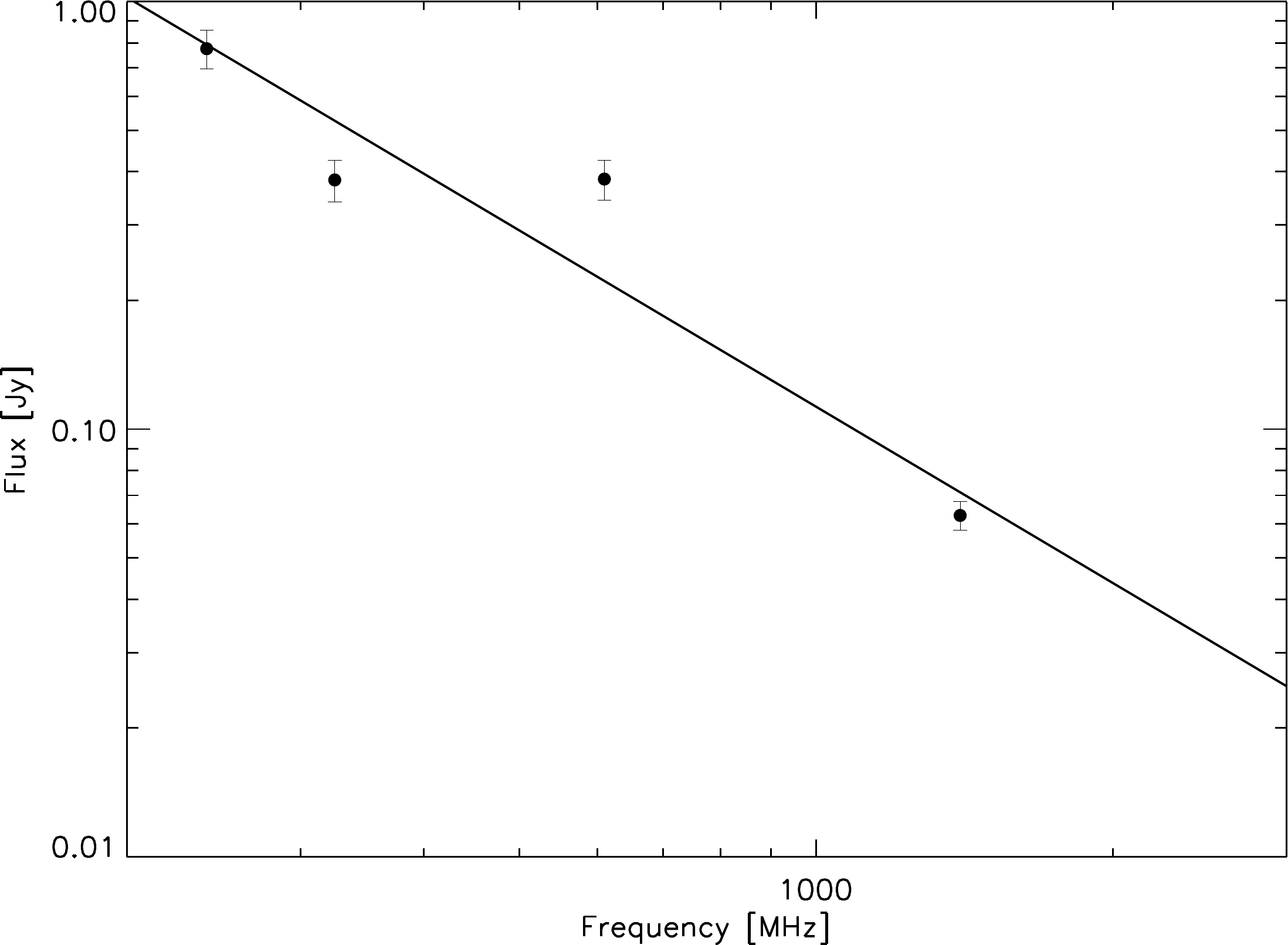}
\end{center}
\caption{Left: WHT V, R, I color image of Abell 1612. Overlaid are the radio contours from Fig.~\ref{fig:a1612_gmrt610} (left panel). Right: A1612 radio relic spectrum. Flux measurements at 241, 325, and 610 MHz are from the GMRT observations. The 1.4 GHz flux is from the NVSS survey image.}
\label{fig:a1612_optical}
\end{figure*}

Abell 1612 is a little studied cluster at $z= 0.179$ \citep{2004A&A...423..449P} with a moderate X-ray luminosity of 
$L_{\rm{X,~0.1-2.4~keV}}=2.41 \times 10^{44}$~erg~s$^{-1}$ \citep{2004A&A...425..367B}.
In the NVSS survey we found an elongated radio source located about 5\arcmin~to the south of the cluster center. 
The source was completely resolved out in the 1.4~GHz 5\arcsec~FIRST survey \citep{1995ApJ...450..559B} indicating diffuse emission on 
scales of about 4\arcmin. In our GMRT 325~MHz image (Fig.~\ref{fig:a1612_gmrt325}, left panel) the source has a total extent of 
4.5\arcmin, which corresponds to a physical size of 780~kpc. We could not identify an optical counterpart for the source (Fig.~\ref{fig:a1612_optical}, left panel).
We therefore classify the source as a radio relic. In the 610~MHz image (Fig.~\ref{fig:a1612_gmrt610}, left panel) the source is connected to a tailed radio galaxy to the north.
Combining flux measurements at 1.4 GHz (NVSS) and GMRT fluxes at 241, 325 and 610~MHz gives $\alpha\approx-1.4$. However, we note  that the individual flux measurements give a large scatter (Fig.~\ref{fig:a1612_optical}, right panel). We checked the absolute flux calibration between the different frequencies by measuring the integrated fluxes of several compact sources. This did not reveal any problems. Some of the short baselines in the 325 and 241~MHz observations were affected by RFI which could have affected the flux measurement since the source is quite extended. In addition, the declination is close to 0\degr which gives a non-optimal uv-coverage.

Probably the relic traces a shock in which particles are accelerated or re-accelerated by 
a merger-related shock wave.  The large extent makes it unlikely that the source traces compressed 
fossil radio plasma \citep{2006AJ....131.2900C}. The connection with the tailed radio source to the north favors a re-acceleration scenario. In this case, the seed relativistic electrons could be supplied by the tailed-radio source.  The elongated galaxy iso-density contours and ICM, see Fig.~\ref{fig:a1612_gmrt325} (right panel), also hint at a merger event along a northwest-southeast axis.

\subsection{Abell 746}
\begin{figure*}
\begin{center}
\includegraphics[angle =90, trim =0cm 0cm 0cm 0cm,width=0.49\textwidth]{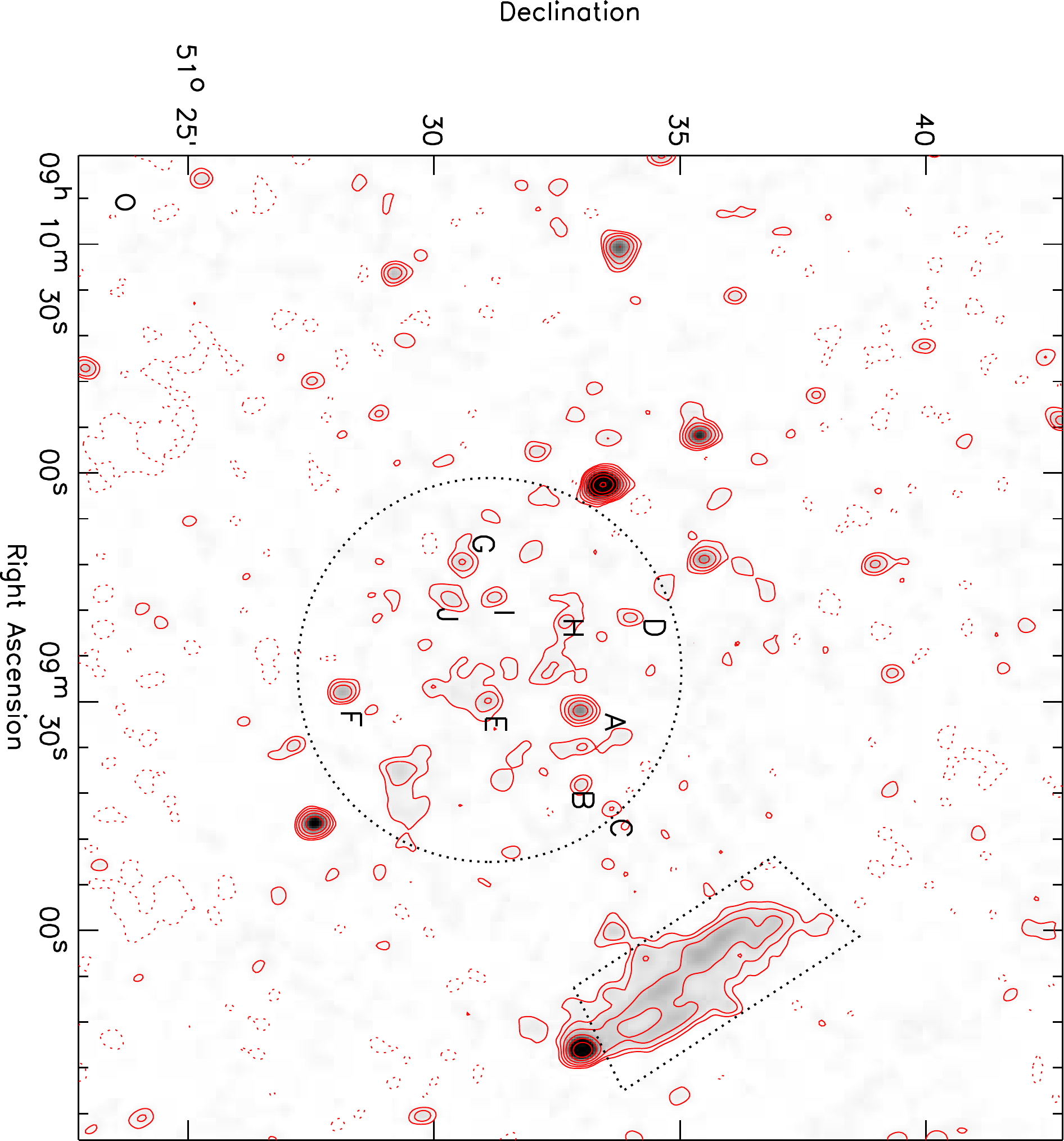}
\includegraphics[angle =90, trim =0cm 0cm 0cm 0cm,width=0.49\textwidth]{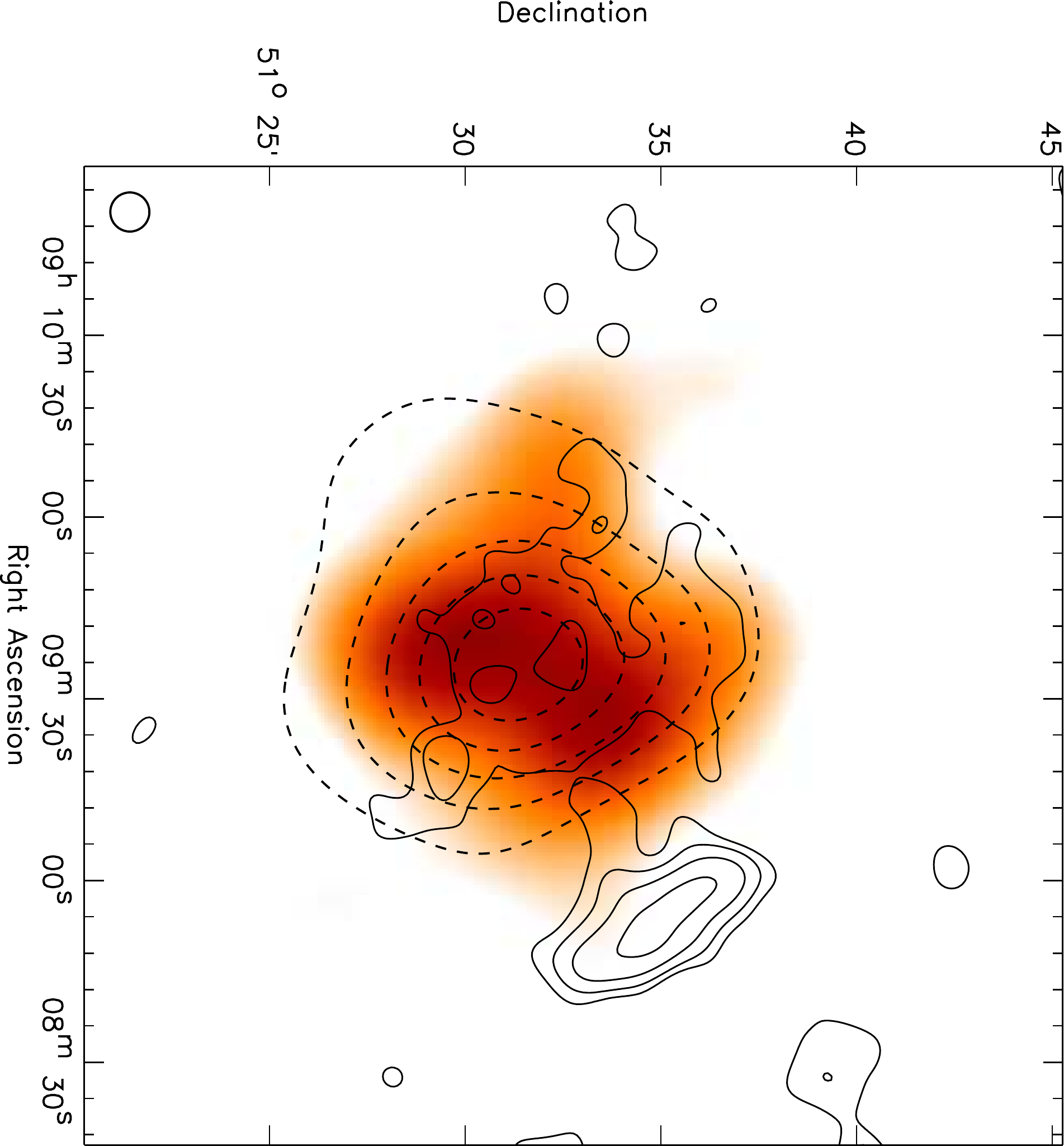}
\end{center}
\caption{Left: WSRT 1382~MHz image of A746. Contour levels are drawn as in Fig.~\ref{fig:a1612_gmrt325}. Black dotted lines indicate the integration areas for the flux measurements. Discrete sources embedded in the diffuse emission are alphabetically labeled, see Table~\ref{tab:compact}. Right: A746 X-ray emission from ROSAT in orange. The original image from the ROSAT All Sky Survey was convolved with a 225\arcsec~FWHM Gaussian.  Solid contours are from the WSRT 1382~MHz natural weighted image convolved to a resolution of 60\arcsec. Compact sources were subtracted and contours are drawn at levels of ${[1, 2, 4, 8, \ldots]} \times 0.4$~mJy~beam$^{-1}$. Dashed contours show the galaxy iso-density distribution derived from the SDSS survey. Contours are drawn at ${[1.0,1.4, 1.8, \ldots]}  \times 0.6$ galaxies arcmin$^{-2}$ selecting only galaxies with  $0.16 < z_{\rm{phot}} <  0.29$.}
\label{fig:a746wsrt21cm}
\end{figure*}

\object{Abel~746} is a Bautz-Morgan 
(B/M) class III type cluster located at $z_{\rm{phot}}=0.232 \pm 0.01$ 
\citep{2007ApJ...660..239K}. Based on the ROSAT flux from  \cite{2000ApJS..129..435B}  
we calculate $L_{\rm{X,~0.1-2.4~keV}} \sim 3.68 \times 10^{44}$~erg~s$^{-1}$.
We discovered an elongated radio source located northwest of the 
cluster center in the NVSS survey. The diffuse 
source is resolved out in the FIRST image, except for an unresolved source located at the southern tip of the diffuse source. In our WSRT image the elongated source is easily detected (Fig.~\ref{fig:a746wsrt21cm}, left panel) . The source has a 
largest angular extent of 5\arcmin, which corresponds to a physical size of 1.1~Mpc at the 
distance of Abell~746. The source has a width of 345~kpc and is located at a distance of $1.7$~Mpc from the cluster center. 

\begin{figure}
\begin{center}
\includegraphics[angle =90, trim =0cm 0cm 0cm 0cm,width=0.49\textwidth]{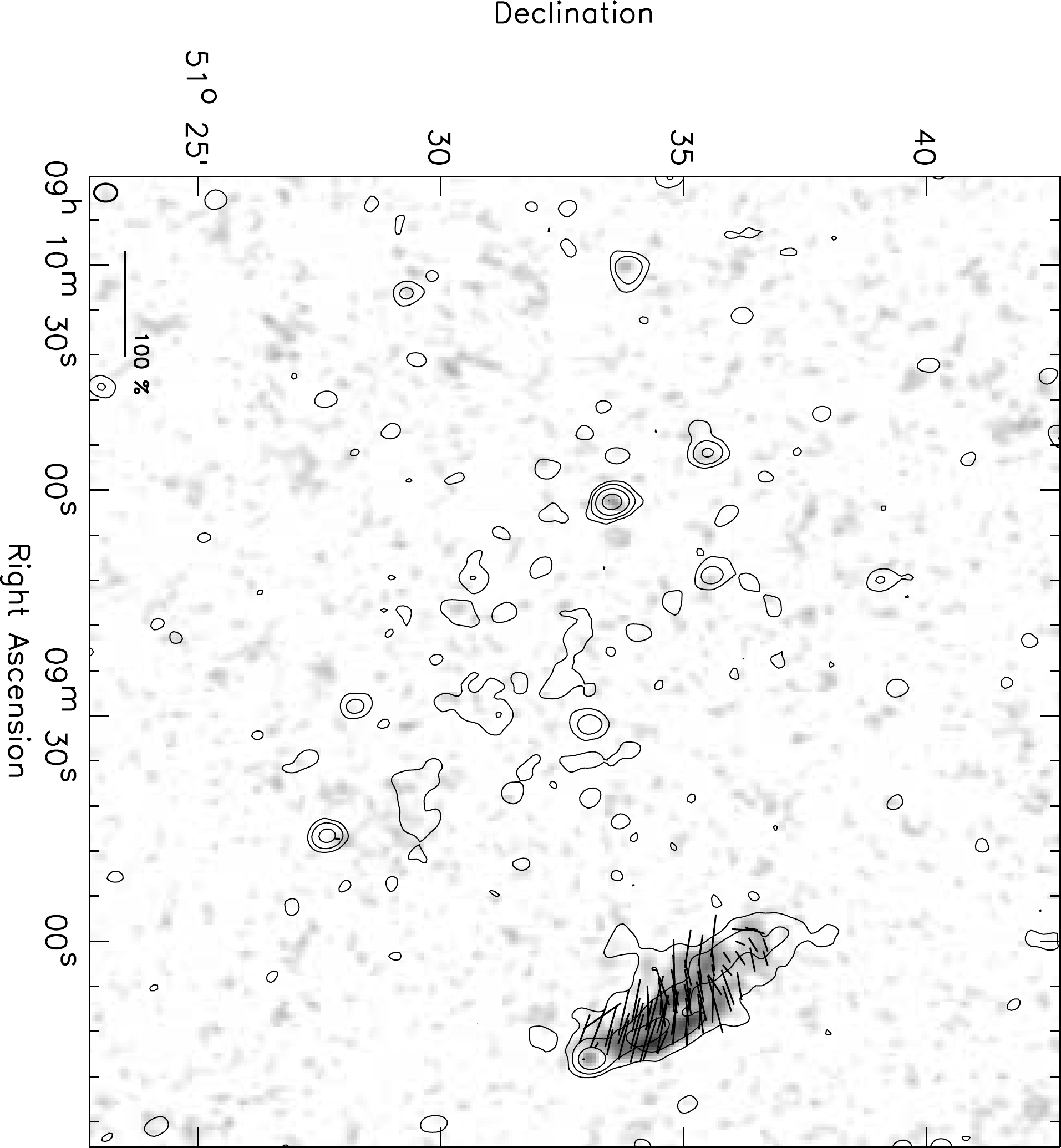}
\end{center}
\caption{WSRT 1382~MHz polarization map of A746. Total polarized intensity 
is shown as a grayscale image.  Vectors depict the polarization E-vectors, 
their length represents the polarization fraction. The length of the E-vectors are corrected 
for Ricean bias \citep{1974ApJ...194..249W}. 
A reference vector for a polarization fraction of 100\% is shown in 
the bottom left corner. No vectors were drawn for pixels with a 
SNR $< 4$ in the total polarized intensity image. 
Contour levels are drawn at ${[1, 4, 16, 64, \ldots]} \times 4\sigma_{\mathrm{rms}}$ 
and are from the Stokes~I image. }
\label{fig:a746pol}
\end{figure}

We do not find it very likely that the point source at the southern end of the elongated source is the hotspot of a radio galaxy because the second hotspot is missing. In addition, we do not detect any radio core. 
The compact source at the southern end of the diffuse source does not have an optical counterpart in SDSS images. The SDSS images are contaminated by the bright star f~Uma (V$_{\rm{mag}}=4.5$). We only conclude that a possible optical counterpart should be located farther away than A746. This limit on the redshift does  not provide any useful constraints to rule it out as a giant radio galaxy on the basis of its physical size or radio luminosity. The morphology of the source is more typical for a radio relic. The size of the sources an its location would also agree with this interpretation. As an additional check, we analyzed the polarization data from the WSRT observation. Radio relics are often polarized at a level of $\sim 20\%$ or more \citep[e.g.,][]{2006AJ....131.2900C, 2009A&A...494..429B, 2010Sci...330..347V}. Indeed, we find that the elongated source is polarized up to the $\sim50\%$ level (Fig.~\ref{fig:a746pol}), which provides additional support for the classification as radio relic.

Galaxy iso-density contours show the cluster 
to be somewhat elongated along a northwest-southeast 
axis, see Fig.~\ref{fig:a746wsrt21cm} (right panel). The ROSAT image reveals 
little structure. However, the number of photons detected from the cluster is low. 

The WSRT image reveals additional 
diffuse emission in the center of the cluster, albeit at a  
low SNR. The emission has an extent of at least 850~kpc which is typical for a giant radio halo. 
The flux of this radio halo is difficult to estimate due to the low SNR, but subtracting the contribution from the compact sources we find a flux of about $18\pm4$~mJy for the radio halo. To better image the diffuse emission we subtracted the clean components from the compact sources using an image made with uniform weighting. We then convolved the image (made with natural weighting) to a resolution of 1\arcmin. The contours from this image are overlaid in Fig.~\ref{fig:a746wsrt21cm} (right panel). The halo is now better detected and the radio emission roughly follows the X-ray emission. The radio power of $3.8 \times 10^{24}$~W~Hz$^{-1}$ is above the $L_{\rm{X}}$--$P_{\rm{1.4GHz}}$ correlation for giant radio halos \citep[e.g.,][]{2000ApJ...544..686L,2006MNRAS.369.1577C}.  The fitted relation from \cite{2006MNRAS.369.1577C} gives a power of $0.64\times 10^{24}$~W~Hz$^{-1}$. We note that both the integrated radio flux and $L_{\rm{X}}$ (from ROSAT) are uncertain, and that the intrinsic scatter in the $L_{\rm{X}}$--$P_{\rm{1.4GHz}}$ relation is quite large \citep{2009A&A...507..661B}. The measured radio power is therefore still marginally consistent with being on the  $L_{\rm{X}}$--$P_{\rm{1.4GHz}}$ correlation.

\subsection{Abell 523}
\begin{figure*}
\begin{center}
\includegraphics[angle =90, trim =0cm 0cm 0cm 0cm,width=0.49\textwidth]{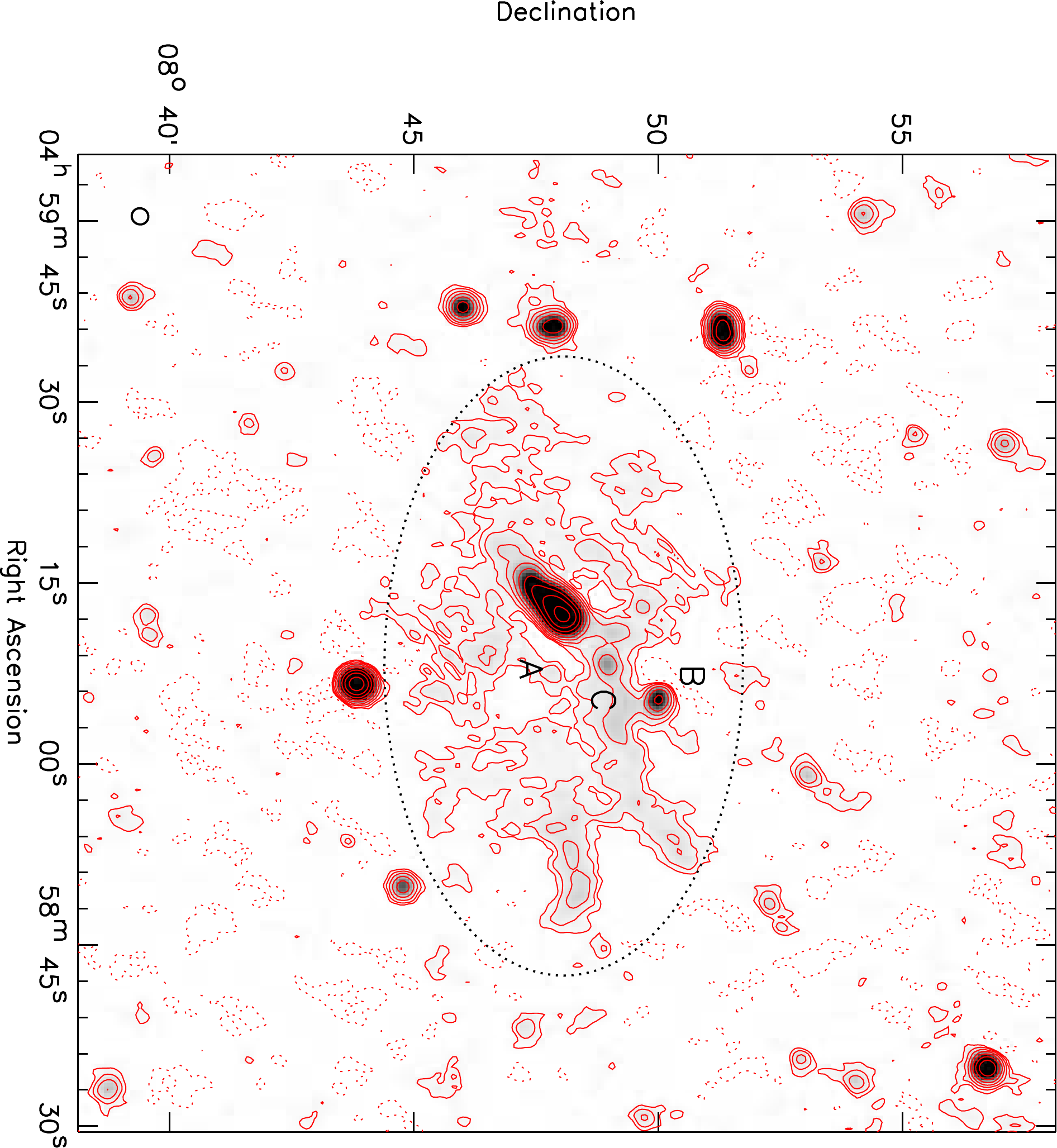}
\includegraphics[angle =90, trim =0cm 0cm 0cm 0cm,width=0.49\textwidth]{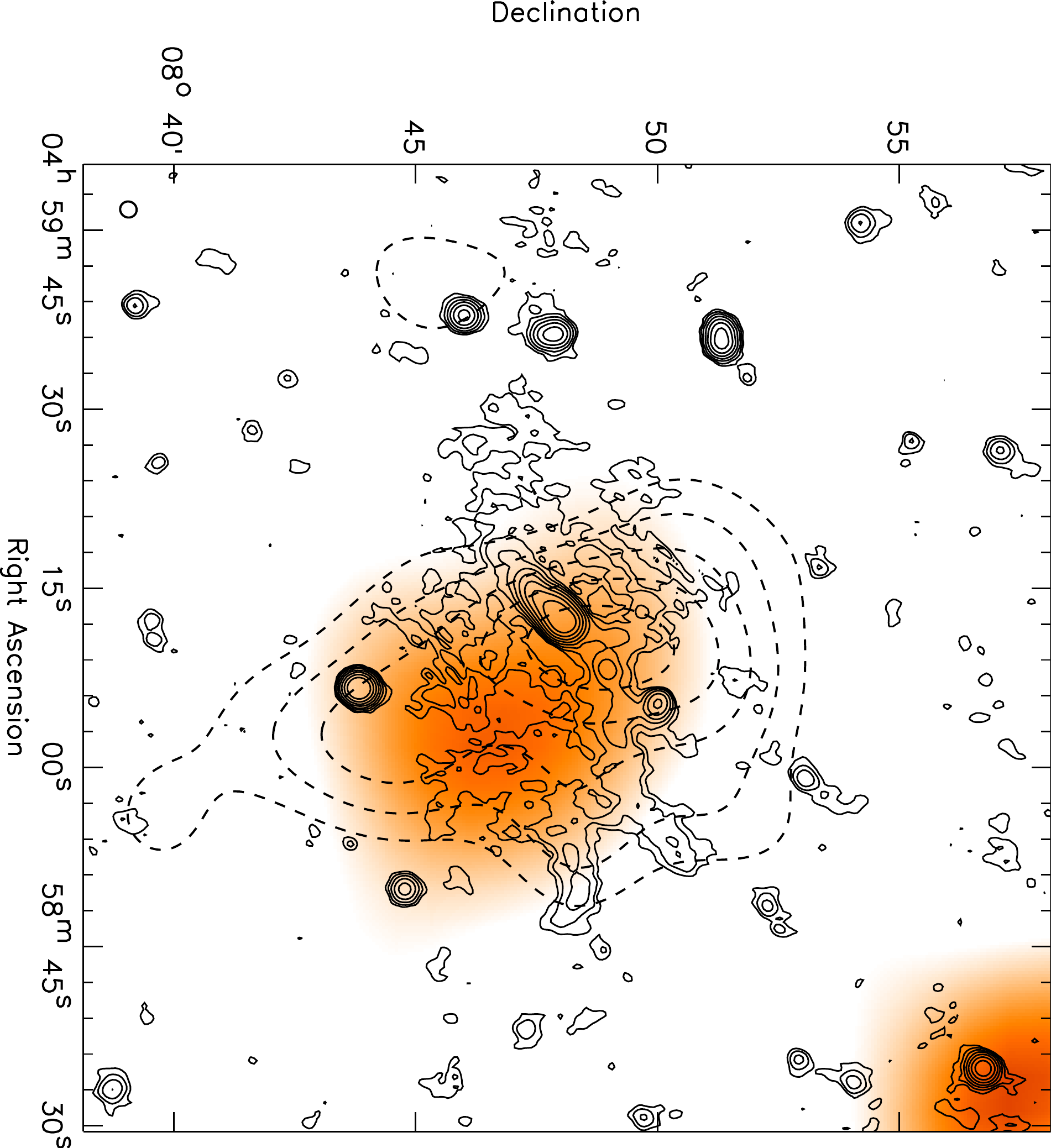}
\end{center}
\caption{Left: A523 VLA 1.4 GHz image. Contour levels are drawn as in Fig.~\ref{fig:a1612_gmrt325}. Black dotted lines indicate the integration area for the flux measurement. Discrete sources embedded in the diffuse emission are alphabetically labeled, see Table~\ref{tab:compact}. Right: A523 X-ray emission from in orange. The original image from the ROSAT All Sky Survey was convolved with a 225\arcsec~FWHM Gaussian. Solid contours are from the VLA 1425~MHz image and drawn at levels of ${[1, 2, 4, 8, \ldots]} \times 4\sigma_{\mathrm{rms}}$.  Dashed contours show the galaxy iso-density distribution derived from INT images. Contours are drawn at ${[1.0,1.2, 1.4, \ldots]}  \times 1.6$ galaxies arcmin$^{-2}$ selecting only galaxies with colors  $1.12 < \rm{V-R} < 1.42$,  $ 0.25 < \rm{R-I} < 0.55$, i.e., within 0.15 magnitudes the V--R and R--I color of the central cD galaxy.}
\label{fig:a523_vla}
\end{figure*}

Abel~523 is a little studied galaxy cluster located at $z=0.10$ \citep{1999ApJS..125...35S}, with 
$L_{\rm{X,~0.1-2.4~keV}}=0.89 \times 10^{44}$~erg~s$^{-1}$ \citep{2000ApJS..129..435B}. Galaxy iso-densities derived from INT images show a north-south elongated cluster consisting of two galaxy clumps. The VLA image (Fig.~\ref{fig:a523_vla}, left panel) reveals a large, irregular and diffuse radio source in the cluster as well as a number of compact sources related to AGN activity. Our radio image and galaxy distribution agree with the recent results from \cite{2011A&A...530L...5G} which showed the presence of diffuse radio emission in this cluster. 

\begin{figure}
\begin{center}
\includegraphics[angle =90, trim =0cm 0cm 0cm 0cm,width=0.5\textwidth]{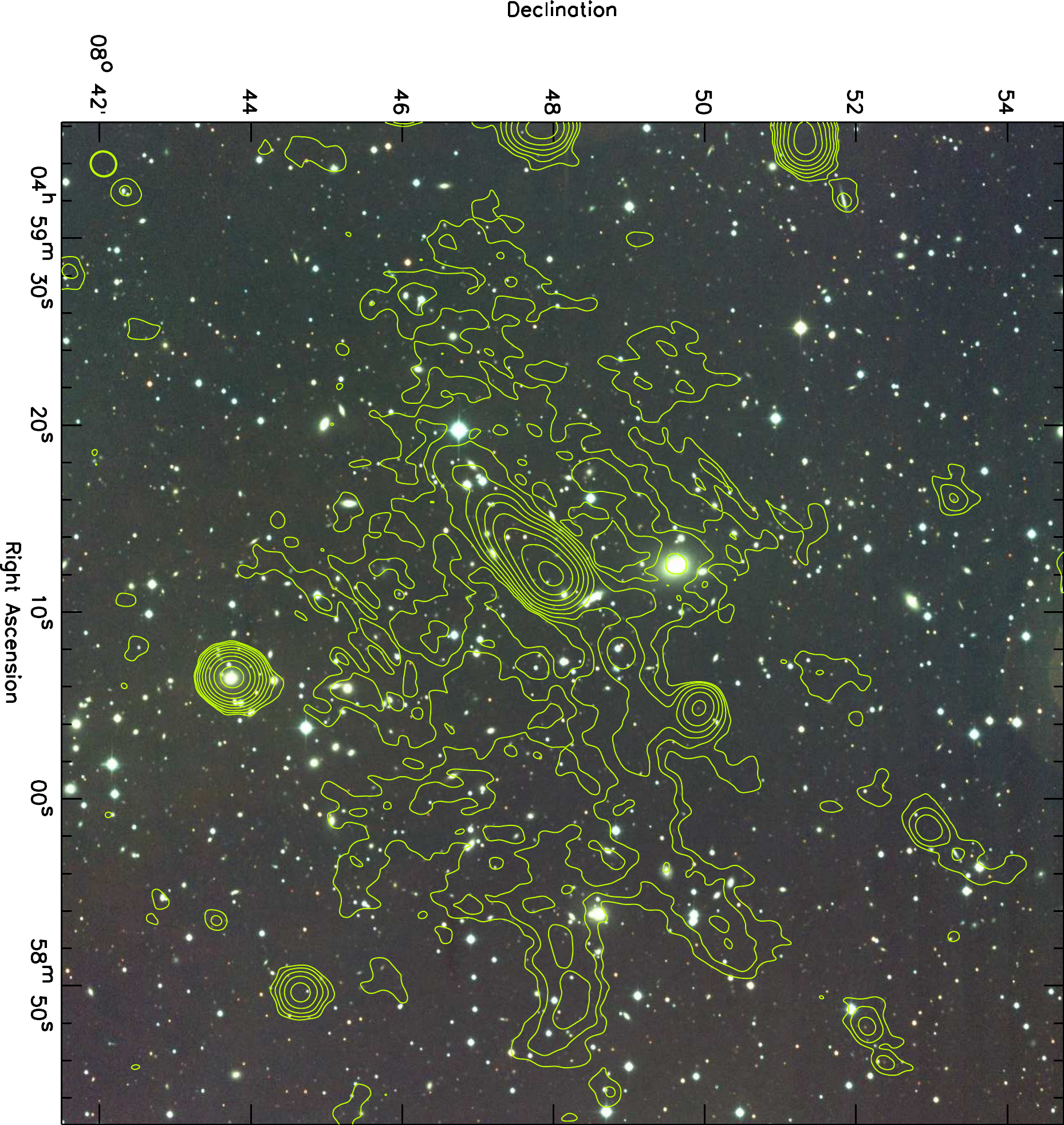}
\end{center}
\caption{INT V, R, I color image of Abell 523. Overlaid are the radio contours from Fig.~\ref{fig:a523_vla}.}
\label{fig:a523_optical}
\end{figure}

The brightest radio source is a tailed radio galaxy projected relatively close to the cluster center. The color of the 
optical counterpart is similar to other galaxies in the cluster making it likely that the  
radio source is associated with cluster. The optical INT image is shown in Fig.~\ref{fig:a523_optical}.
We also detect some radio emission from the largest cD galaxy in the cluster located north of the tailed radio source.
In the southern part of the cluster there is a brighter compact source associated with another 
large elliptical galaxy.

 The diffuse source has a patchy morphology, with the brightest part of the diffuse source located to the northwest of the tailed radio source.
 To the west the diffuse source extends into two filamentary structures. The total flux of the diffuse radio source, minus the point sources and head-tail galaxy, is $61 \pm 7$~mJy. The diffuse source has a largest extent of 1.35~Mpc. Both numbers are consistent with the result from \cite{2011A&A...530L...5G}.

The large extent and morphology make it unlikely that the diffuse sources is directly related to the tailed radio galaxy. \cite{2011A&A...530L...5G} classify the source as a radio halo. They note that the radio emission permeates both galaxy clumps. However, the possibility that the source is a radio relic projected onto central region of the cluster should also be considered. The source has a very patchy morphology unlike typical radio halos for which the surface brightness follows that of the X-ray emission. Currently we have too little constraints to completely rule out the relic scenario. Deep X-ray observations will be needed to characterize the dynamical state of the cluster. Polarization observations can be used to distinguish between the radio halo and relic scenarios as radio relics are usually strongly polarized (at the 10--20\% level or more) while halos are mostly unpolarized.

\subsection{Abell 697}
\label{sec:a697}
Abell~697 is a massive  Bautz-Morgan (B/M) type II-III cluster, located at $z=0.282$, with 
a high X-ray luminosity of  
$L_{\rm{X,~0.1-2.4~keV}}=19.42 \times 10^{44}$~erg~s$^{-1}$ \citep{2004A&A...423..449P}.
The ICM has an elliptical shape and an overall temperature of $8.8^{+0.7}_{-0.6}$~keV determined from Chandra 
observations by \cite{2008ApJS..174..117M}. \cite{2009ApJS..182...12C} reported a slightly higher 
temperature of 9.52~keV. 
\cite{2006A&A...455...45G} found a velocity dispersion of $1334^{+114}_{-96}$~km~s$^{-1}$ for the cluster.  
They noted this is expected in the case of energy-density equipartition between galaxies and gas.
They suggested that the cluster has undergone a complex cluster merger event,
occurring mainly along the LOS, with a transverse component in the SSE-NNW direction. 

\cite{2001ApJ...548..639K} first suggested the presence of diffuse radio emission in the cluster.
\cite{2008A&A...484..327V} showed the presence of a radio halo using 610~MHz GMRT observations, the halo was also reported by 
\cite{2009ApJ...697.1341R} and \cite{2009A&A...507.1257G}. \cite{2010A&A...517A..43M} presented a more detailed study which included GMRT 325~MHz observations. They found the radio halo to have a an ultra-steep spectrum ($\alpha_{325 \rm{~MHz}}^{1.4 \rm{~GHz}}$) with $\alpha$ about -1.7 to -1.8.

\begin{figure}
\begin{center}
\includegraphics[angle =90, trim =0cm 0cm 0cm 0cm,width=0.5\textwidth]{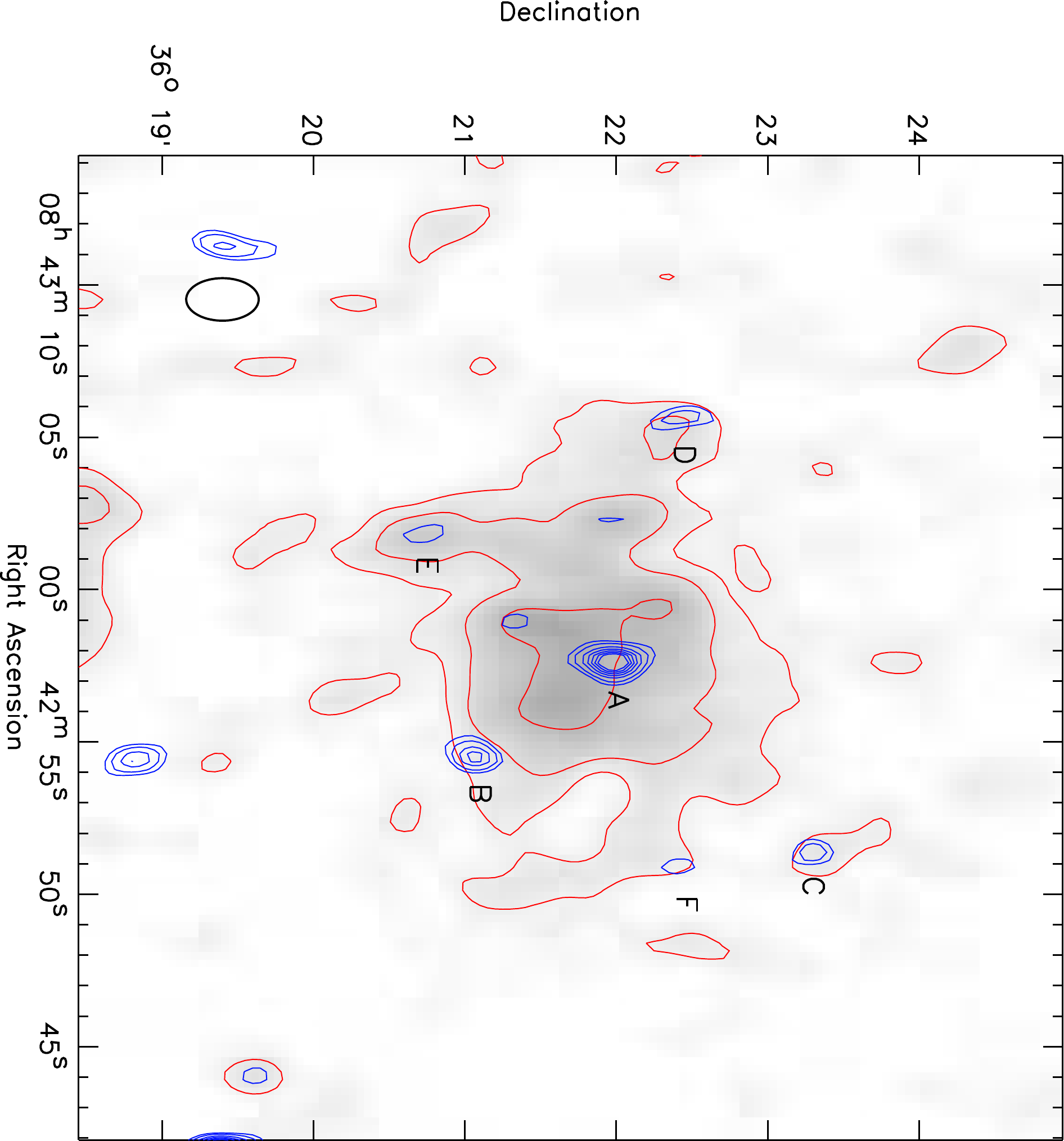}
\end{center}
\caption{WSRT combined 1382 and 1714~MHz image of the radio halo in Abel~697, compact sources were subtracted from the uv-data. Red contour levels are drawn at ${[1, 2, 4, 8 \ldots]} \times 60$~$\mu$Jy~beam$^{-1}$. Blue contours display the compact sources at 1382~MHz. This high-resolution image ($19\arcsec~\times~10\arcsec$) was made with uniform weighting and data $< 2.5$k$\lambda$ was excluded. Contour levels are drawn at $\sqrt{[1, 2, 4, 8 \ldots]} \times 100$~$\mu$Jy~beam$^{-1}$. Discrete sources embedded in the diffuse emission are alphabetically labeled, see Table~\ref{tab:compact}.}
\label{fig:a697_halo}
\end{figure}

We detect the radio halo at both 1382 and 1714~MHz with better SNR than the previous observations at 1.4~GHz. In our 1382~MHz image we find a total extent of about 750~kpc for the radio halo which is lower than the 1.3~Mpc reported by \cite{2010A&A...517A..43M} at 325~MHz. The lower 1382~MHz extent is expected due to the steep radio spectrum of the halo. To better image the diffuse emission and remove the contribution from compact sources, we made images with uniform weighting and excluded data $< 2.5$k$\lambda$ at both 1382 and 1714~MHz. We then subtracted the clean components of the compact sources from the uv-data before re-imaging. To increase the SNR we combined the 21 and 18 cm images after convolving them to a common resolution of $29\arcsec \times 17\arcsec$, see Fig.~\ref{fig:a697_halo}. 

\begin{figure*}
\begin{center}
\includegraphics[angle =90, trim =0cm 0cm 0cm 0cm,width=0.3\textwidth]{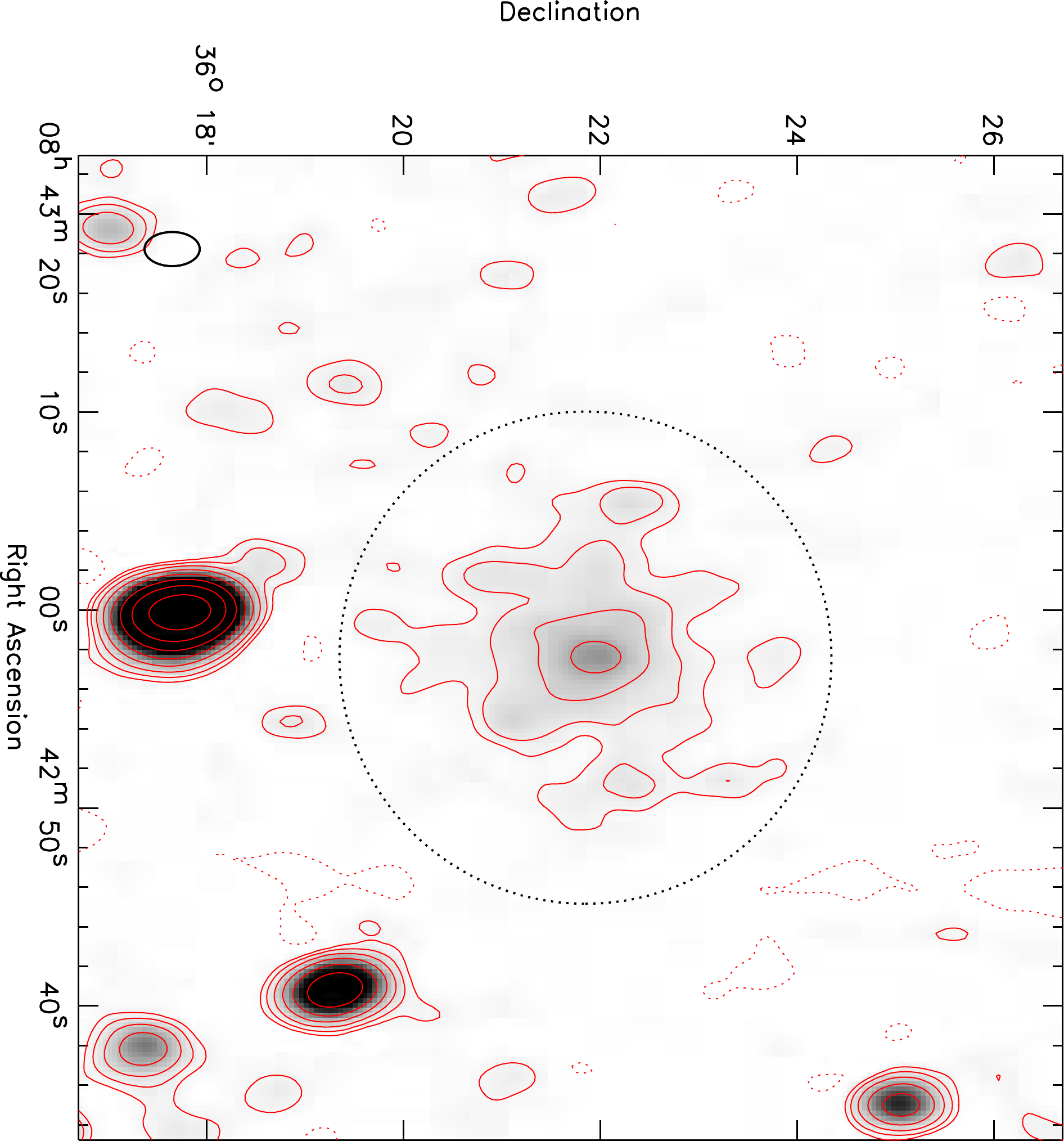}
\includegraphics[angle =90, trim =0cm 0cm 0cm 0cm,width=0.3\textwidth]{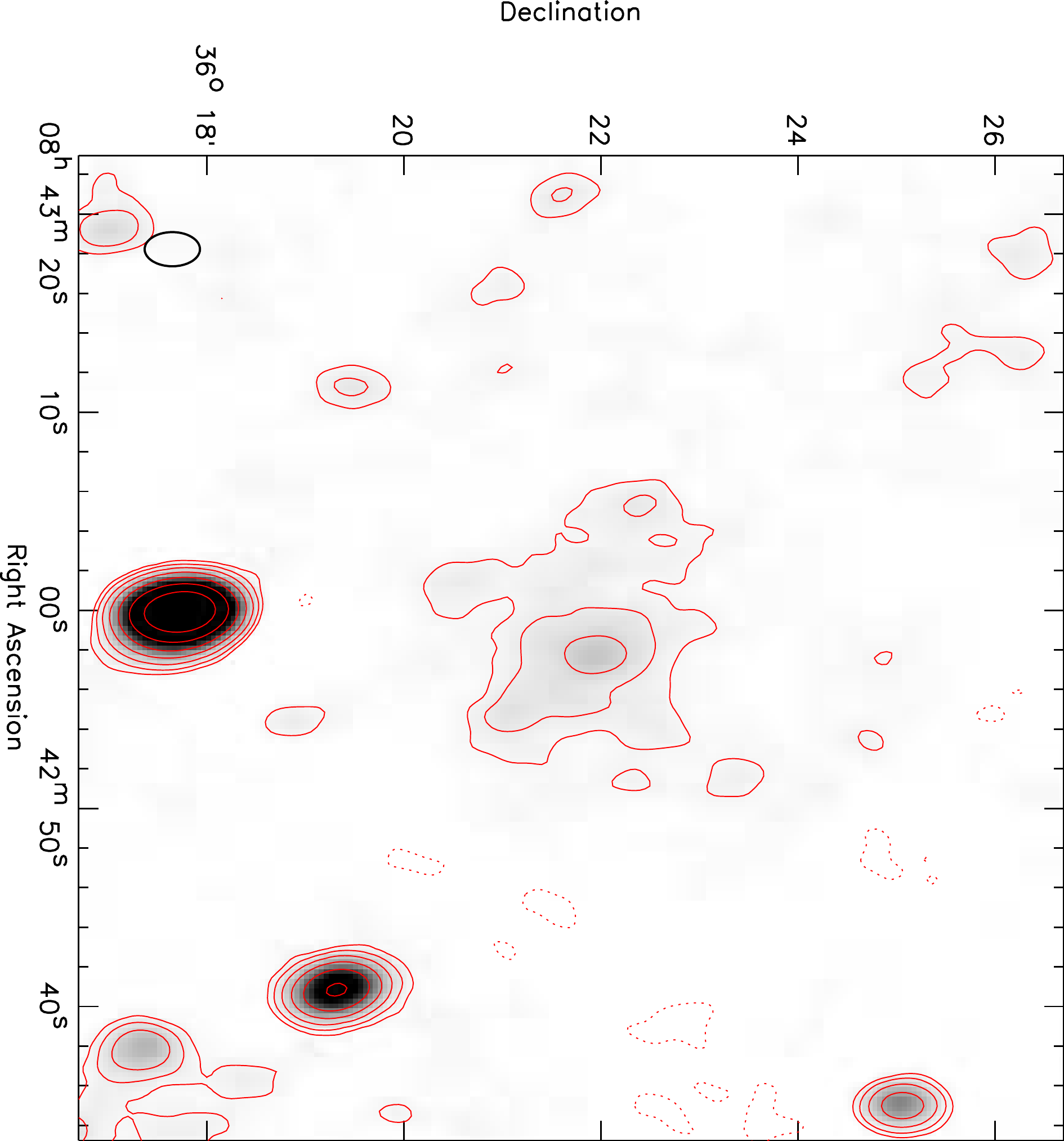}
\includegraphics[angle =90, trim =0cm 0cm 0cm 0cm,width=0.35\textwidth]{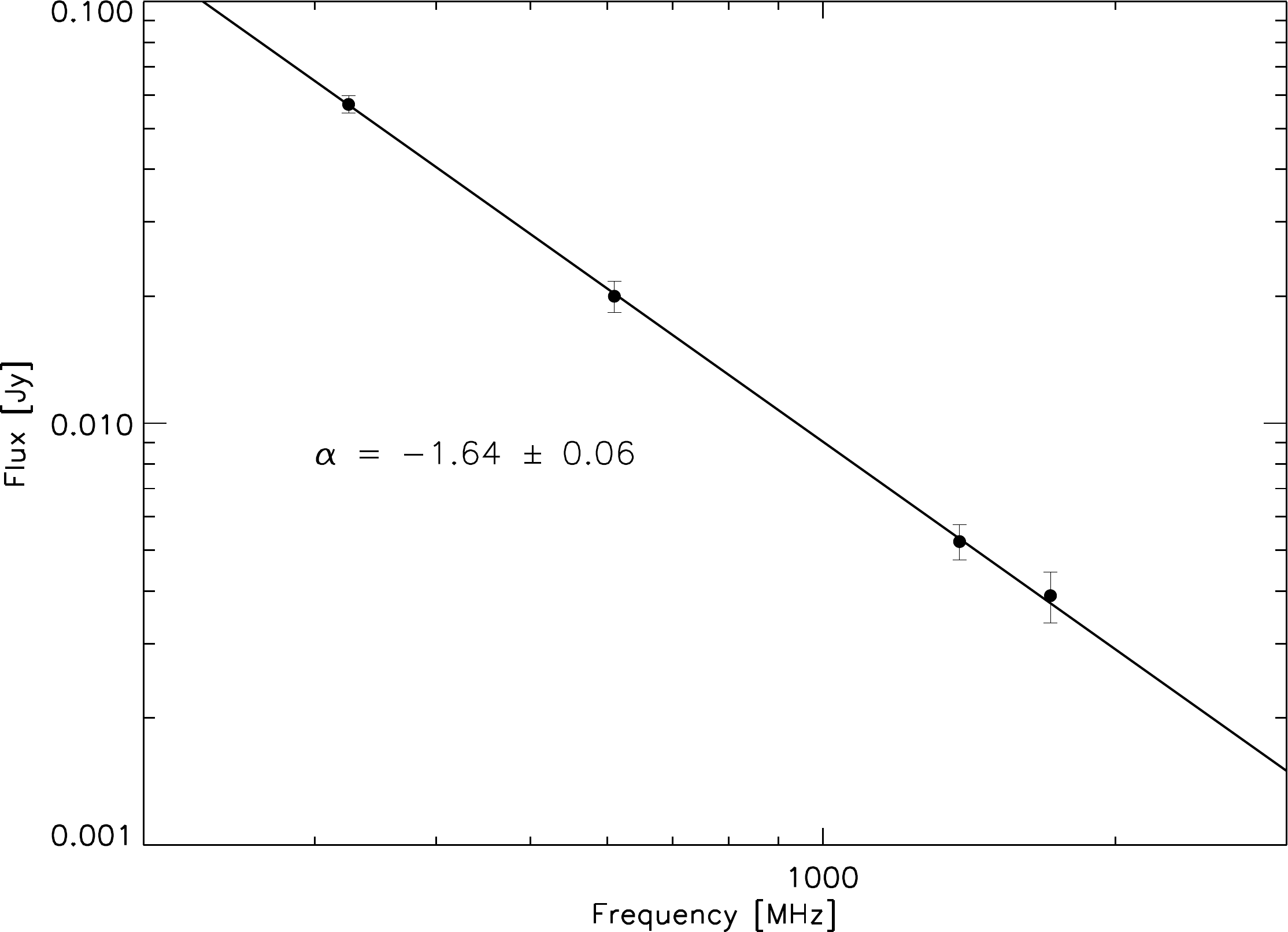}
\end{center}
\caption{Left: WSRT 1382~MHz image of Abel~697. Contour levels are drawn as in Fig.~\ref{fig:a1612_gmrt325}. Black dotted lines indicate the integration area for the flux measurements. Middle: WSRT 1714~MHz image of Abell~697. Contour levels are drawn as in Fig.~\ref{fig:a1612_gmrt325}. Right: A697 radio halo spectrum. Flux measurements at 325 and 610~MHz are taken from \cite{2010A&A...517A..43M}. }
\label{fig:a697wsrt21cm}
\end{figure*}

The discrete sources D, E, and F (see Fig.~\ref{fig:a697_halo}) were not detected with a SNR~$>3$. By using the positions from the 1382~MHz image,  we could still get an approximate estimate of their fluxes. We subtracted the flux from the discrete sources (Table~\ref{tab:compact}) and took the fluxes reported by \cite{2010A&A...517A..43M} to fit a power-law radio spectrum through the flux measurement with ${\alpha =-1.64 \pm 0.06}$, see Fig.~\ref{fig:a697wsrt21cm}. This confirms that the radio halo has a very steep spectral index, although it is marginally flatter than the $-1.7$ to $-1.8$ reported by \cite{2010A&A...517A..43M}.

We do not detect any polarized flux from the radio halo in our WSRT observations.  We set an upper limit on the radio halo polarization fraction of 6\% at 1382~MHz.

\subsection{Abell 2061}
\begin{figure}
\begin{center}
\includegraphics[angle =90, trim =0cm 0cm 0cm 0cm,width=0.5\textwidth]{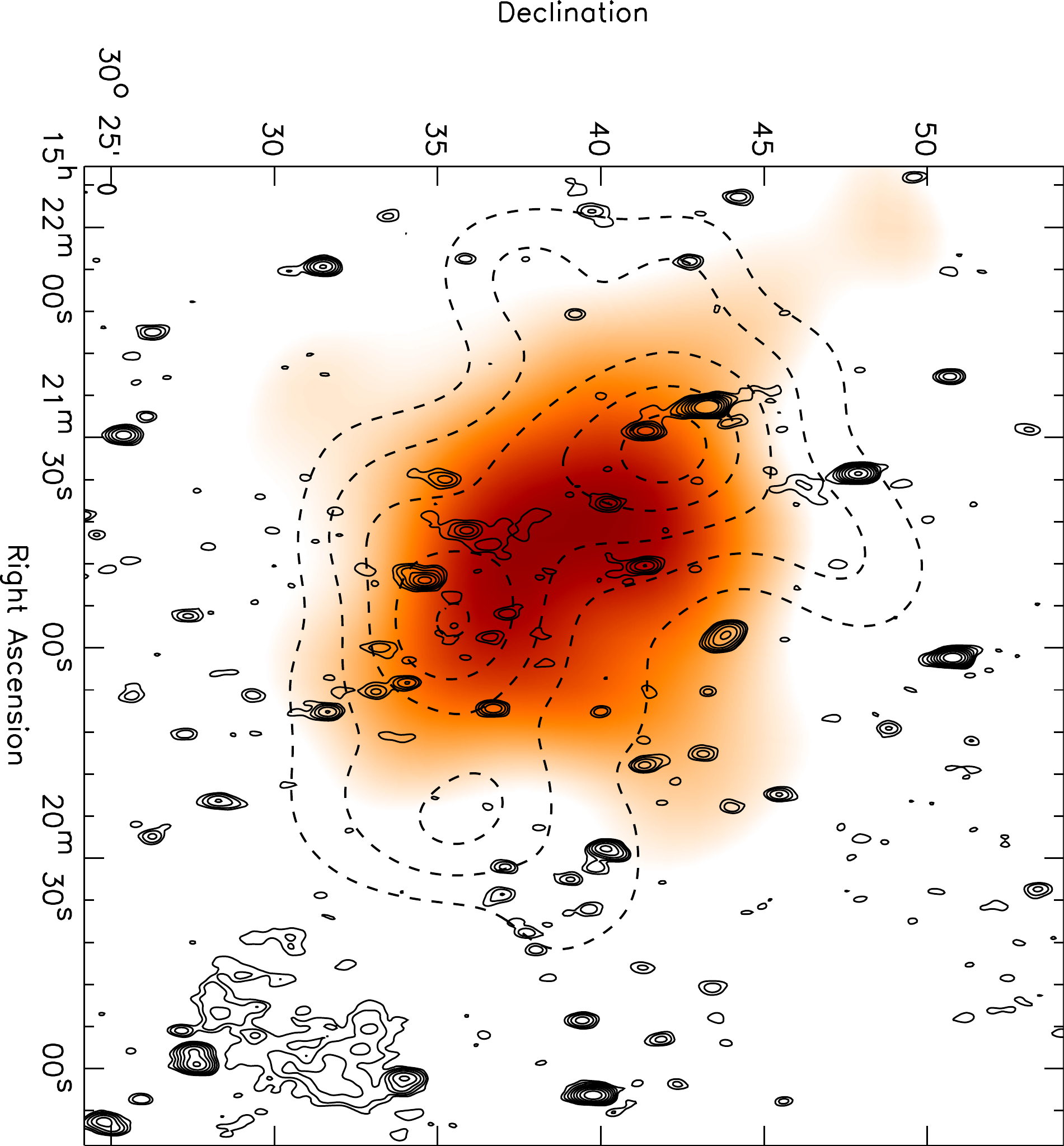}
\end{center}
\caption{A2061 X-ray emission from ROSAT in orange. The original image from the ROSAT All Sky Survey was convolved with a 225\arcsec~FWHM Gaussian. Solid contours are from the WSRT 1382~MHz image and drawn at levels of ${[1, 2, 4,  8, \ldots]} \times 4\sigma_{\mathrm{rms}}$.  Dashed contours show the galaxy iso-density distribution derived from the SDSS survey. Contours are drawn at ${[1.0,1.4, 1.8, \ldots]}  \times 0.55$ galaxies arcmin$^{-2}$ selecting only galaxies with  $0.05 < z_{\rm{phot}} <  0.1$.}
\label{fig:a2061_xray}
\end{figure}

Abell 2061 is a Bautz-Morgan (B/M) type III cluster located 
at $z=0.0784$. The cluster has a X-ray luminosity of 
$L_{\rm{X,~0.1-2.4~keV}}=3.95 \times 10^{44}$~erg~s$^{-1}$ \citep{1998MNRAS.301..881E}. 
The cluster has a  velocity dispersion of 673$^{+49}_{\rm{-40}}$~km~s$^{-1}$ \citep{2001AJ....122.2858O}. 
The ROSAT image shows the cluster to be extended in the 
northeast-southwest direction, see Fig.~\ref{fig:a2061_xray}. Previous studies report that the cluster is part of the Corona Borealis supercluster 
\citep{1998ApJ...492...45S,2004MNRAS.353.1219M}.
The galaxy iso-density 
contours display three substructures located roughly
along a northeast southwest axis.

Abell~2067 is located at a projected distance 
of only 1.8~$h^{-1}$~Mpc north from Abell~2061, and  
the cluster's systematic velocities are separated by $\sim 1600$~km~s~$^{-1}$ 
\citep{2001AJ....122.2858O}. Therefore, they probably form a bound 
system \citep{2004MNRAS.353.1219M, 2006AJ....132.1275R} consisting of 
a massive cluster (A2061) with a smaller in-falling group/cluster (A2067). 
A2061 also contains an X-ray extension in the direction of 
A2067 (towards the north-east) which also suggests a dynamical connection 
between the two systems \citep{2004MNRAS.353.1219M}. According 
to \citeauthor{2004MNRAS.353.1219M}, the interaction between the clusters is in the 
phase in which the cores have not yet met and in which the formation 
of a shock is expected.   
The global temperature for A2061 is reported to be $4.53^{+0.48}_{-0.38}$~keV 
from BeppoSax observations \citep{2004MNRAS.353.1219M}. 
A region with a higher temperature is found in the northern 
part of A2061 with a temperature of $10.67^{+3.90}_{-2.47}$~keV. 
This region could correspond to the presence of an internal shock. 

\cite{2001ApJ...548..639K} reported a possible relic in the 
southwest periphery of A2061 in WENSS images. 
They found a flux of $104\pm15$ and $19 \pm 3$~mJy at 327~MHz 
and 1.4~GHz, respectively. This would give a spectral 
index of $\alpha= -1.17 \pm 0.23$. 
\cite{2009ApJ...697.1341R} also listed the presence of 
the diffuse peripheral source and measured a flux of 120~mJy in the WENSS image. 
They also claimed the presence of additional diffuse emission 
in the center of the cluster which could make up a radio halo.

\begin{figure*}
\begin{center}
\includegraphics[angle =90, trim =0cm 0cm 0cm 0cm,width=0.48\textwidth]{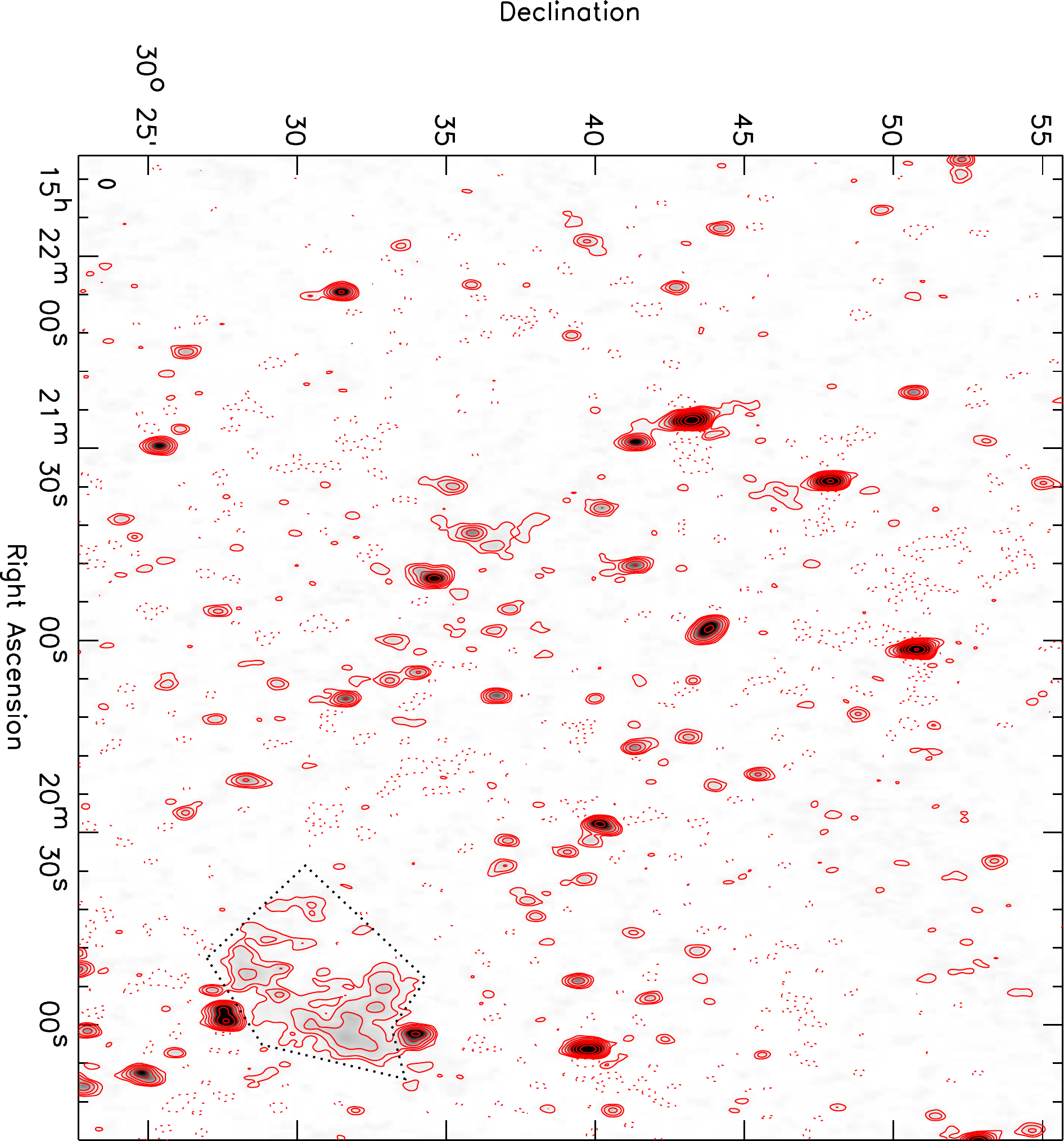}
\includegraphics[angle =90, trim =0cm 0cm 0cm 0cm,width=0.48\textwidth]{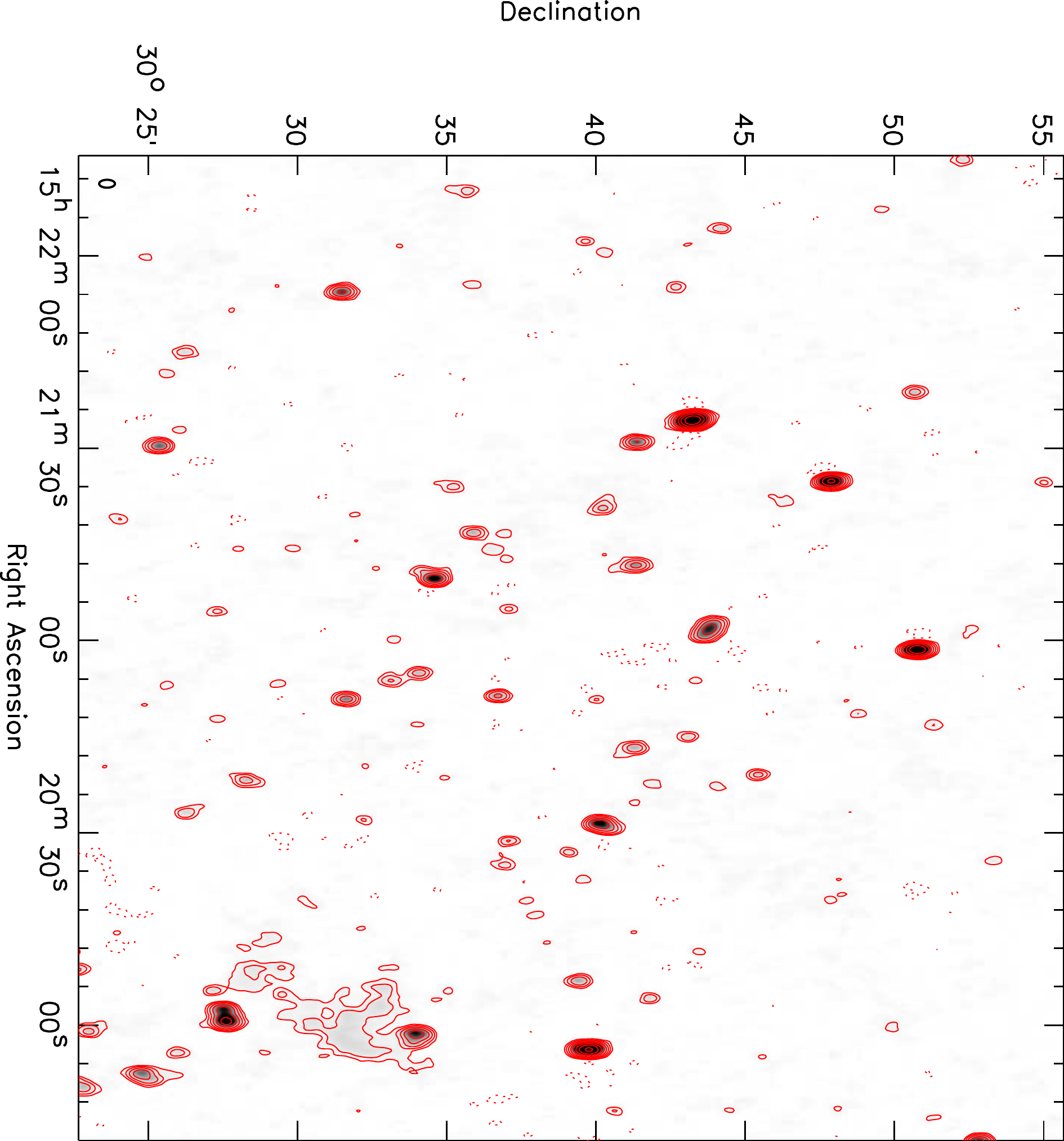}
\end{center}
\caption{Left: A2061 WSRT 1382~MHz image. Contour levels are drawn as in Fig.~\ref{fig:a1612_gmrt325}. Black dotted lines indicate the integration area for the flux measurements. Right: A2061 WSRT 1714~MHz image. Contour levels are drawn as in Fig.~\ref{fig:a1612_gmrt325}.}
\label{fig:a2061wsrt21cm}
\end{figure*}

With our WSRT observations we confirm the presence of the diffuse radio source in 
the southwestern periphery of A2061 
and we also classify it as radio relic. We cannot confirm the presence of a radio halo. 
The radio relic is clearly seen in both the 21 and 18~cm WSRT 
images, Fig.~\ref{fig:a2061wsrt21cm}.  
The relic is located at a distance of  
1.5~ Mpc from the cluster center and has a largest angular 
extent of 7.7\arcmin, corresponding to a physical size of 675~kpc. 
In the direction towards the cluster center the relic has an 
extent of $\sim 320$~kpc. 
The western outer boundary of the relic is somewhat more pronounced, while the 
emission fades more slowly inwards to the cluster center. 
The relic consist of a northern and fainter southern component. 
Two compact sources are found directly to the north and south of the relic. 

\begin{figure}
\begin{center}
\includegraphics[angle =90, trim =0cm 0cm 0cm 0cm,width=0.5\textwidth]{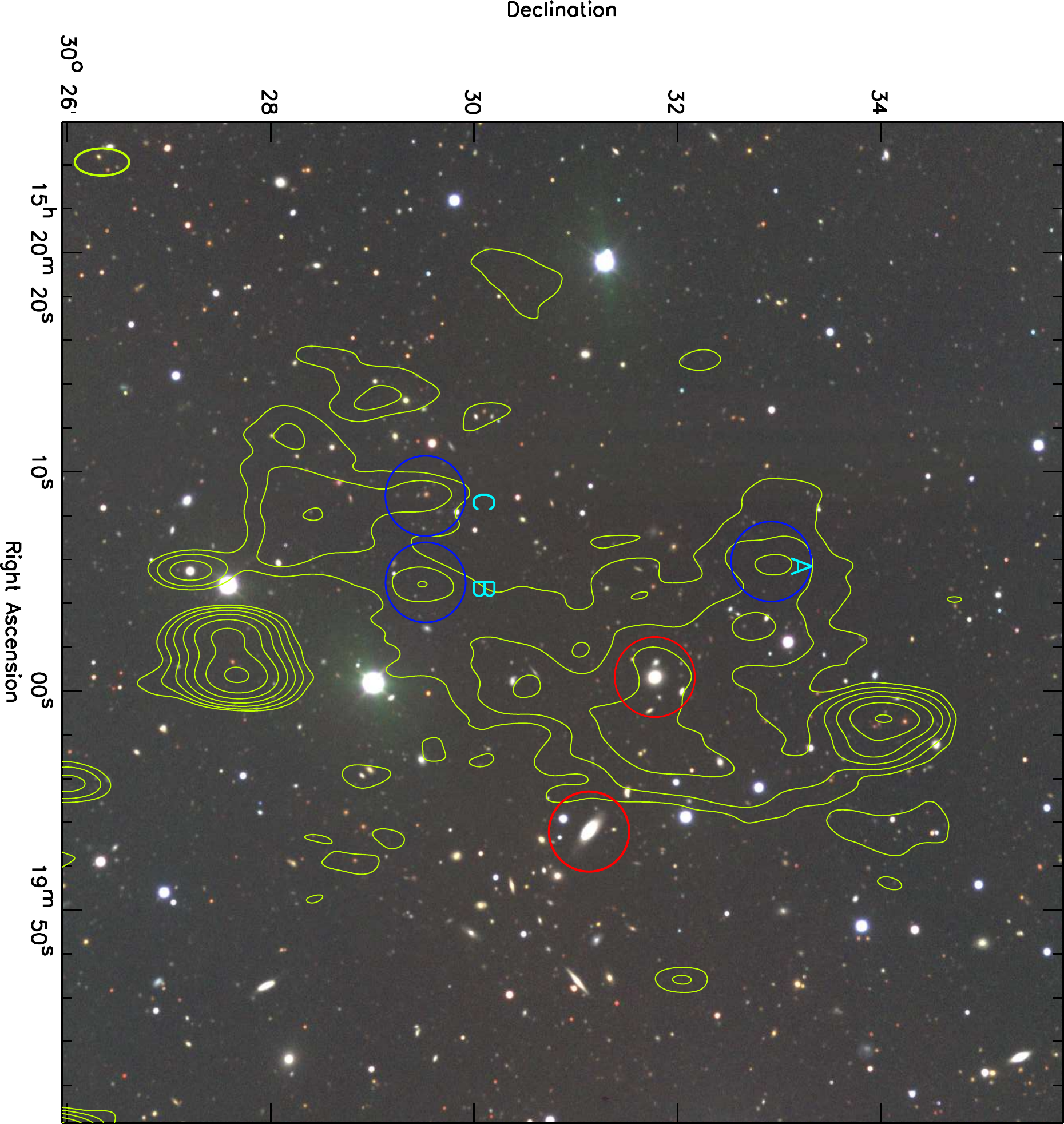}
\end{center}
\caption{A2061 optical WHT V, R, I color image at the location of the radio relic. Radio contours are from Fig.~\ref{fig:a2061wsrt21cm}. The two red circles indicate galaxies with spectroscopic redshifts of 0.0815 (west) and 0.061 \citep{2009ApJS..182..543A}. Discrete sources embedded in the diffuse emission are alphabetically labeled, see Table~\ref{tab:compact}.}
\label{fig:a2061_optical}
\end{figure}

An optical color image at the location of 
the relic is shown in Fig.~\ref{fig:a2061_optical}. This image does
not reveal any obvious counterparts to the radio relic. One galaxy
is located at the brightest region of the relic, but no compact
radio source is associated with that galaxy. The bright compact sources
to the north and south of the relic are both associated with
background galaxies unrelated to A2061 because of their small angular sizes ($\lesssim 3\arcsec$) and very red color. For comparison, we have marked two galaxies located approximately at the distance of A2061 (Fig.~\ref{fig:a2061_optical}).

\begin{figure}
\begin{center}
\includegraphics[angle =90, trim =0cm 0cm 0cm 0cm,width=0.5\textwidth]{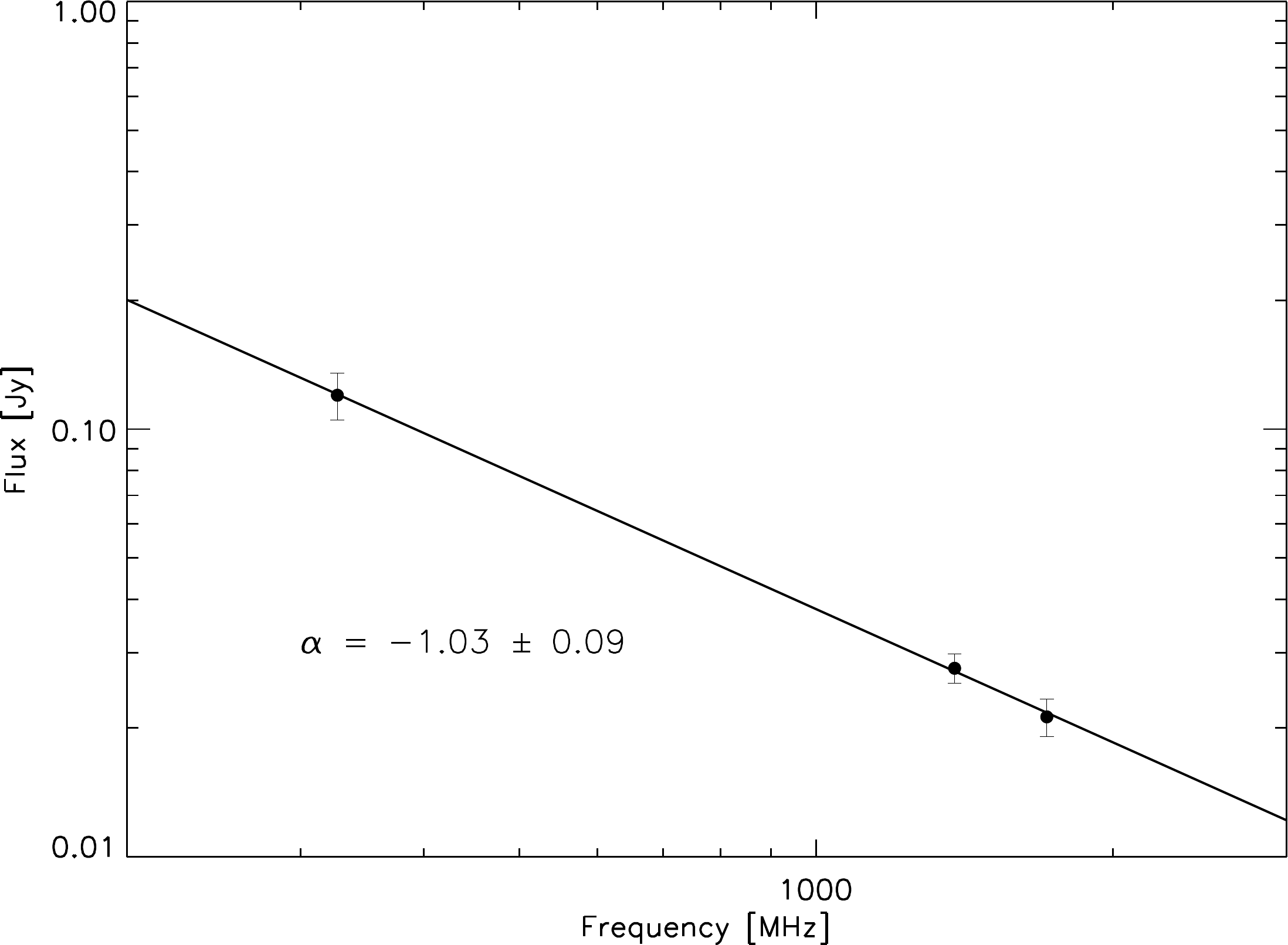}
\end{center}
\caption{A2061 radio relic spectrum. Flux measurement at 325~MHz is taken from \cite{2009ApJ...697.1341R}.}
\label{fig:a2061_flux}
\end{figure}

The radio spectrum is fitted by a single power-law 
spectrum, see Fig.~\ref{fig:a2061_flux}. We find $\alpha = -1.03 \pm 0.09$. 
The 325~MHz WENSS flux is from \citeauthor{2009ApJ...697.1341R}. 
The 325~MHz measurement is however uncertain as the SNR on the relic is low. 
To better constrain the spectral index an additional low-frequency flux measurement is needed.

The picture that emerges from our observations is that of a radio relic 
tracing a shock wave from a cluster merger event. The merger event does not seem to be directly related 
to the cluster A2067, located north of A2061. Instead the shock wave 
is more likely related to the observed substructures seen in the SDSS galaxy iso-density 
contours.

\subsection{Abell 3365}
\begin{figure*}
\begin{center}
\includegraphics[angle =90, trim =0cm 0cm 0cm 0cm,width=0.49\textwidth]{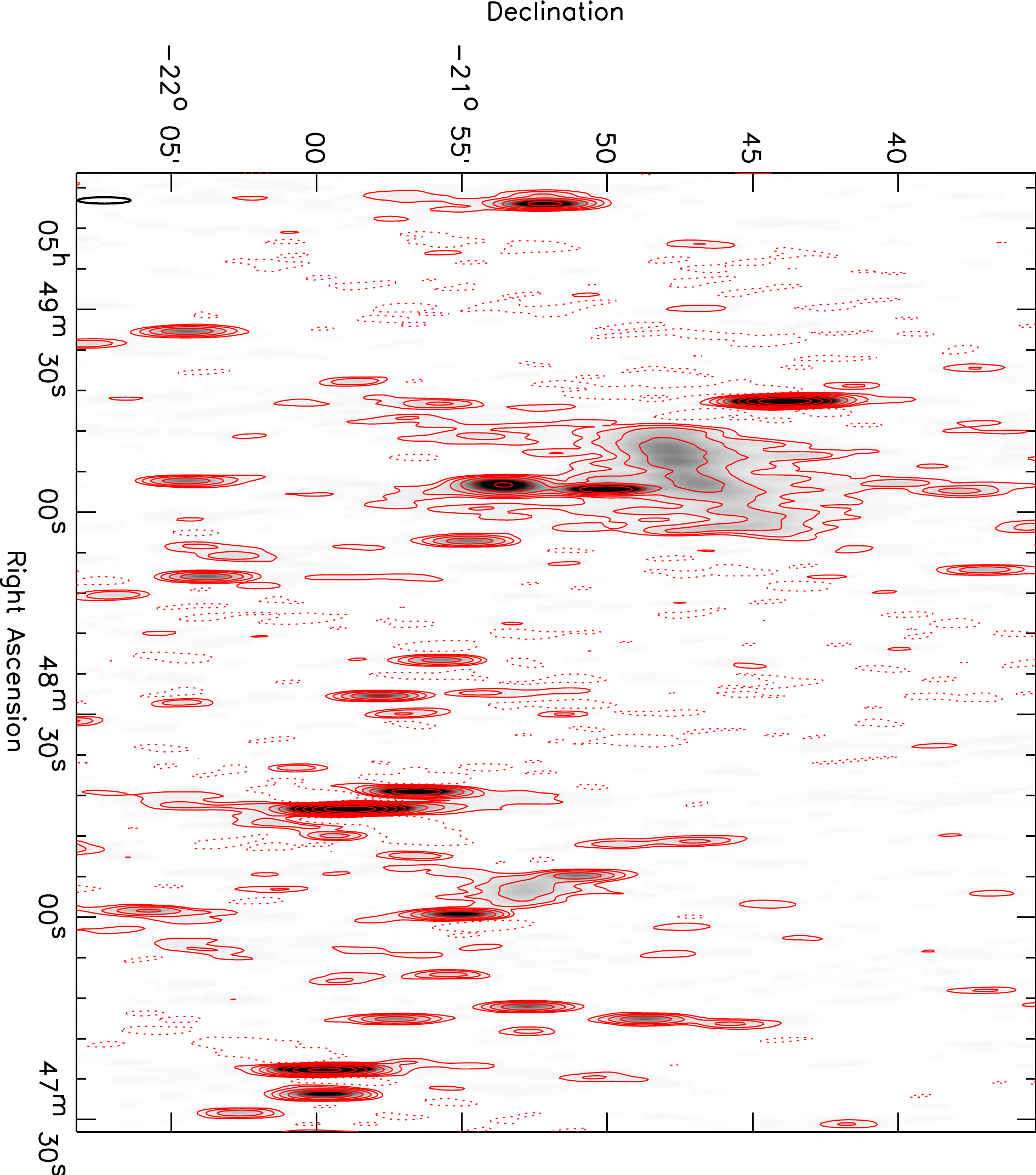}
\includegraphics[angle =90, trim =0cm 0cm 0cm 0cm,width=0.49\textwidth]{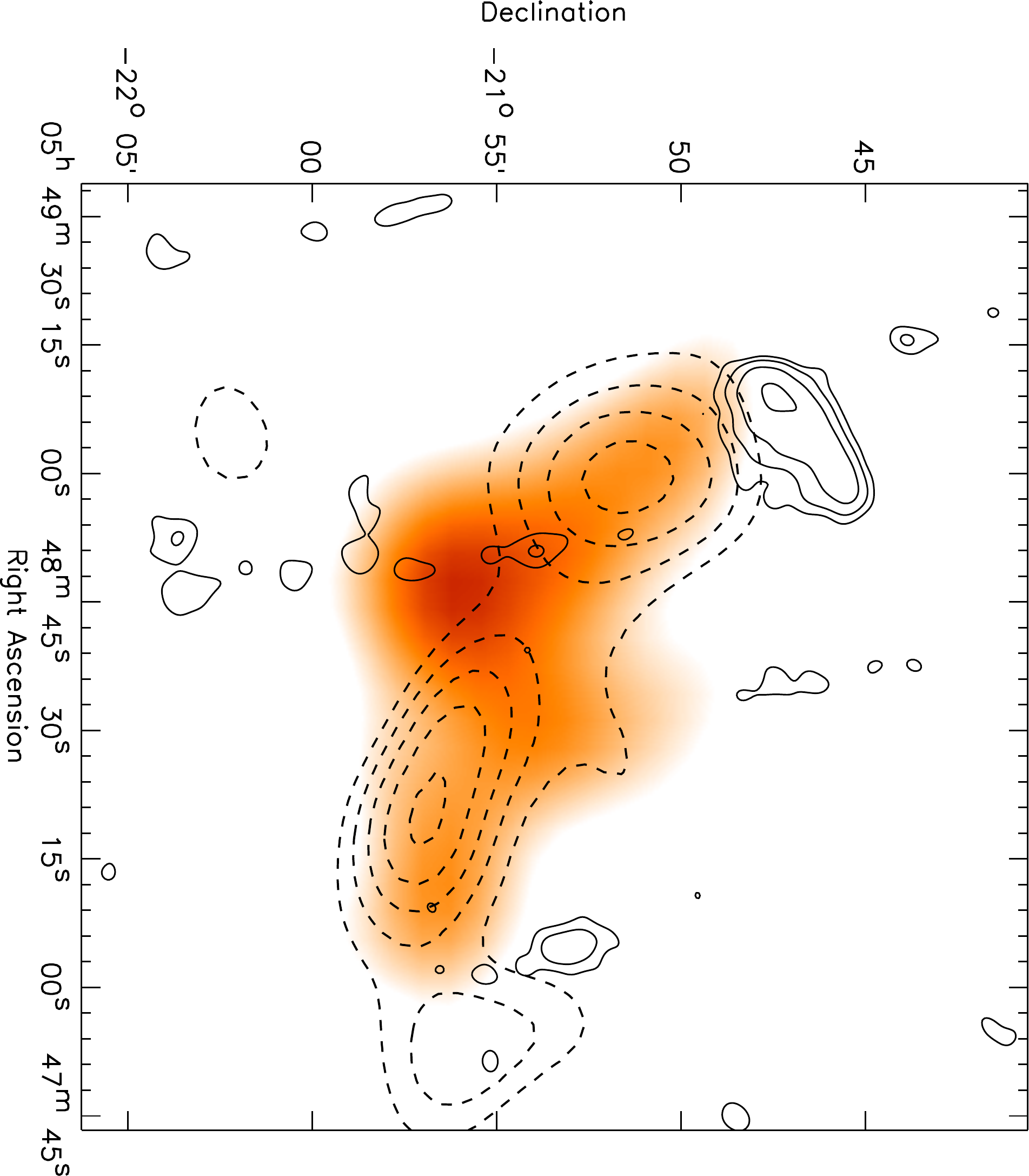}
\end{center}
\caption{Left: A3365 WSRT 1382~MHz image. Contour levels are drawn as in Fig.~\ref{fig:a1612_gmrt325}. Right: A3365 X-ray emission from ROSAT in orange. The original image from the ROSAT All Sky Survey was convolved with a 225\arcsec~FWHM Gaussian. Solid contours are from a VLA 1.4~GHz image with compact sources subtracted from the uv-data. This DnC array image has a resolution of $47\arcsec~\times~42\arcsec$. Contours are drawn at levels of ${[1, 2, 4, 8,  \ldots]} \times 0.4$~mJy~beam$^{-1}$.  Dashed contours show the galaxy iso-density distribution derived from INT images. Contours are drawn at ${[1.0,1.1, 1.2, \ldots]}  \times 0.78$ galaxies arcmin$^{-2}$ selecting only galaxies with colors  $0.6 < \rm{V-R} < 0.9$,  $ 0.54 < \rm{R-I} < 0.84$, i.e., within 0.15 magnitudes the V--R and R--I color of the largest cD galaxy.}
\label{fig:rxc_xray}
\end{figure*}

Abell~3365 \citep{1989ApJS...70....1A} is located at a 
redshift of $z=0.0926$ \citep{1999ApJS..125...35S}. The cluster is listed as a 
Bautz-Morgan (B/M) type II cluster. The ESO Nearby Abell 
Cluster Survey \citep[ENACS,][]{1996A&A...310....8K} measured a velocity dispersion 
of 1153~km~s$^{-1}$ \citep{1996A&A...310...31M} for A3365,
quite high compared to other clusters in the sample. Galaxy cluster \object{RXC~J0548.8-2154} 
is located 9\arcmin~to the west of the NED listed position for A3365 at a redshift of $z=0.0928$. 
The galaxy distribution, see Fig.~\ref{fig:rxc_xray} (right panel), peaks at the center of A3365, while the 
X-ray peak is located at the position of RXC~J0548.8-2154. 
The galaxy distribution around A3365 is complex, with two main concentrations along 
an east-west axis and a smaller concentration at the far west. The ICM distribution is also complex
and elongated in the east-west direction. Based on the very similar redshift, and complex X-ray emission and galaxy distribution, we conclude that A3365 and RXC~J0548.8-2154 belong to the same merging system to which we will simply refer as A3365.

\begin{figure*}
\begin{center}
\includegraphics[angle =90, trim =0cm 0cm 0cm 0cm,width=0.49\textwidth]{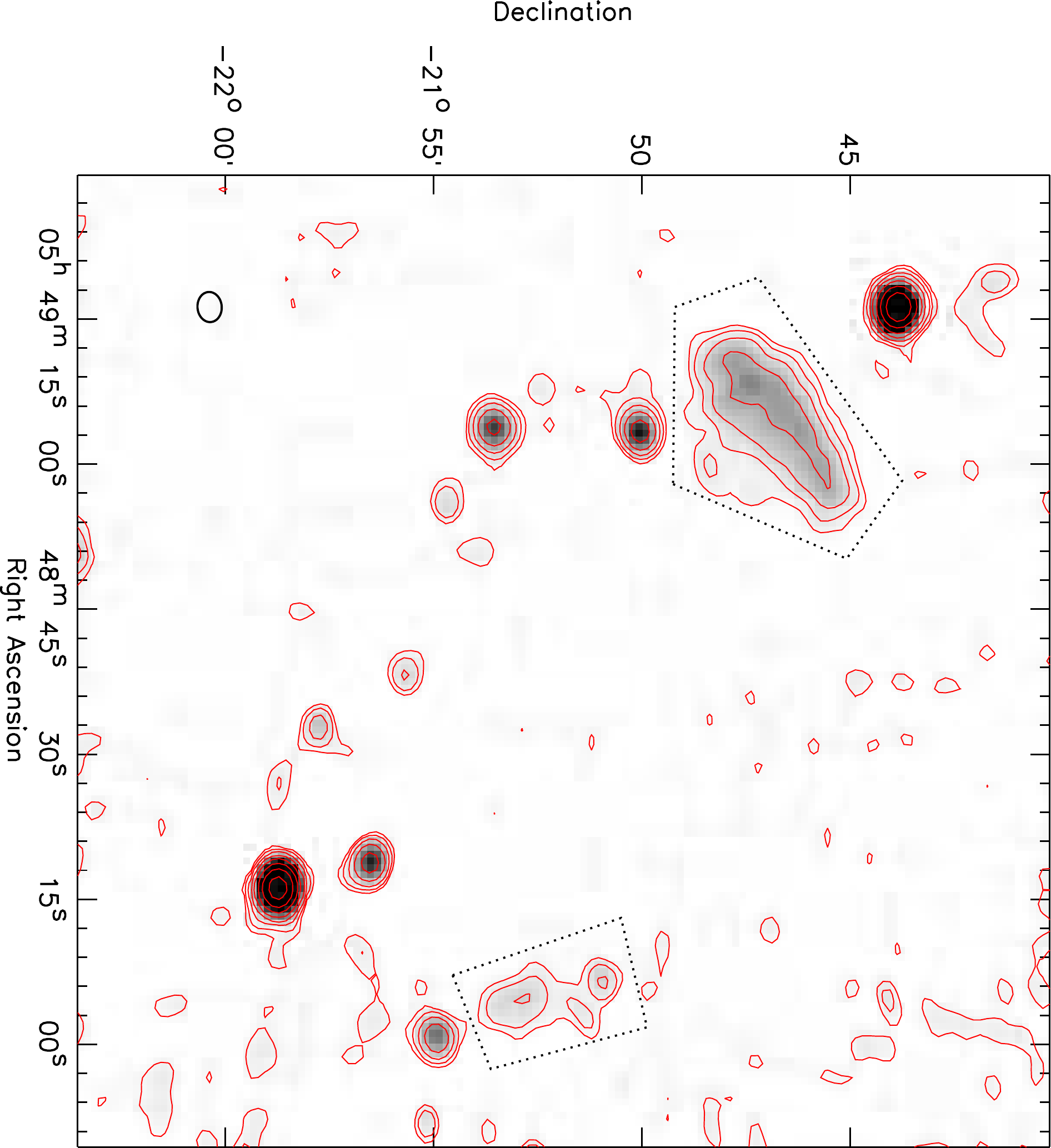}
\includegraphics[angle =90, trim =0cm 0cm 0cm 0cm,width=0.49\textwidth]{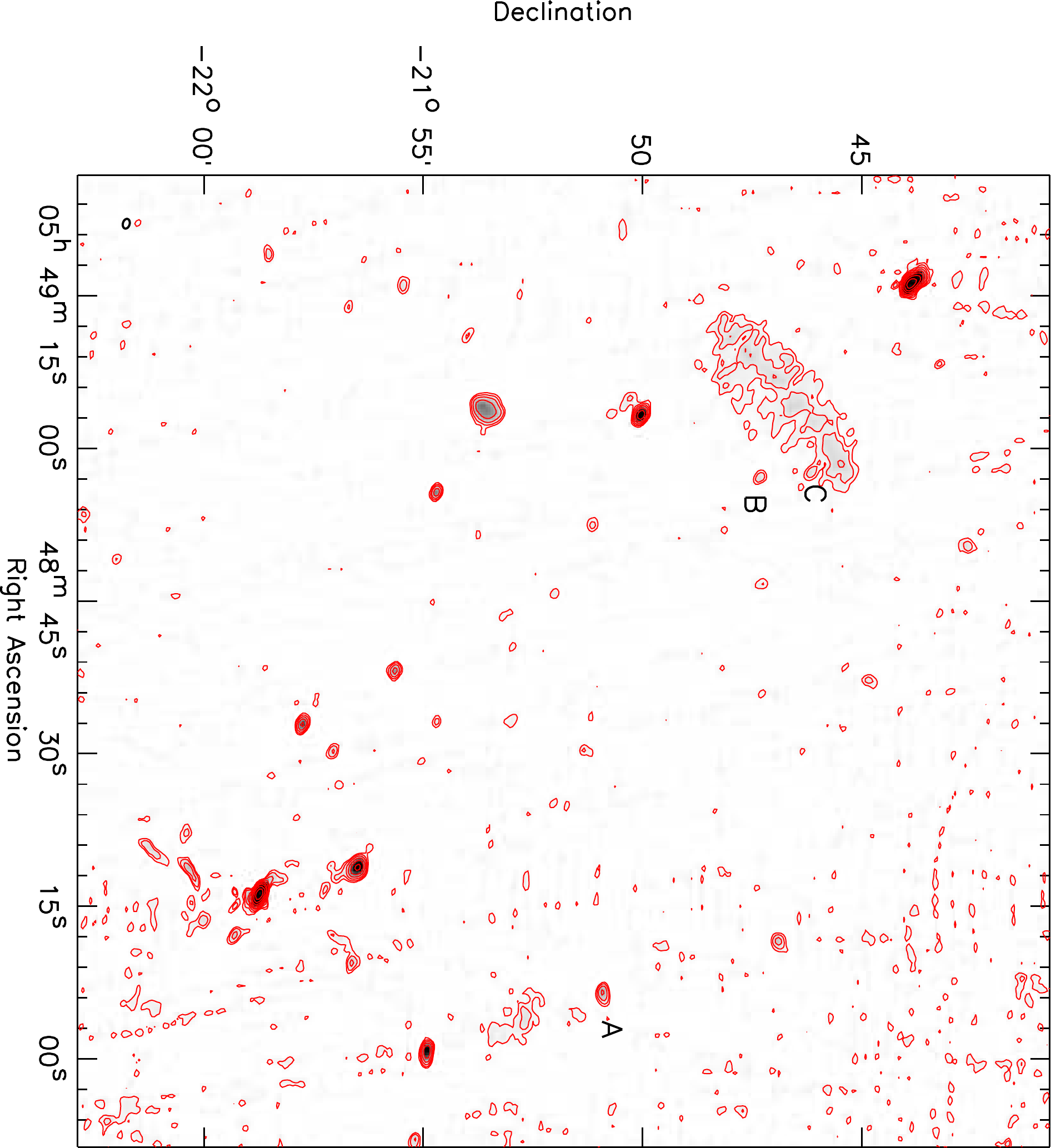}
\end{center}
\caption{A3365 VLA 1.4~GHz DnC (left) and CnB (right) array images. Contour levels are drawn as in Fig.~\ref{fig:a1612_gmrt325}. Black dotted lines indicate the integration areas for the flux measurements. Discrete sources embedded in the diffuse emission are alphabetically labeled, see Table~\ref{tab:compact}.}
\label{fig:a3365_vla}
\end{figure*}

\begin{figure*}
\begin{center}
\includegraphics[angle =90, trim =0cm 0cm 0cm 0cm,width=0.49\textwidth]{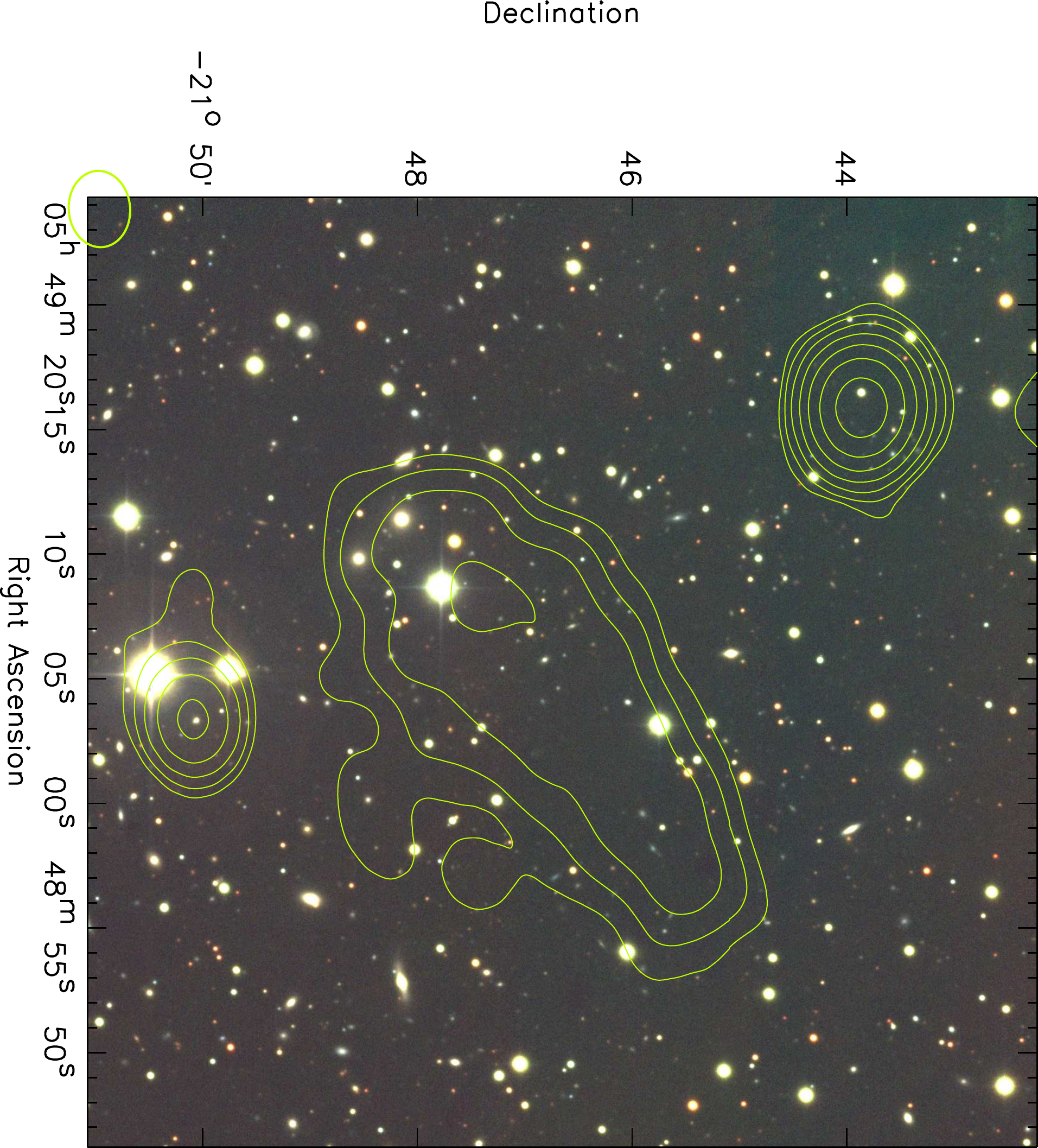}
\includegraphics[angle =90, trim =0cm 0cm 0cm 0cm,width=0.49\textwidth]{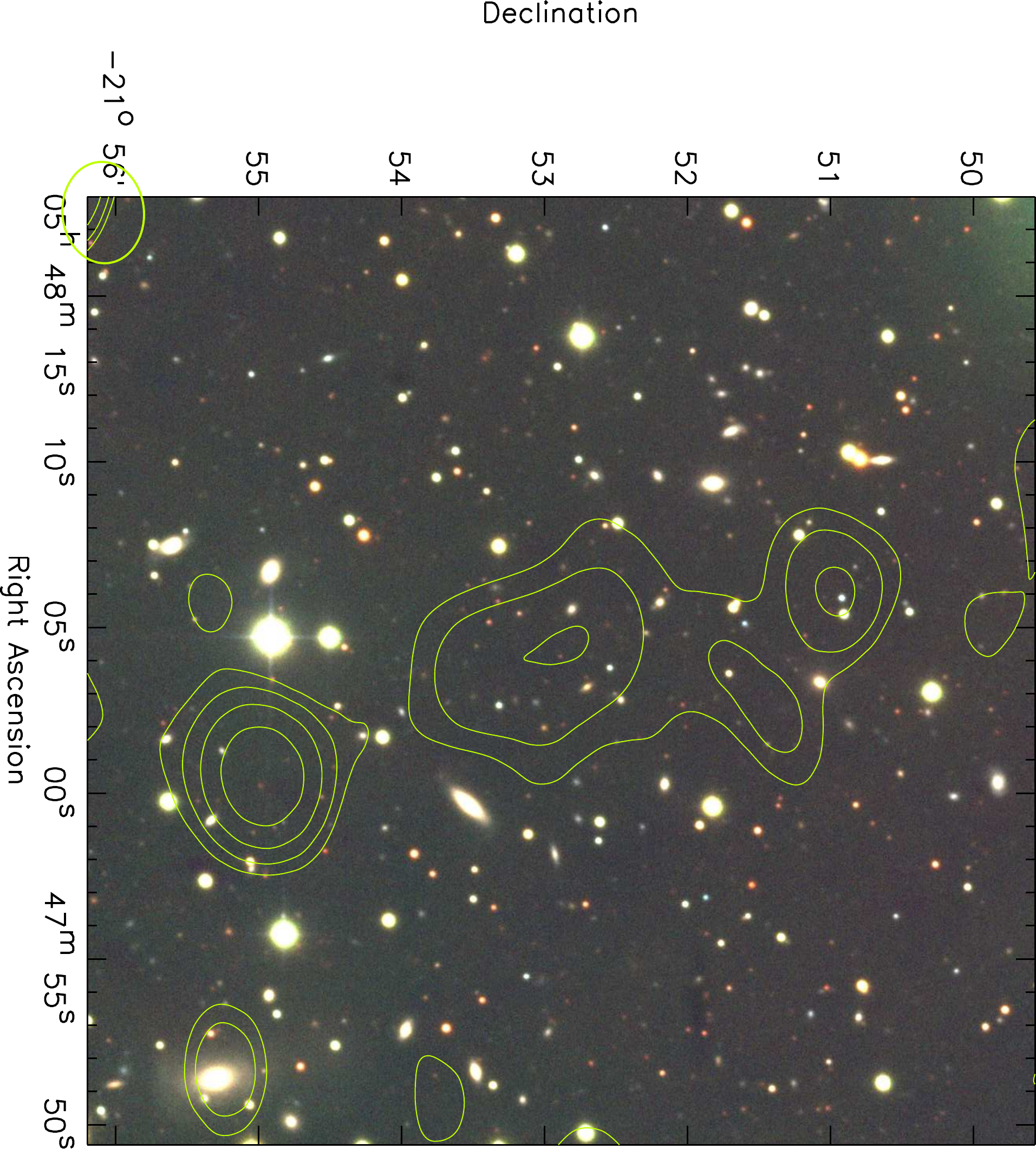}
\end{center}
\caption{A3365 optical INT V, R, I color images at the location of the radio relics. Radio contours are from Fig.~\ref{fig:a3365_vla} (right panel).}
\label{fig:a3365_optical}
\end{figure*}

In the NVSS image we noted the presence of a peripheral elongated radio source to the east of the cluster center. The WSRT image (Fig.~\ref{fig:rxc_xray}, left panel) reveals a second smaller diffuse source on the west side of the cluster. The diffuse sources are also detected in the VLA images (Fig.~\ref{fig:a3365_vla}). 
For both sources we cannot identify obvious optical counterparts in our INT images (Fig.~\ref{fig:a3365_optical}).  The east and western sources have an angular extent of 5.5\arcmin~and 2.3\arcmin, which correspond to 560 and 235~kpc at the distance of A3365. We classify the eastern source as a radio relic. Most likely, the western source is another radio relic, which means that A3365 hosts a double radio relic system, but more observations are needed to confirm this classification. 
This interpretation is supported by the elongated X-ray and galaxy distribution which suggests a merger event along an east-west axis (Fig.~\ref{fig:rxc_xray} right panel). We also made an image where we subtracted the compact sources from the uv-data. To make this image, we subtracted the clean components from a uniformly weighted image (CnB array), from the DnC array data. This image is overlaid in Fig.~\ref{fig:rxc_xray}.

\subsection{CIZA J0649.3+1801}
\object{CIZA~J0649.3+1801} is a little studied galaxy 
cluster located at $z=0.064$ discovered by \cite{2002ApJ...580..774E} at 
a galactic latitude of $b=7.668\degr$. The cluster has a moderate X-ray luminosity of 
$L_{\rm{X,~0.1-2.4~keV}}=2.38 \times 10^{44}$~erg~s$^{-1}$. 
The cluster forms part of a supercluster in the Zone of 
Avoidance  hidden by the Milky Way \citep{2007ApJ...662..224K}.

\begin{figure*}
\begin{center}
\includegraphics[angle =90, trim =0cm 0cm 0cm 0cm,width=0.48\textwidth]{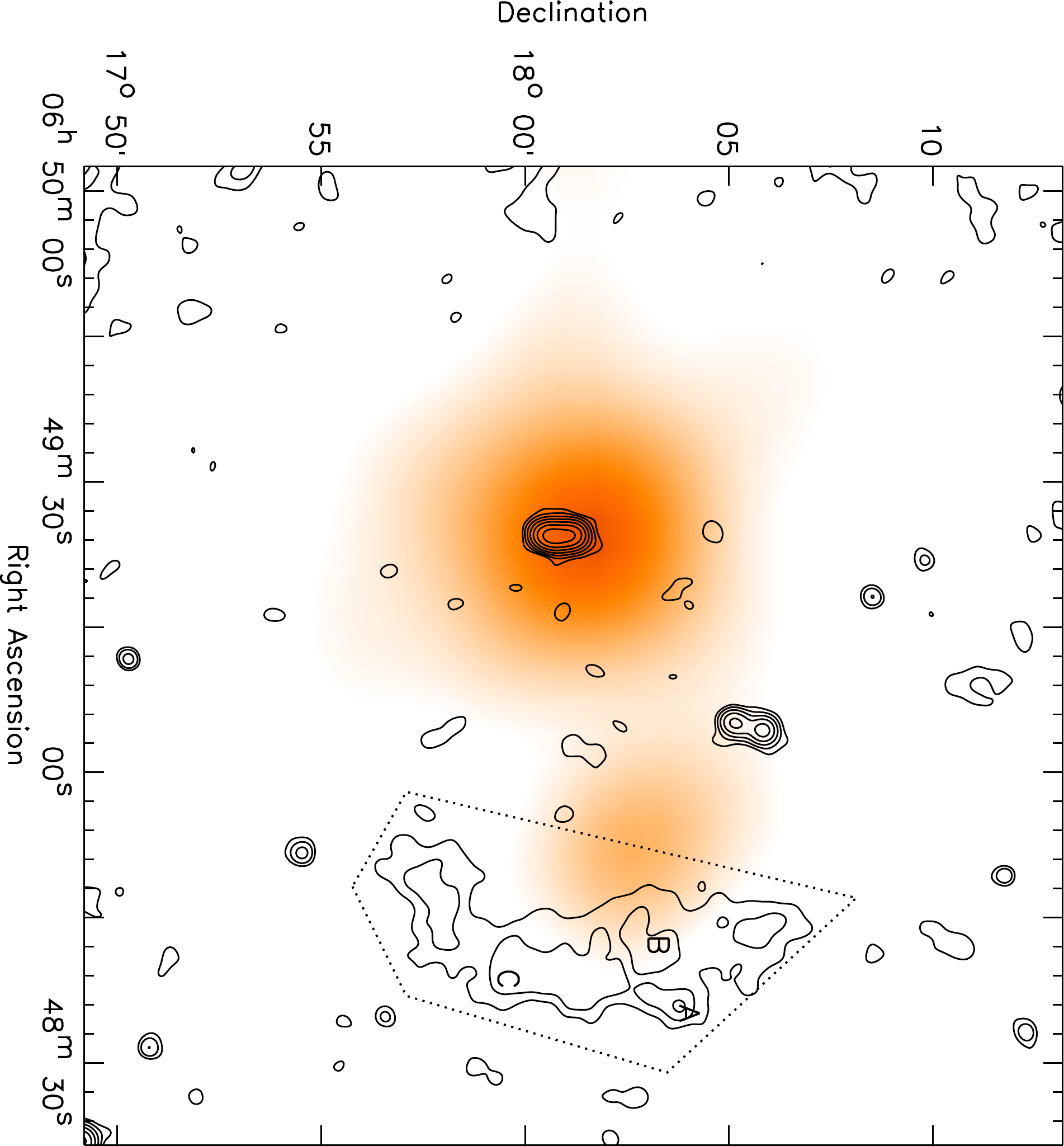}
\includegraphics[angle =90, trim =0cm 0cm 0cm 0cm,width=0.48\textwidth]{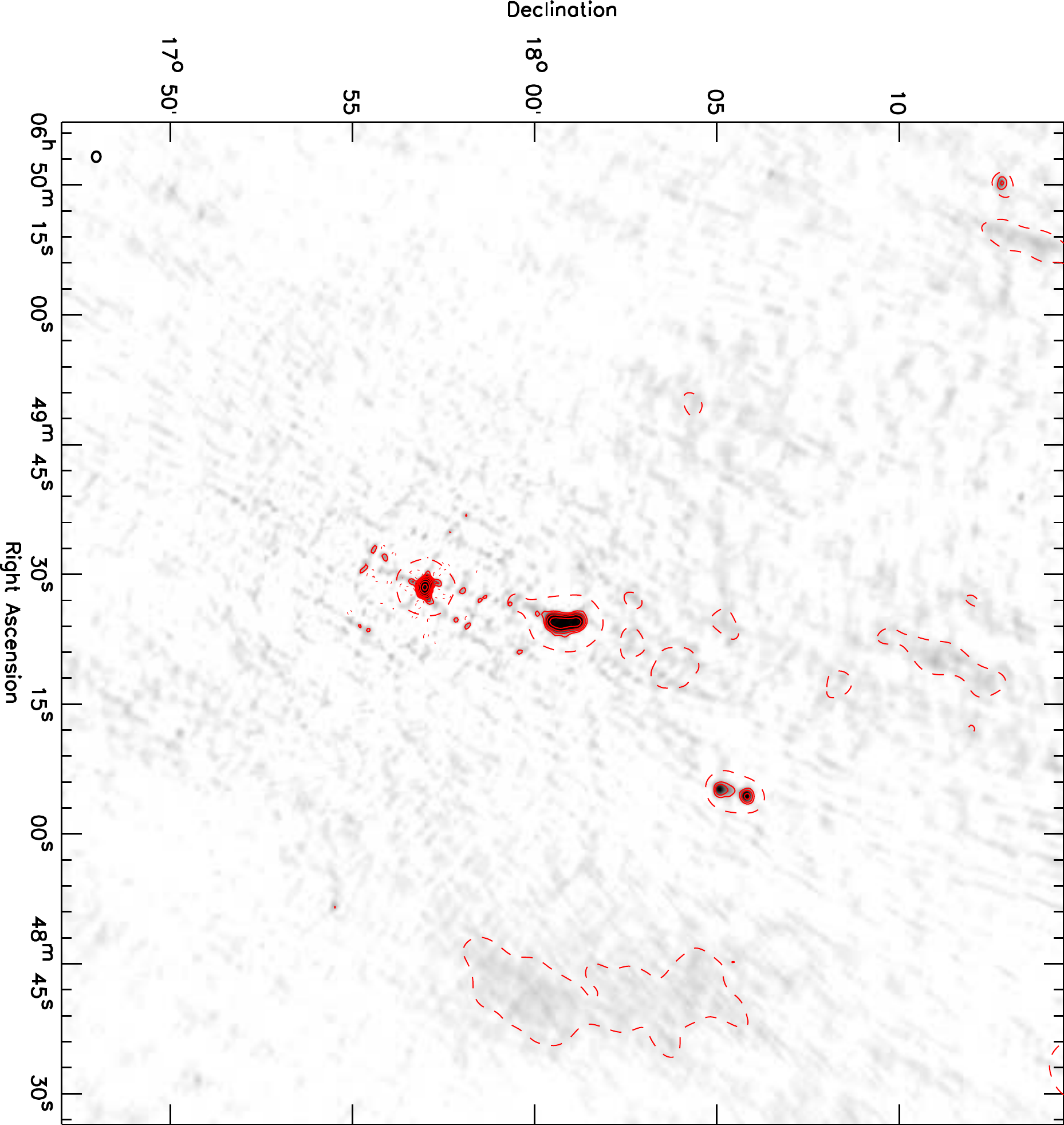}
\end{center}
\caption{Left: CIZA~J0649.3+1801 X-ray emission from ROSAT in orange. The original image from the ROSAT All Sky Survey was convolved with a 225\arcsec~FWHM Gaussian. Solid contours are from the GMRT 610~MHz image convolved to a resolution of 25\arcsec~and drawn at levels of ${[1, 2, 4, 8,  \ldots]} \times 4\sigma_{\mathrm{rms}}$. The source \object{NVSS~J064928+175700} was removed using the ``peeling''-method \citep[e.g.,][]{2004SPIE.5489..817N}. Black dotted lines indicate the integration area for the flux measurement. Discrete sources embedded in the diffuse emission are alphabetically labeled, see Table~\ref{tab:compact}. Right: CIZA~J0649.3+1801 GMRT 241~MHz image. Contour levels are drawn as in Fig.~\ref{fig:a1612_gmrt325}. The dashed line is the 15 mJy~beam$^{-1}$contour of the 241~MHz images convolved to a circular beam of 45\arcsec.}
\label{fig:rx18_xray}
\end{figure*}

We discovered the presence of a diffuse $\sim 10\arcmin$~elongated source to the west of the cluster in the NVSS survey. We do not find an optical counterpart for the source in our WHT images. In the GMRT image, see Fig.~\ref{fig:rx18_xray} (left panel), the source has a total extent of 10.6\arcmin~which corresponds to a size of about 800~kpc at the distance of CIZA~J0649.3+1801. We therefore classify the source as a peripheral radio relic located at 0.8~Mpc from the cluster center. The relic is also visible in the 241~MHz image, although the SNR on the relic is very low (Fig.~\ref{fig:rx18_xray} right panel).

The ROSAT image reveals another fainter X-ray source located between the radio relic and the cluster center. The source is not resolved and therefore it could be unrelated to the cluster. Also we do not detect any group or cluster of galaxies associated with this source in our optical WHT images. Therefore, this X-ray source seems to be unrelated to CIZA~J0649.3+1801.

\subsection{CIZA J0107.7+5408}
\begin{figure*}
\begin{center}
\includegraphics[angle =90, trim =0cm 0cm 0cm 0cm,width=0.49\textwidth]{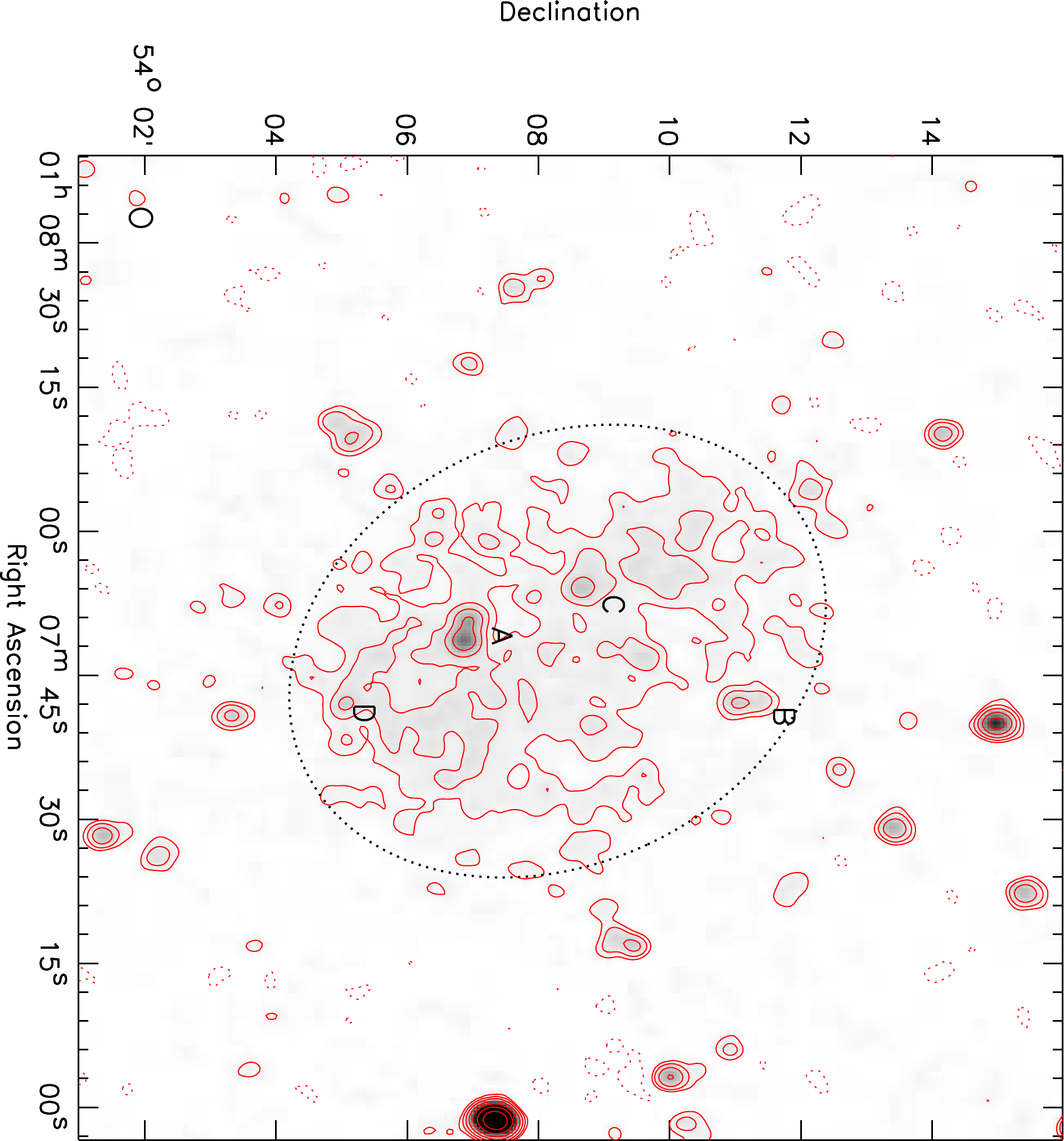}
\includegraphics[angle =90, trim =0cm 0cm 0cm 0cm,width=0.49\textwidth]{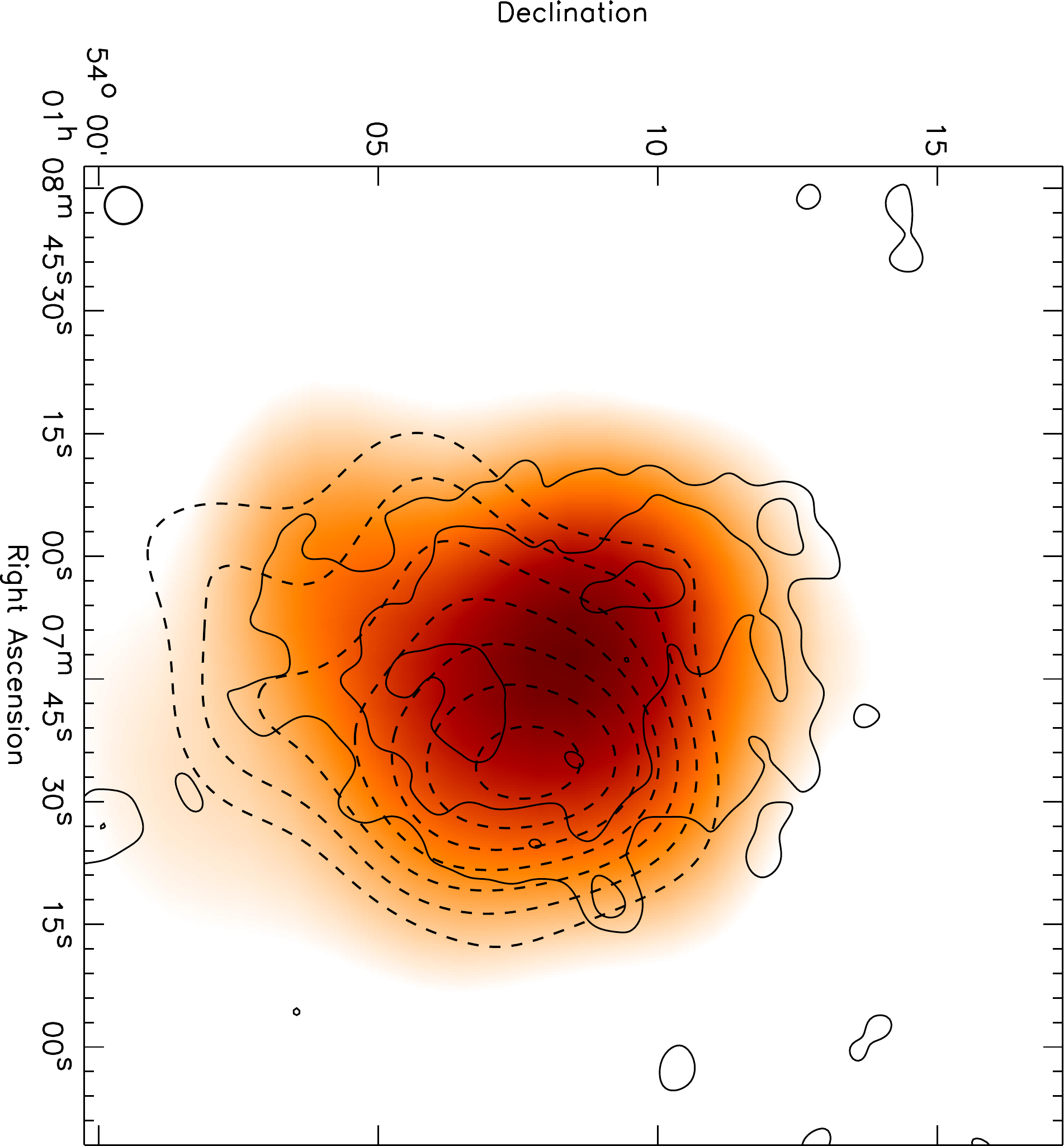}
\end{center}
\caption{Left: CIZA J0107.7+5408 WSRT 1382~MHz image. Contour levels are drawn as in Fig.~\ref{fig:a1612_gmrt325}.  Black dotted lines indicate the integration area for the flux measurement. Discrete sources embedded in the diffuse emission are alphabetically labeled, see Table~\ref{tab:compact}. Right: CIZA J0107.7+5408 X-ray emission from ROSAT in orange. The original image from the ROSAT All Sky Survey was convolved with a 225\arcsec~FWHM Gaussian. Solid contours are from the WSRT 1382~MHz image convolved to a resolution of 40\arcsec. Compact sources were subtracted and contours are drawn at levels of ${[1, 2, 4, 8, \ldots]} \times 0.3$~mJy~beam$^{-1}$.  Dashed contours show the galaxy iso-density distribution derived from INT images. Contours are drawn at ${[1.0,1.2, 1.4, \ldots]}  \times 0.3$ galaxies arcmin$^{-2}$ selecting only galaxies with colors  $0.85 < \rm{V-R} < 1.15$,  $ 0.6 < \rm{R-I} < 0.9$, i.e., within 0.15 magnitudes the V--R and R--I color of the central cD galaxy.}
\label{fig:cizah_xray}
\end{figure*}

CIZA~J0107.7+5408 is located at $z=0.1066$ \citep{1995MNRAS.274...75C} and has quite a high X-ray luminosity of $L_{\rm{X,~0.1-2.4~keV}}=5.42 \times 10^{44}$~erg~s$^{-1}$ \citep{2002ApJ...580..774E}. The cluster is projected relatively close to the galactic plane with $b=-8.65\degr$. Both the NVSS and WENSS survey images display an extended diffuse radio source located roughly at the cluster center. Our WSRT image clearly reveals the presence of a  somewhat elongated radio halo with a largest extent in the north-south direction of 1.1~Mpc. The galaxy and ICM distribution are also elongated along the major axis of the radio halo, see Fig.~\ref{fig:cizah_xray}, which supports the scenario that the cluster is currently undergoing a merger event. An image of the radio halo with the discrete sources subtracted (using the same technique as described in Sect.~\ref{sec:a697}) is overlaid with contours. The radio power of $3.8 \times 10^{24}$~W~Hz$^{-1}$ is consistent with the $L_{\rm{X}}$--$P_{\rm{1.4GHz}}$ correlation for giant radio halos. 

\subsection{Abell 2034}
Abell 2034 is a merging galaxy cluster as revealed by a Chandra X-ray study from \cite{2003ApJ...593..291K}. \cite{2000MNRAS.312..663W} reported a 
global cluster temperature of 7.6~keV in agreement with the value of $7.9 \pm 0.4$~keV found by \cite{2003ApJ...593..291K}. 
\cite{2001ApJ...548..639K} suggested the presence of a radio relic on the basis of 
a WENSS image. The diffuse emission is located close to the position of a cold front found by \cite{2003ApJ...593..291K}.
The presence of diffuse emission was confirmed by \cite{2009A&A...507.1257G} and was classified as an irregular elongated radio halo.

\begin{figure*}
\begin{center}
\includegraphics[angle =90, trim =0cm 0cm 0cm 0cm,width=0.49\textwidth]{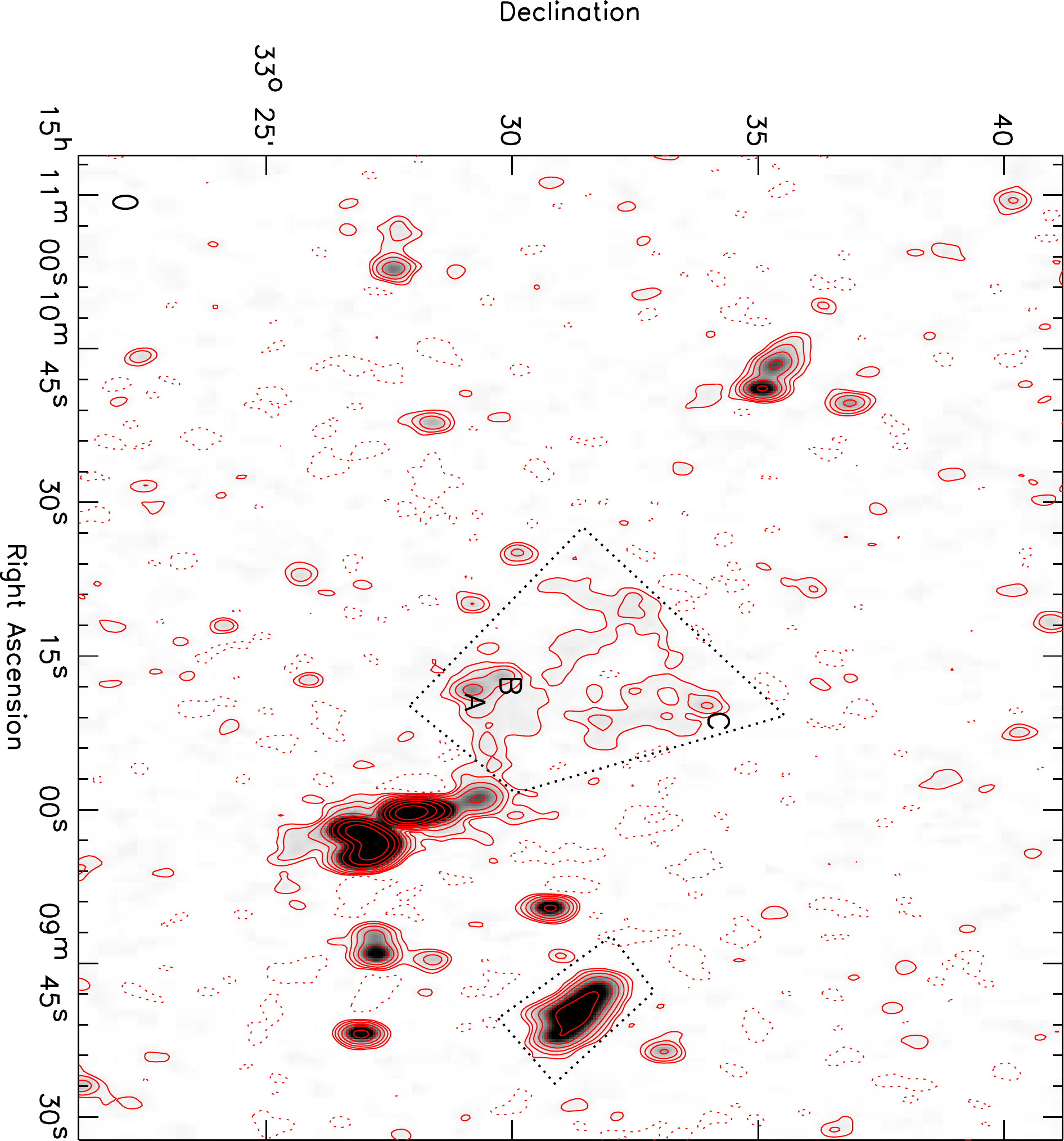}
\includegraphics[angle =90, trim =0cm 0cm 0cm 0cm,width=0.49\textwidth]{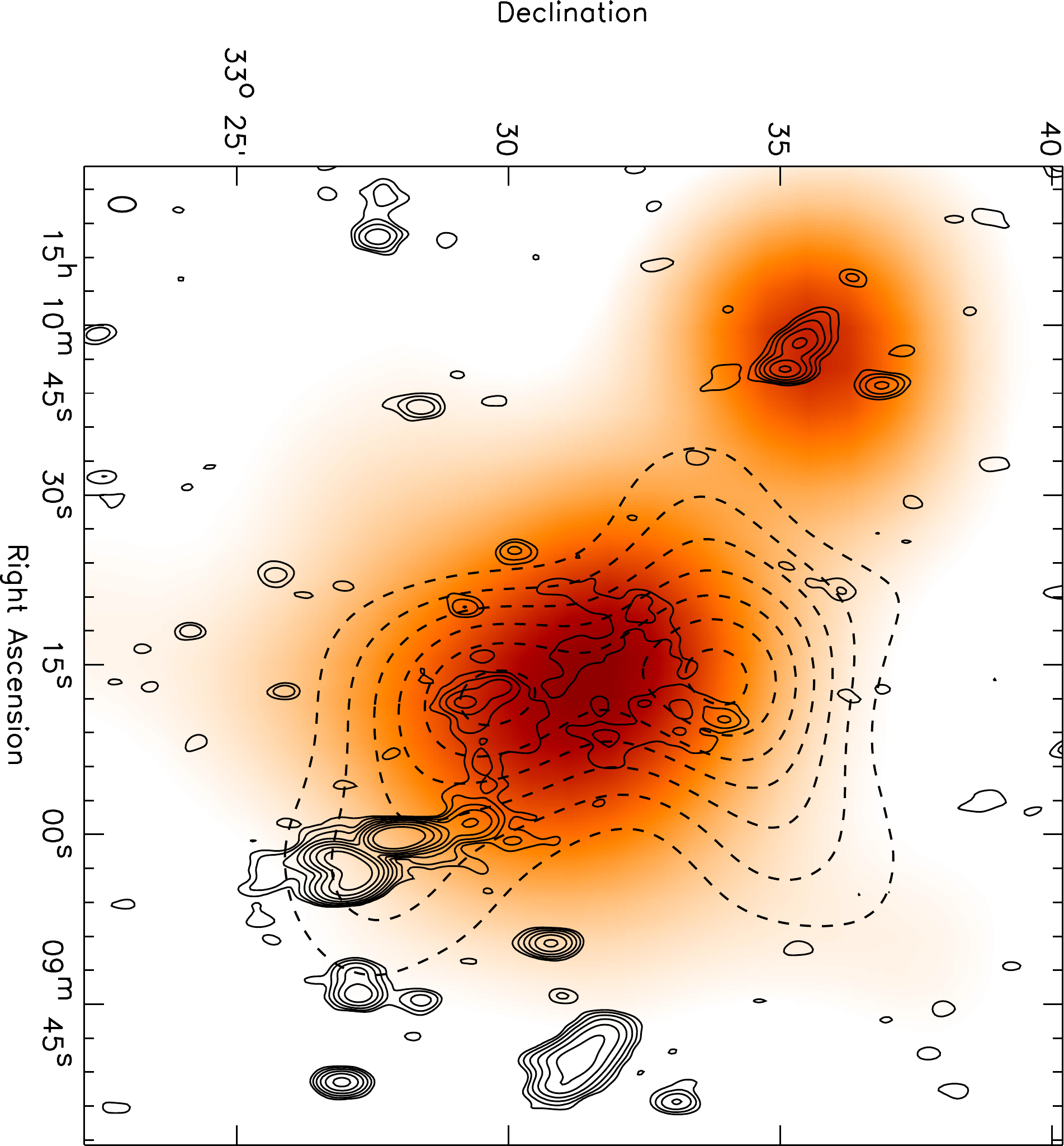}
\end{center}
\caption{Left: A2034 WSRT 1382~MHz image. Contour levels are drawn as in Fig.~\ref{fig:a1612_gmrt325}.  Black dotted lines indicate the integration areas for the flux measurements. Discrete sources embedded in the diffuse emission are alphabetically labeled, see Table~\ref{tab:compact}. Right: A2034 X-ray emission from ROSAT in orange. The original image from the ROSAT All Sky Survey was convolved with a 180\arcsec~FWHM Gaussian. Solid contours are from the WSRT 1382~MHz image and drawn at levels of ${[1, 2, 4, 8, \ldots]} \times 4\sigma_{\mathrm{rms}}$.  Dashed contours show the galaxy iso-density distribution derived from the SDSS survey. Contours are drawn at ${[1.0,1.2, 1.4, \ldots]}  \times 1.1$ galaxies arcmin$^{-2}$ selecting only galaxies with  $0.07 < z_{\rm{phot}} <  0.16$.}
\label{fig:a2034_xray}
\end{figure*}

We also observe the diffuse emission   
to brighten at the position of the cold front in the WSRT image. The SDSS galaxy distribution (Fig.~\ref{fig:a2034_xray}, right panel) 
is bimodal which supports the scenario that the cluster is undergoing a merger event. We consider the classification of the source uncertain, since the radio emission does not show a very clear correlation with the X-rays. It could therefore also be a radio relic. 
The radio power (${P_{\rm{1.4GHz}} = 0.28 \times 10^{24}}$~W~Hz$^{-1}$) though is in agreement with the $L_{\rm{X}}$--$P_{\rm{1.4GHz}}$ correlation for giant radio halos \citep[e.g.,][]{2000ApJ...544..686L,2006MNRAS.369.1577C}.  The total flux we find is $7.3 \pm 2.0$~mJy, lower than the $13.6\pm1.0$~mJy reported by \cite{2009A&A...507.1257G}.  The reason for the difference is unclear because \cite{2009A&A...507.1257G} do not report the integration area and the fluxes of the discrete sources that were subtracted. Deeper observations which sufficient resolution are needed to classify the nature of the diffuse emission. Polarimetric observations could be very helpful here as radio relics are usually highly polarized.

\begin{figure}
\begin{center}
\includegraphics[angle =90, trim =0cm 0cm 0cm 0cm,width=0.5\textwidth]{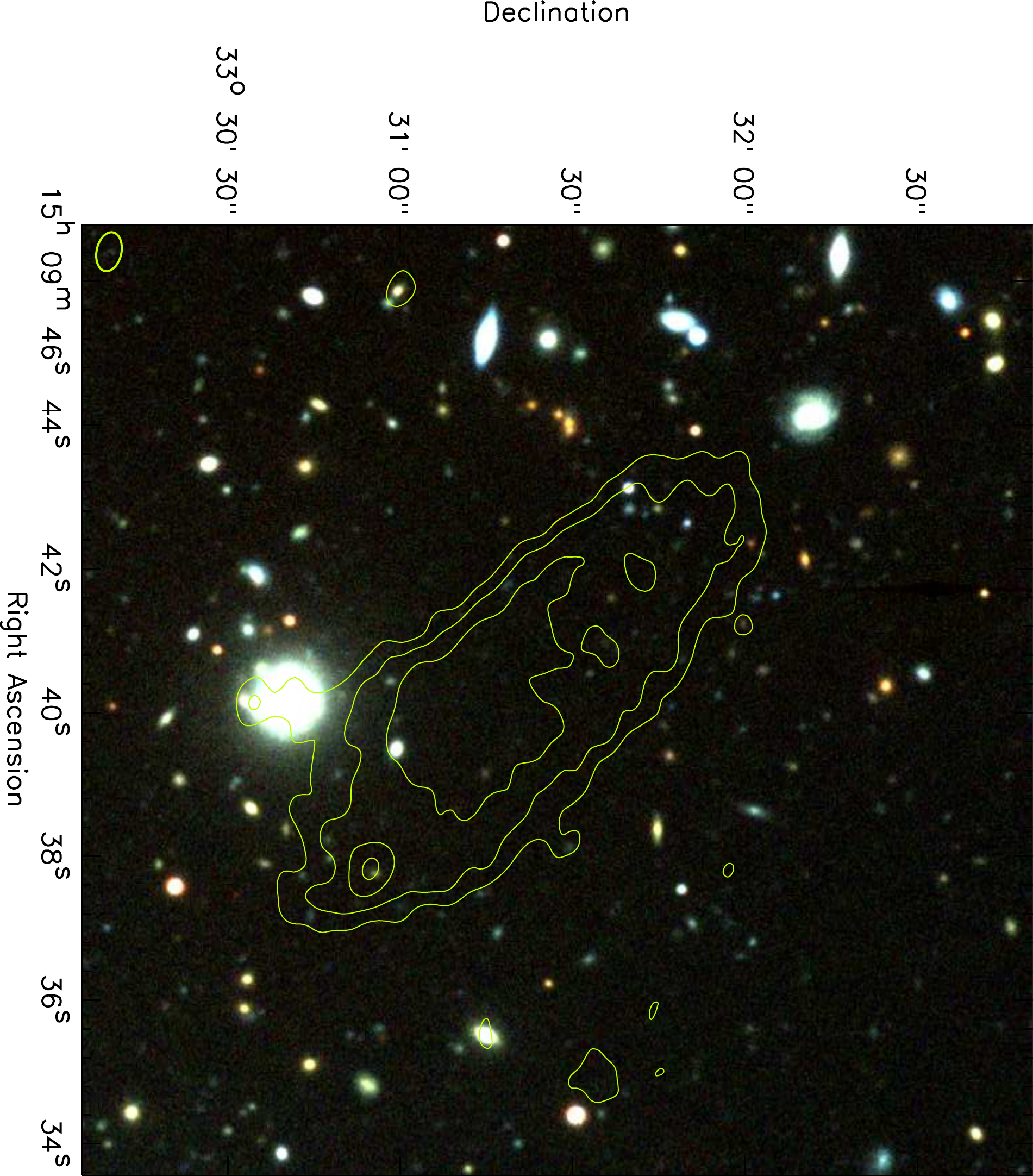}
\end{center}
\caption{A2034 WHT V, R, I color image around the small radio relic. Contours at 610~MHz from the GMRT are overlaid and drawn at levels of ${[1, 2, 4, 8, \ldots]} \times 4\sigma_{\mathrm{rms}}$. The noise level in the image is $41\mu$~Jy~beam$^{-1}$ and the beam size is $6.8\arcsec~\times~4.3\arcsec$.}
\label{fig:a2034optical}
\end{figure}

We report the detection of a new relatively small 
radio relic located west of the cluster center (\object{NVSS~J150940+333119}, \object{WN~B1507.6+3342}, \object{7C~150739.39+334252.00}). This relic is already visible in the previous studies mentioned above but was not recognized as a radio relic probably because it is 
quite compact. The source has a size about 220 by 75~kpc and a spectral index 
$\approx -1.2$, including flux measurements from the NVSS, WENSS and 7C \citep{2007MNRAS.382.1639H} surveys . A high-resolution 610~MHz GMRT of the relic overlaid on an optical WHT image is shown in Fig.~\ref{fig:a2034optical}.
 The source does not have an optical counterpart which should have easily been visible in the WHT image. The radio plasma could  have originated from the compact radio source located immediately south of the relic. On the other hand the spectral index is more typical of a relic directly tracing a shock wave.

\subsection{RXC J1053.7+5452}
\begin{figure*}
\begin{center}
\includegraphics[angle =90, trim =0cm 0cm 0cm 0cm,width=0.49\textwidth]{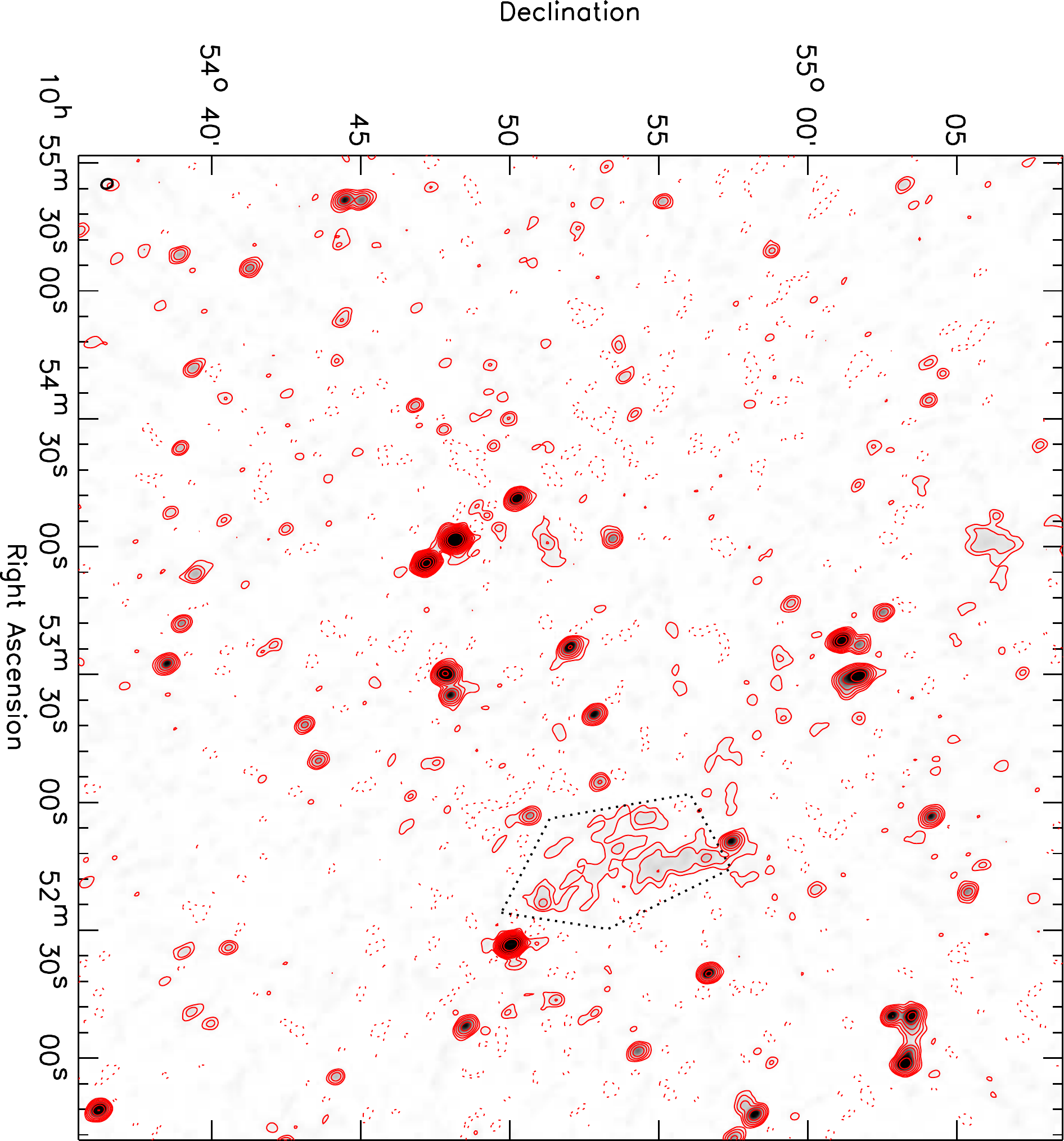}
\includegraphics[angle =90, trim =0cm 0cm 0cm 0cm,width=0.49\textwidth]{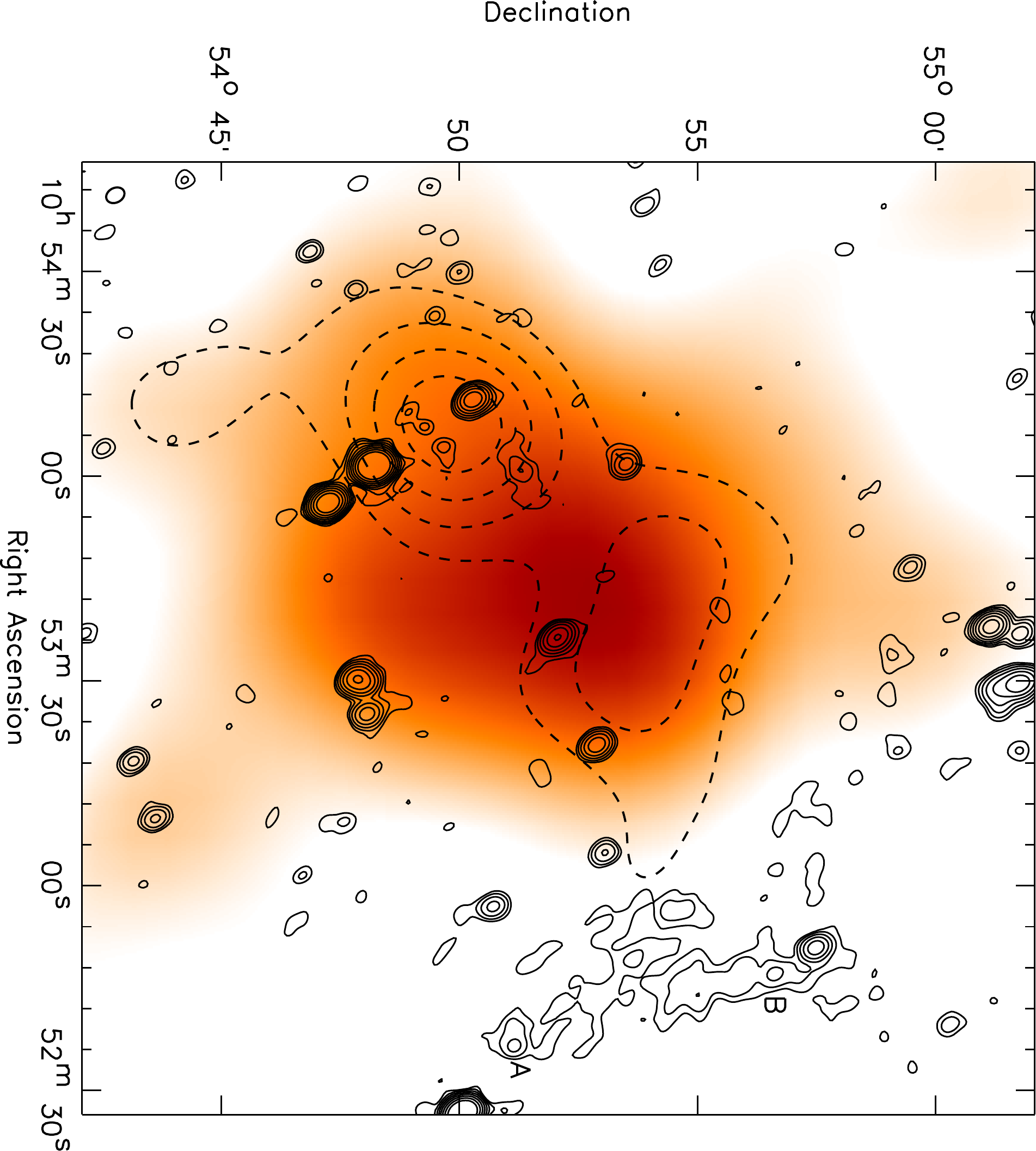}
\end{center}
\caption{Left: RXC~J1053.7+5452 WSRT 1382~MHz image. Contour levels are drawn as in Fig.~\ref{fig:a1612_gmrt325}.  Black dotted lines indicate the integration area for the flux measurement. Right: RXC~J1053.7+5452 X-ray emission from ROSAT in orange. The original image from the ROSAT All Sky Survey was convolved with a 225\arcsec FWHM Gaussian. Solid contours are from the WSRT 1382~MHz image and drawn at levels of ${[1, 2, 4, 8, \ldots]} \times 4\sigma_{\mathrm{rms}}$.  Dashed contours show the galaxy iso-density distribution derived from the SDSS survey. Contours are drawn at ${[1.0,1.2, 1.4, \ldots]}  \times 0.25$ galaxies arcmin$^{-2}$ selecting only galaxies with  $0.05 < z_{\rm{phot}} <  0.1$. Discrete sources embedded in the diffuse emission are alphabetically labeled, see Table~\ref{tab:compact}.}
\label{fig:rx1054_xray}
\end{figure*}

\object{RXC~J1053.7+5452} is located at $z=0.0704$. \cite{2007A&A...471...17A} reported a velocity dispersion of $665^{+51}_{-45}$~km~s$^{-1}$ and an $r_{200}$ radius of 1.52~Mpc based on SDSS data. 
\cite{2009ApJ...697.1341R} mentioned the presence of a diffuse radio relic with an extent of about 1~Mpc on the west side of the cluster. \citeauthor{2009ApJ...697.1341R} reported a total flux of 0.36~Jy at 325~MHz. The  low-surface brightness radio relic is also visible in the WSRT image. We find a total extent of 600~kpc, lower than that of \cite{2009ApJ...697.1341R}. The ROSAT image overlaid with SDSS galaxy iso-density contours and the 1382~MHZ WSRT image are shown in Fig.~\ref{fig:rx1054_xray}. The galaxy distribution seems irregular with the main peak to the southeast of the X-ray center and a second peak to the northwest. The radio relic is roughly located along a line connecting these two galaxy concentrations. This is expected for a cluster merger event, with the shock waves traveling outwards along the merger axis. Deeper radio observations are necessary to better map the extent of this faint radio relic.

\section{Discussion}
\begin{figure*}
\begin{center}
\includegraphics[angle =90, trim =0cm 0cm 0cm 0cm,width=0.45\textwidth]{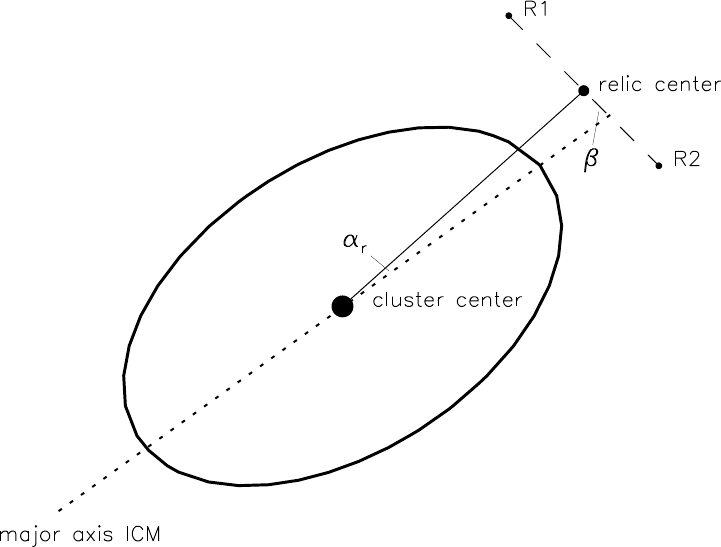}
\includegraphics[angle =90, trim =0cm 0cm 0cm 0cm,width=0.45\textwidth]{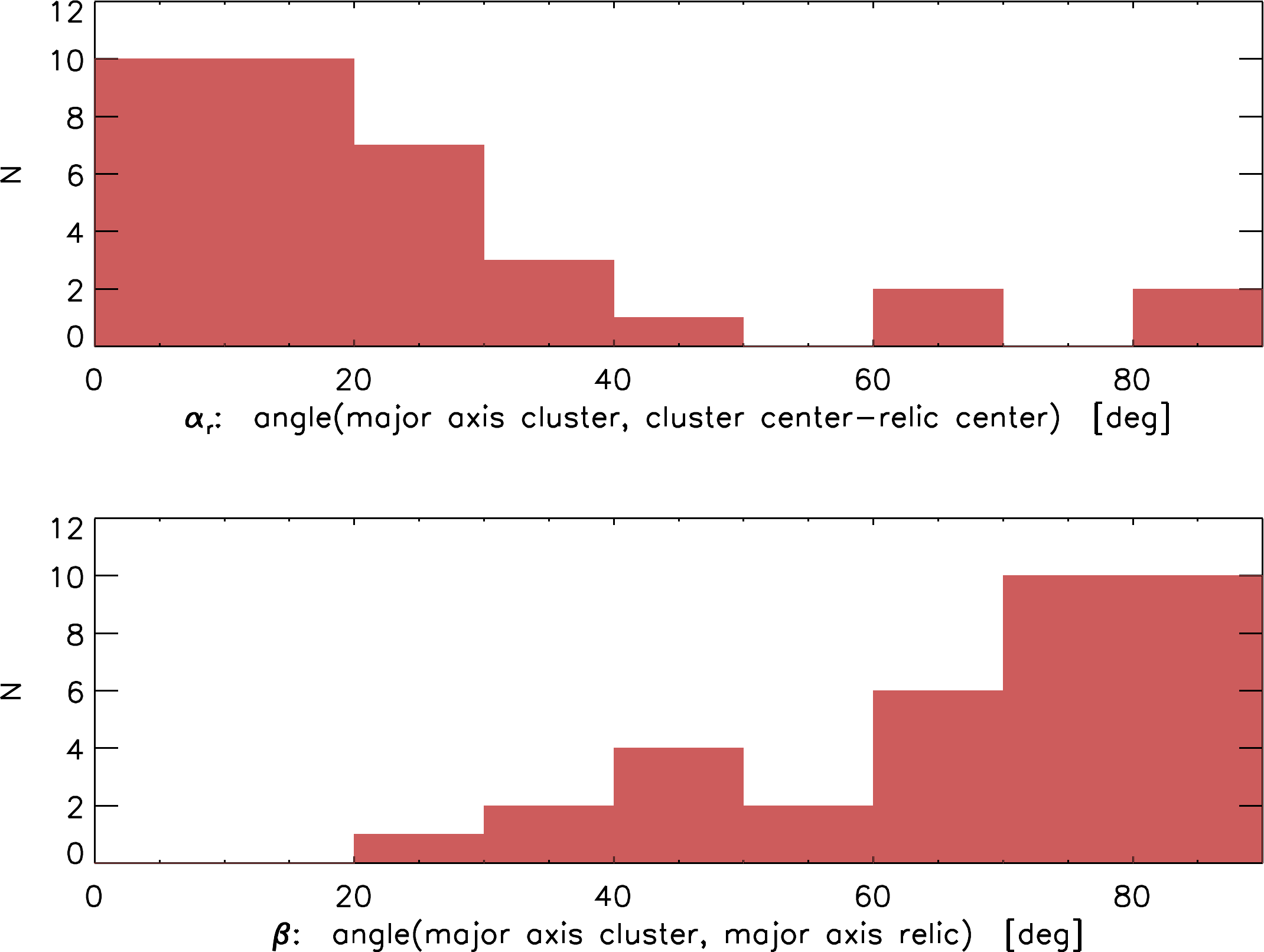}
\end{center}
\caption{Left: Schematic illustration of the angle between the major axis of the ICM and the line relic center--cluster center ($\alpha_{\rm{r}}$), and the angle between the relic orientation and major axis of the ICM ($\beta$), see Table~\ref{tab:samplefull}. Top Right: Histogram of angles between the major axis of the X-ray emission and the line connecting the cluster center with the center of the relic. Bottom Right: Histogram of angles between the major axis of the X-ray emission and the relic's major axis.}
\label{fig:angles}
\end{figure*}
\label{sec:discussion}

Giant radio relics are proposed to directly trace merger shock waves in galaxy clusters. Simulations show that in the case of a binary merger event two shell-like shock waves form at the moment of core passage \citep[e.g.,][]{1999ApJ...518..603R, 2001ApJ...561..621R}. These shock waves then move outwards into the lower density ICM of the cluster's outskirts. Double radio relics are thought to trace these binary merger events, with two symmetric shock waves on each side of the cluster center. However, often merger events are more complex with multiple substructures merging, so that relics are not necessarily symmetric structures and not always come in pairs. Also, the shock structures may break up when they interact with the galaxy filaments connected to the cluster \citep[e.g., ][]{2010arXiv1001.1170P}. Still, it is expected that relics are mainly found along the main axis of a merger event, while their orientation is perpendicular to this axis.

We have tested this prediction by constructing a sample of 35 relics, taking relics from the literature and those presented in this paper (see Table~\ref{tab:samplefull}). We did not include any radio relics classified as AGN relics or radio phoenices and selected only relics in clusters which are detected in the ROSAT All Sky Survey (RASS) images.  

For every relic we record the end positions of their largest spatial extent (R1, R2), see Fig.~\ref{fig:angles}. The line connecting these two points we define as the relic's major axis. The midpoint between the two extrema we take as a the relic's center position. The RASS image can be used to estimate the position of a possible merger axis. We convolve the ROSAT X-ray images for each cluster to the same spatial resolution of 650~kpc. We then fit a 2-dimensional elliptical Gaussian to the X-ray emission. We record both the position of the major axis and the center of the fitted Gaussian. For the merger axis we take as a proxy the major axis of the fitted X-ray emission. 
Finally, we compute the angle ($\alpha_{\rm{r}}$) between the major axis of the ICM and the line cluster center--radio relic center. The resulting histogram is shown in Fig. ~\ref{fig:angles}. From this histogram we see that relics are preferably found along the major axis of the ICM. This is in line with the simple picture that shock waves propagate outwards along the merger axis. We also calculate the angle ($\beta$) between the major axis of the ICM and the relic's major axis. We find that most relics are oriented perpendicular to the ICM major axis, also in agreement with a proposed shock origin for radio relics.

 \begin{table*}
\begin{center}
\caption{Clusters with radio relics and measured orientations}
\begin{tabular}{llllll}
\hline
\hline
Cluster name & $z$  &   $\alpha_{\rm{r}}$ &  $\beta$  \\ 
                          &    &   ``location PA''  (deg)& ``orientation PA'' (deg)\\
\hline
Abell 115 &0.197 &45&78  \\
Abell 521 &0.247 & 6&64 \\
Abell 523$^{a}$ & 0.100 &23&39\\
Abell 548b$^{b}$ &0.042&87(A), 40(B)&84(A), 48(B)\\
 
Abell 746 & 0.232 &25&56\\

Abell 1240 &0.159 &13(N), 15(S)&85(N), 85(S)\\
Abell 1612 & 0.179 &62&69\\
Abell 1664 &0.128 &3&78\\
Abell 2061 & 0.078 &15&77\\
Abell 2163 &0.203 &1&67\\

Abell 2255 & 0.081 &61&38\\
Abell 2256 & 0.059 &26&77\\

Abell 2345 &0.176 &13(W), 24(E)&87(W), 43(E)\\
Abell 2744 &0.308 &20&81\\
Abell 3365 & 0.093 &20(W), 141(E)&82(W), 43(E)\\
Abell 3376 &0.046& 3(W), 8(E)&85(W), 67(E)\\
Abell 3667 &0.055 &13(W), 0(E)&75(W), 75(E)\\

Coma cluster &0.023 &22&60\\
CIZA~J2252.8+5301 & 0.192 & 27 (N), 4(S) & 67(N), 84(S) \\ 

MACS J0717.5+3745 &0.555 &2&79\\
RXC J1053.7+5452  & 0.070 &19&76\\
RXCJ1314.4-2515 & 0.244 &30(W), 99(E) &26(W), 46(E)\\
ZwCl 0008.8+5215 & 0.104 &10(W), 11(E)& 76(W), 85(E)\\
ZwCl 2341.1+0000  & 0.270 &14(N), 8(S)& 88(N), 63(S)\\ 
\hline
\hline
\end{tabular}
\label{tab:samplefull}
\end{center}
- for double relics two values are measured (N=north, S=south, E=east, W=west)\\
- references: Abell~2255 (main relic only) \cite{2009A&A...507..639P, 2005A&A...430L...5G}, MACS~J0717.5+3745 \cite{2009A&A...505..991V, 2009A&A...503..707B}, Abell~548b (2 relics) \cite{2006MNRAS.368..544F}, Abell~2256 \cite{2006AJ....131.2900C}, Abell~521 \cite{2008A&A...486..347G}, Coma Cluster \cite{1991A&A...252..528G, 2011MNRAS.412....2B}, Abell~2163 \cite{2001A&A...373..106F}, Abell~2744 \cite{2001A&A...376..803G}, Abell~115 and Abell~1664  \cite{2001A&A...376..803G}. Symmetric double relics included from the literature are: Abell~1240 and Abell~2345 \cite{2009A&A...494..429B}, Abell~3376 \cite{2006Sci...314..791B}, Abell~3667 \cite{1997MNRAS.290..577R}, ZwCl~0008.8+5215 \cite{2011A&A...528A..38V},  CIZA~J2242.8+5301 \cite{2010Sci...330..347V}, ZwCl~2341.1+0000 \cite{2009A&A...506.1083V, 2010A&A...511L...5G}, RXC~J1314.4-2515 \cite{2007A&A...463..937V}. \\
$^{a}$ identification uncertain (either relic or halo)\\
$^{b}$A548b: A and B refer to the names used in \cite{2006MNRAS.368..544F}\\
\end{table*}

\subsection{Comparison with the REFLEX and NORAS X-ray clusters} 
\begin{figure}
\begin{center}
\includegraphics[angle =90, trim =0cm 0cm 0cm 0cm,width=0.49\textwidth]{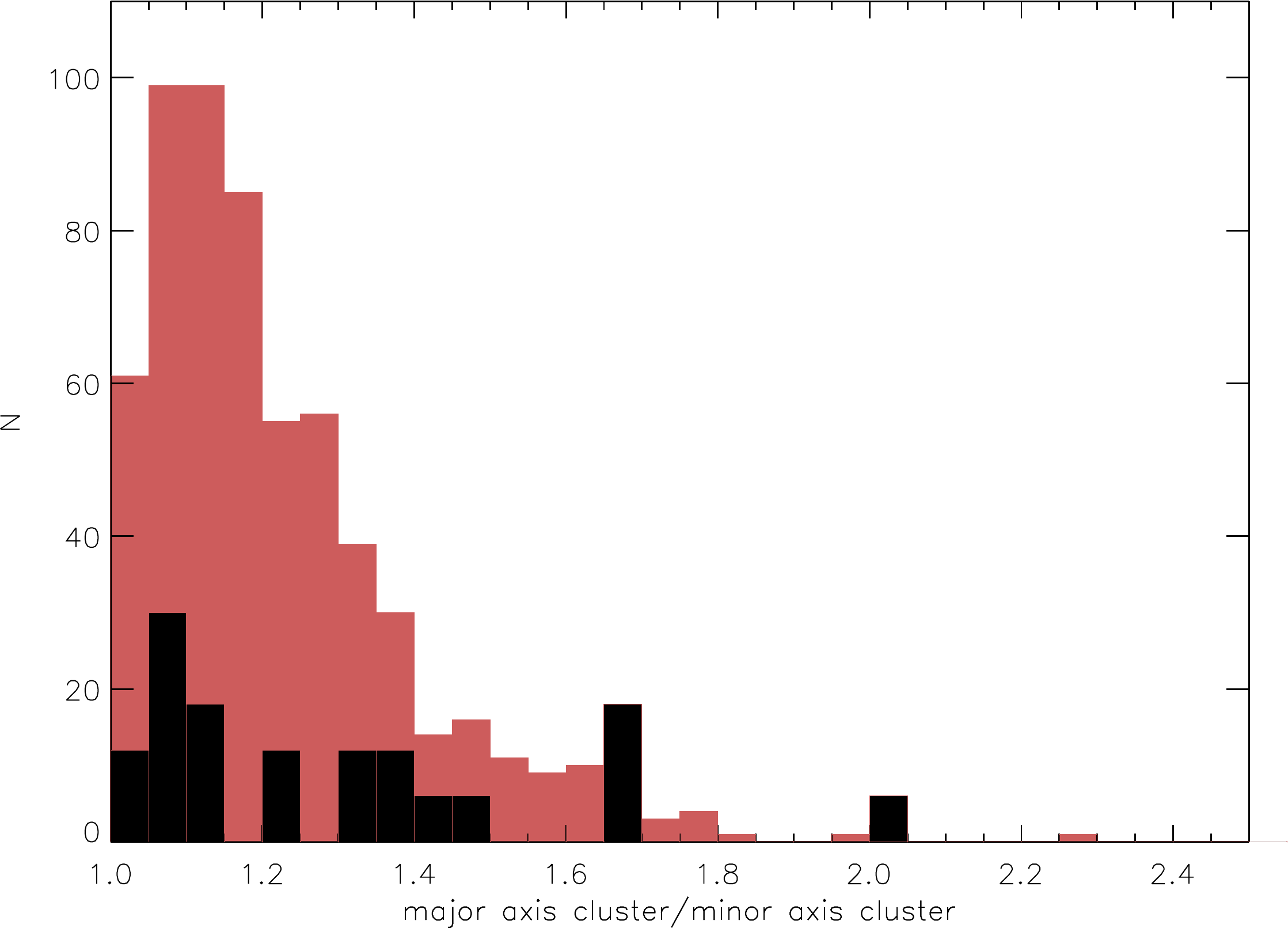}
\end{center}
\caption{Histograms showing the distribution of the major and minor axis ratio. The red histogram is for clusters from the NORAS and REFLEX surveys, the black histogram is for clusters containing giant radio relics (see the caption of Fig.~\ref{fig:angles}). The black histogram was scaled by a factor of six for easier comparison with the NORAS-REFLEX sample. Clusters for which the 2 dimensional Gaussian fit did not converge were not included.}
\label{fig:histo}
\end{figure}

We also compared the properties of cluster hosting radio relics with the X-ray clusters from the NORAS \citep{2000ApJS..129..435B} and REFLEX \citep{2004A&A...425..367B} surveys. The NORAS survey contains 378 galaxy clusters and has an estimated  completeness of about 50\% at an X-ray Flux of $F_{\rm{X,~0.1-2.4~keV}}=3.0 \times 10^{-12}$~erg~s$^{-1}$~cm$^{-2}$. The REFLEX sample contains 447 galaxy clusters and has and a  completeness of about 90\% at the same flux level as the NORAS survey. For each cluster in this sample, we fitted a 2-dimensional elliptical Gaussian to the X-ray emission from the RASS image, using the same procedure as for the clusters hosting giant relics. For some clusters the fitting procedure did not converge because of nearby bright confusing X-ray sources, these cluster were not included in the analysis.  The resulting histogram is displayed in Fig.~\ref{fig:histo}. The distribution of the major-minor axis ratios for the relic cluster sample is broader. This is expected since relics should be found in merging  cluster which are typically more elongated. 

\begin{figure}
\begin{center}
\includegraphics[angle =0, trim =0cm 0cm 0cm 0cm,width=0.49\textwidth]{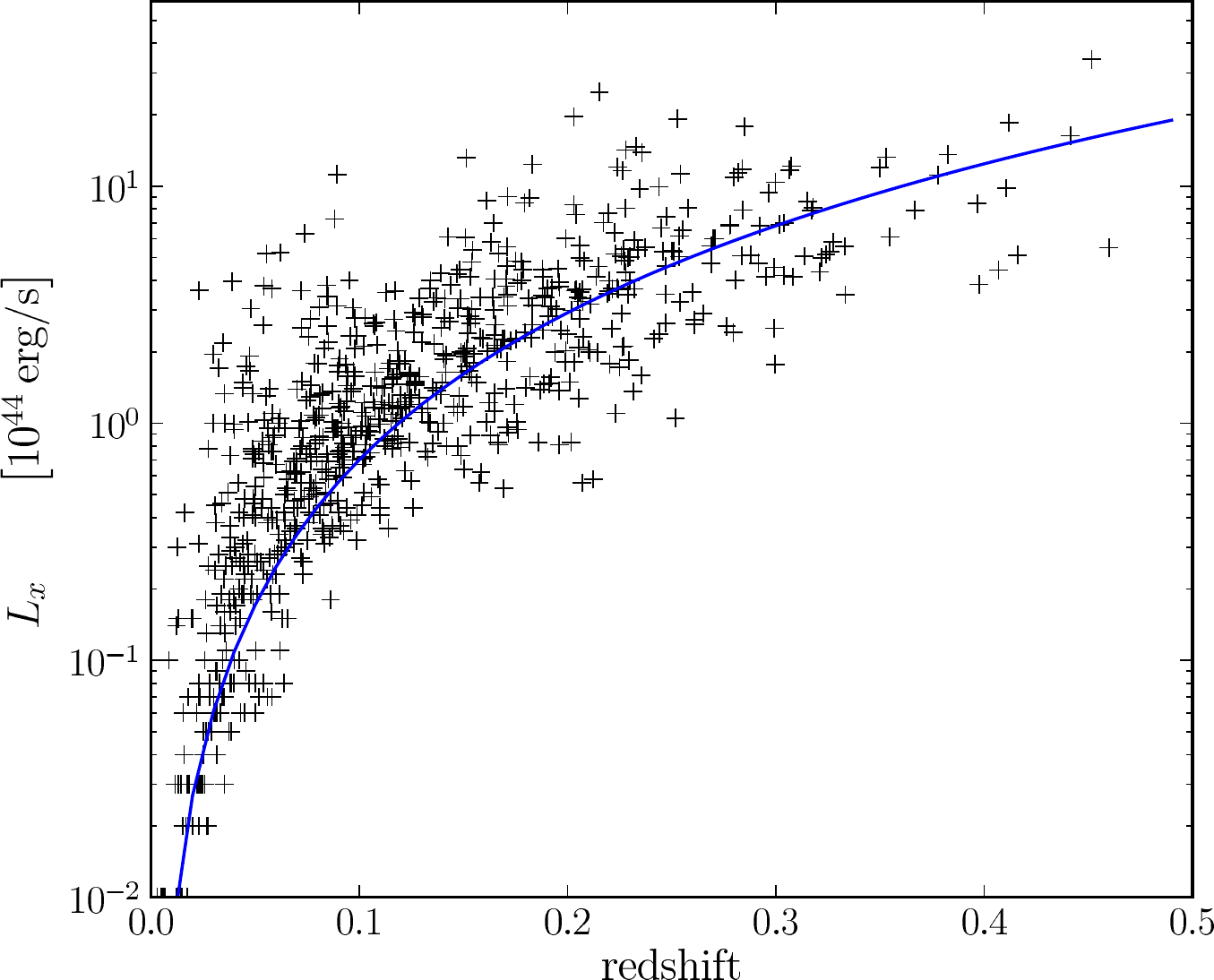}
\end{center}
\caption{$L_{\rm{X}}$-redshift distribution for the NORAS and REFLEX surveys. The solid blue line is the flux cutoff of $3.0 \times 10^{-12}$~erg~s$^{-1}$~cm$^{-2}$ we use for comparison with the relic cluster sample.}
\label{fig:zlx}
\end{figure}

\begin{figure}
\begin{center}
\includegraphics[angle =90, trim =0cm 0cm 0cm 0cm,width=0.49\textwidth]{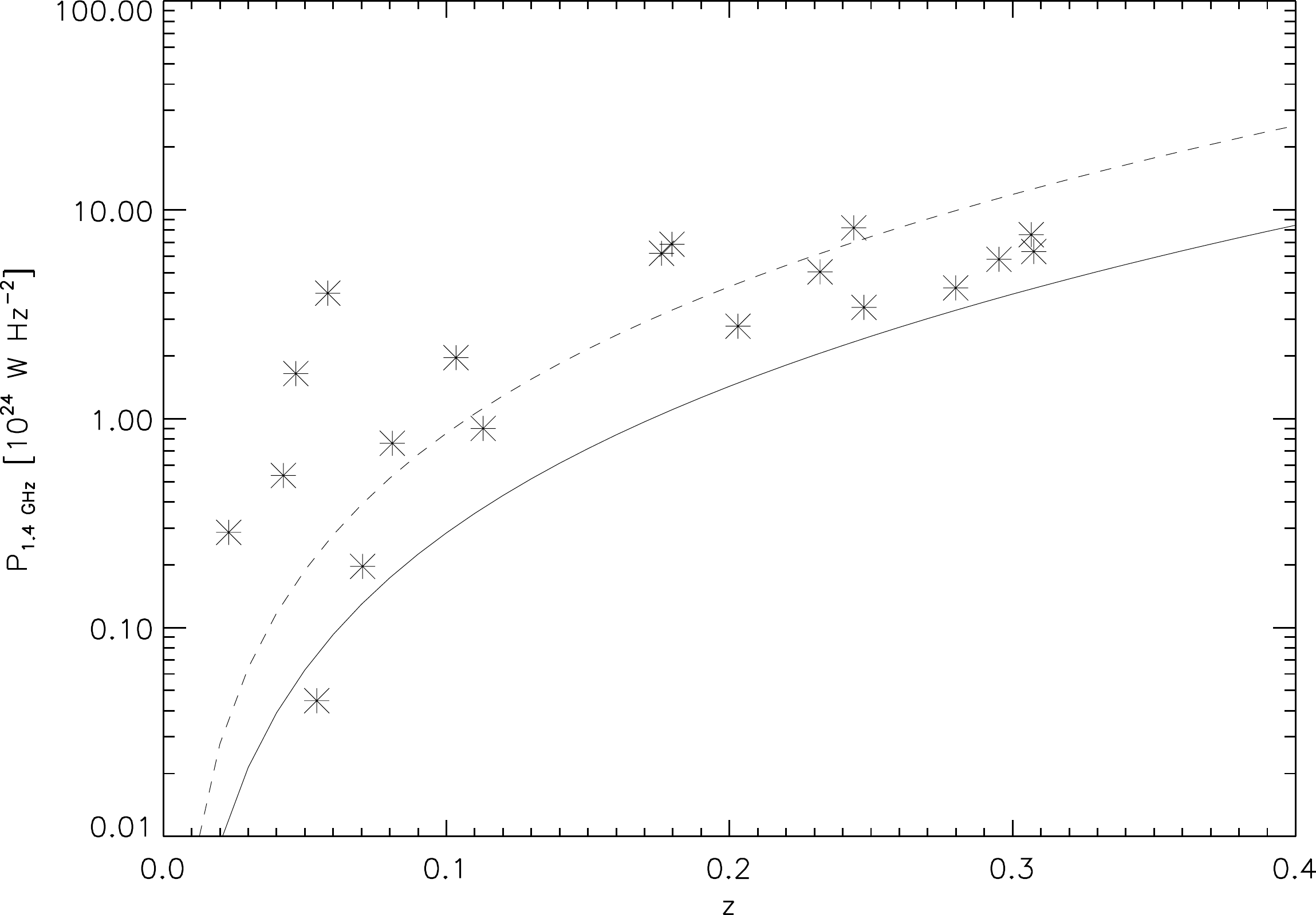}
\end{center}
\caption{$P_{\rm{1.4 GHz}}$-redshift distribution for the relics in Table~\ref{tab:samplenorasreflex}. The solid line is for a relic flux limit of $10$~mJy, the dashed line for $30$~mJy. Radio phoenices are not included. }
\label{fig:zP}
\end{figure}

In addition, we compare both the X-ray luminosity and redshift distribution of the NORAS-REFLEX sample with the relic sample. We selected all clusters
with a flux larger than $3.0 \times 10^{-12}$~erg~s$^{-1}$~cm$^{-2}$ in the ROSAT band. Below this flux limit the NORAS is more than 50\% incomplete, see Fig.~\ref{fig:zlx}.  The total number of clusters above this flux limit is $540$, and the number of clusters with relics is 16 (using the same flux cutoff). From this we find that the currently observed fraction of clusters hosting relics is 3\% in this sample. The list of relics is given in Table~\ref{tab:samplenorasreflex} and the radio power are plotted as function of redshift are plotted in Fig.~\ref{fig:zP}.

\begin{figure*}
\begin{center}
\includegraphics[angle =0, trim =0cm 1cm 0cm 0cm,width=0.45\textwidth]{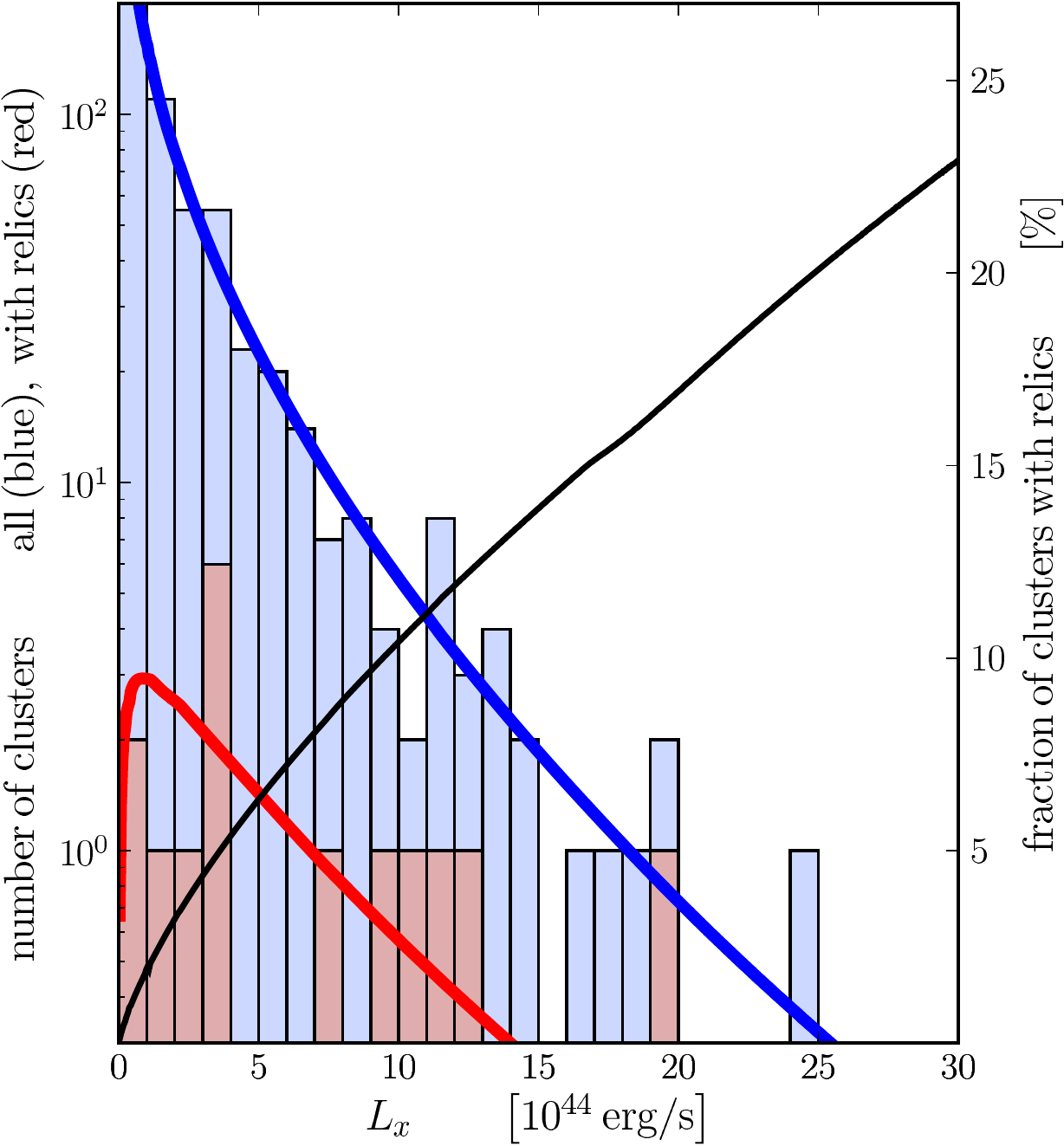}
\includegraphics[angle =0, trim =0cm 1cm 0cm 0cm,width=0.45\textwidth]{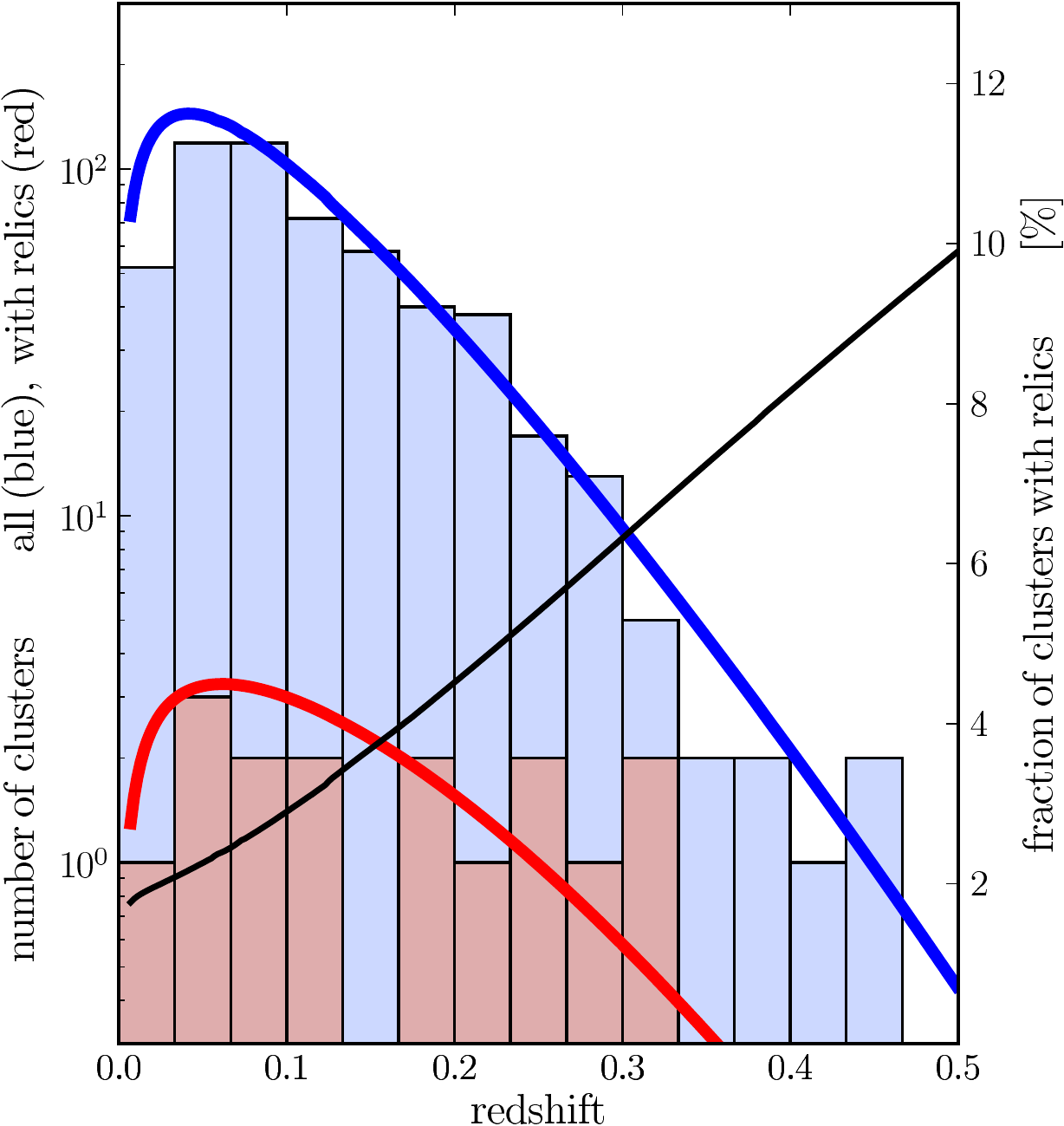}
\end{center}
\caption{Histograms showing the X-ray luminosity (left) and redshift (right) distribution. Blue histograms shows the NORAS-REFLEX sample, red the relic cluster  sample. The solid blue line displays the predicted luminosity/redshift distributions from \cite{nuza} for clusters with fluxes $> 3.0 \times 10^{-12}$~erg~s$^{-1}$~cm$^{-2}$ , while the solid red line is the prediction for clusters hosting relics in the simulation. 
The fraction of clusters with relics is given by the black solid line (the ratio of the blue and red lines).}
\label{fig:lxhisto}
\end{figure*}

The resulting histograms are displayed in Fig.~\ref{fig:lxhisto}. Although the number of known relics is rather small, and the relics were selected using various methods, there are few interesting trends visible. Apparently, the fraction of clusters with relics increases with the X-ray luminosity, from about a percent at $L_{\rm{X,~0.1-2.4~keV}} = 1 \times 10^{44}$~erg~s$^{-1}$ to  more than 10\% above $\sim1\times 10^{45}$~erg~s$^{-1}$. Also, the redshift distribution for cluster with relics is somewhat broader than the corresponding NORAS-REFLEX sample. Therefore, the chance of finding a relic above a certain flux density increases with redshift for clusters selected from flux-limited X-ray surveys. 
Since the number of clusters with relics is small this might be a statistical
fluctuation. However, given that the average fraction of clusters with relics
is 3 \%, the probability to find four (or even more) in the 26 clusters with
$L_{\rm{X}} >  10^{45}$~erg~s$^{-1}$ is 0.7\%. Hence, a pure statistical
fluctuation is rather unlikely. This rises the question if selection effects may
causes this trend. 

As discussed above the NORAS-REFLEX sample is more or less complete up to $3
\times 10^{-12}$~erg~s$^{-1}$~cm$^{-2}$. The relic sample in contrast is
probably not complete up to a specific flux limit. For example, large low-surface brightness relics could be missed, as are relics in more distant clusters because they are barely resolved in the NVSS and WENSS surveys images and therefore not easily recognizable. 
The construction of a flux-limited relic sample is therefore challenging as selection effects due to angular size, morphology and surface brightness need to be properly taken into account. In addition, one needs to properly identify other diffuse radio sources (such as radio halos and giant radio galaxies) to prevent them from ending up in the relic sample. \cite{2009ApJ...697.1341R} give a good overview of some of the problems involved in constructing these samples. 
The important question is whether there is any systematic
effect which leads to a preferential detection of relics in luminous clusters. 
Since the luminous clusters are on average at higher redshift, see Fig.~\ref{fig:zlx}, relic
detection is less affected by a too low surface brightness. However, it is
possibly affected by resolution effects. This will only decrease the detection probability at higher redshift (luminosity) and hence there is no evident reason why
relics in more luminous clusters should have a better chance of being discovered.

\cite{2011ApJ...735...96S} found in simulations, based on DSA, that the radio power of 
clusters with relics scatters largely for a given X-ray luminosity. In addition, 
they found that the mean radio power strongly correlates  with the X-ray luminosity. Hence, 
the fraction of clusters with relics should increase with X-ray luminosity. 
To compute the actual fractions many factors need to be considered, namely 
the X-ray flux limit, the discovery probability for radio relics, the radio power 
distribution of relics and the abundance of clusters as function of X-ray luminosity 
and redshift. In \cite{nuza} we cary out this analysis in detail. 
We postulate a radio relic probability density based on 
numerical simulations $p (P, M_{\rm vir}, 
z, \nu_{\rm obs})$, where $P$ is the radio luminosity, $M_{\rm vir}$ is 
the virial mass of the cluster and $\nu_{\rm obs}$ is the observing 
frequency.  We convolve this with the cosmological abundance of dark 
matter halos. As a result we indeed find that the fraction of radio relics in the NORAS-REFLEX sample should 
increase with both the X-ray luminosity and the redshift, see Fig.~\ref{fig:lxhisto} (solid lines). 

The reason why in simulations more massive clusters show on average much brighter
radio relics is unclear. We speculate that multiple aspects contribute: temperature and
density are higher and shock fronts are larger in more massive clusters. It seems
also likely that mergers of more massive clusters result in higher Mach numbers. Since
according to the sub-grid models used in the simulation the radio emission strongly increases
with Mach number \citep{2007MNRAS.375...77H}, a rather small increase of the Mach numbers
would have a large effect on the resulting radio luminosity. (We note that the simulation of \cite{2007MNRAS.375...77H} is 
also based on DSA.) Finally, the merger rate increases
with redshift. However, it needs to be clarified from simulations why an increase of the radio power
with X-ray luminosity is expected. The fractions of relics in X-ray selected cluster 
samples is therefore a powerful tool to put constraints on the models used in the simulations, and
hence on the evolution of magnetic fields and on particle acceleration in the ICM.

 \begin{table*}
\begin{center}
\caption{Clusters with radio relics in the NORAS and REFLEX surveys}
\begin{tabular}{llllll}
\hline
\hline
Cluster name & $z$  &   $P_{\rm{1.4 GHz}}$ (total for cluster) & $L_{\rm{X,~0.1-2.4~keV}}$ \\
                          &    &   $10^{24}$ W~Hz$^{-1}$ & $10^{44}$~erg~s$^{-1}$   \\
\hline

A3376    &0.0468   &1.6&   1.01   \\
A2744    & 0.3066  &7.7&   11.68   \\
A4038$^{c}$    & 0.0292   &0.1&   1.00   \\
A0548W$^{b}$&  0.0424   &0.5&   0.10 \\
 RXC J1314.4-2515  & 0.2439  &8.2&    9.92 \\%
 A133$^{c}$ &  0.0569   &1.1&   1.40   \\
 A1300 &  0.3075  &6.3&   12.12   \\
A13$^{c}$   & 0.0940   &0.9&   1.24   \\
A2345 &  0.1760  &6.2&    3.91  \\
A521   &0.2475   &3.4&   7.44  \\
A754  & 0.0542  &0.04&    3.79\\
A85$^{c}$  &0.0555    &0.3&  5.18   \\
A2163 & 0.2030   &2.8&  19.62   \\
A1612 &  0.1797  &7.9&    2.41  \\
A523$^{a}$  &   0.1034  &1.8&    0.89  \\
Coma cluster& 0.0231  &0.29&    3.63  \\
A781$^{b}$  &0.2952     &5.8& 4.15   \\
A2034$^{a}$  &  0.1130   &1.17&   3.56  \\
A1758   & 0.2799   &4.1&  10.90   \\
A746$^{b}$   &  0.232   &6.8&  3.68  \\
 RXC J1053.7+5452  & 0.0704  &0.2&    0.44 \\                 
A2255 &  0.0809  &0.8&    3.08 \\ 
A2256 &  0.0581   &4.2&   3.69 \\  
\hline
\hline
\end{tabular}
\label{tab:samplenorasreflex}
\end{center}
references: Abell~754 \cite{2011ApJ...728...82M}; Abell~1758 \cite{2009A&A...507.1257G}; Abell~781 \cite{2011MNRAS.tmpL.251V}; Abell 1300 \cite{1999MNRAS.302..571R}; Abell 13, 85, 133, and 4038 \cite{2001AJ....122.1172S}; for other clusters see Table~\ref{tab:samplefull} \\
$^{a}$ identification uncertain (either relic or halo)\\
$^{b}$ below X-ray flux limit of $F_{\rm{X,~0.1-2.4~keV}}=3 \times 10^{-12}$~erg~s$^{-1}$~cm$^{-2}$\\
$^{c}$ radio phoenix, not included in sample
\end{table*}

\subsection{X-ray peak and galaxy distribution separation}
Cluster mergers are thought to decouple the baryonic matter component from the dark matter (DM). This causes  and offset between the gravitational center (measured from lensing)  and X-ray center of the cluster. A clear example is the ``Bullet cluster'' \cite[e.g.,][]{2006ApJ...648L.109C}, but the effect has also been measured for other clusters \citep[e.g.,][]{2010MNRAS.406.1134S}. Radio relics and halos are mostly found in merger cluster and are thus good candidates to measure this effect. We do not have lensing measurements for the clusters in our sample, but simply taking the galaxy distribution as a proxy for the dark matter distribution we note offsets between the X-ray emission from ROSAT and the galaxy distribution in the clusters A2034,  A3365, A2061, A523 and RXC J1053.7+5452  for example. These clusters would therefore make interesting targets for future X-ray and lensing observations.

\section{ Conclusions}
We have presented WSRT, VLA and GMRT observations of galaxy clusters with diffuse emission selected from the WENSS and NVSS surveys. We find  peripheral radio relics in the clusters Abell 1612, Abell 746 and CIZA~J0649.3+1801 and a smaller relic Abell~2034. Abell~3365 seems to host a double radio relic system. Our observations reveal radio halos in the clusters Abell~746 and RX~J0107.8+5408. We confirm the presence of radio relics in Abell~2061, RXC~J1053.7+5452 and diffuse emission in Abell~523 and Abell~2034 (for which the classification is uncertain). In addition, we detect the radio halo Abell~697 providing additional flux measurements around 1.4~GHz.

By constructing a sample of 35 relics we have found that relics are generally located along the major axis (which can be used as a proxy for the merger axis) of the cluster's elongated ICM. Their orientation is mostly perpendicular to this major axis. The distribution of the major-minor axis ratios for the relic cluster sample is broader than that of the NORAS-REFLEX sample. These results are consistent with the scenario that relics trace shock waves which form along the merger axis of clusters. We also compared the redshift and X-ray luminosity distributions of clusters with relics to a sample of clusters from the NORAS and REFLEX surveys. Interestingly, we find indications that the observed fraction of clusters hosting relics increases with X-ray luminosity and redshift. However, selection biases for radio relics play a role and this needs to be investigated further.  
The significantly improved sensitivity of upcoming radio telescopes (e.g., the EVLA and LOFAR)
will allow to find many more radio relics. Correlating this larger relic sample with cluster
samples selected at X-rays wavelengths will provide a powerful tool to constrain the evolution of relics in galaxy clusters.

\label{sec:conclusion}

\begin{acknowledgements}

We would like to thank the anonymous referee for useful comments. 
We thank the staff of the GMRT who have made these observations possible. 
The GMRT is run by the National Centre for Radio Astrophysics of the 
Tata Institute of Fundamental Research. The Westerbork Synthesis 
Radio Telescope is operated by ASTRON (Netherlands Institute 
for Radio Astronomy) with support from the Netherlands Foundation 
for Scientific Research (NWO). The National Radio Astronomy Observatory 
is a facility of the National Science Foundation operated under 
cooperative agreement by Associated Universities, Inc.  The 
William Herschel Telescope and Isaac Newton Telescope are operated 
on the island of La Palma by the Isaac Newton Group in the Spanish 
Observatorio del Roque de los Muchachos of the Instituto de Astrofsica 
de Canarias. The National Radio Astronomy Observatory is a facility 
of the National Science Foundation operated under cooperative agreement 
by Associated Universities, Inc.
RJvW acknowledges funding from the Royal Netherlands Academy of Arts and Sciences. MB and MH acknowledge support by the research group FOR 1254 funded by the Deutsche Forschungsgemeinschaft.

\end{acknowledgements}

\bibliographystyle{aa}
\bibliography{ref_filaments}

\begin{thebibliography}{121}
\expandafter\ifx\csname natexlab\endcsname\relax\def\natexlab#1{#1}\fi

\bibitem[{{Abazajian} {et~al.}(2009){Abazajian}, {Adelman-McCarthy},
  {Ag{\"u}eros}, {Allam}, {Allende Prieto}, {An}, {Anderson}, {Anderson},
  {Annis}, {Bahcall}, \& et~al.}]{2009ApJS..182..543A}
{Abazajian}, K.~N., {Adelman-McCarthy}, J.~K., {Ag{\"u}eros}, M.~A., {et~al.}
  2009, \apjs, 182, 543

\bibitem[{{Abell} {et~al.}(1989){Abell}, {Corwin}, \&
  {Olowin}}]{1989ApJS...70....1A}
{Abell}, G.~O., {Corwin}, Jr., H.~G., \& {Olowin}, R.~P. 1989, \apjs, 70, 1

\bibitem[{{Aguerri} {et~al.}(2007){Aguerri}, {S{\'a}nchez-Janssen}, \&
  {Mu{\~n}oz-Tu{\~n}{\'o}n}}]{2007A&A...471...17A}
{Aguerri}, J.~A.~L., {S{\'a}nchez-Janssen}, R., \& {Mu{\~n}oz-Tu{\~n}{\'o}n},
  C. 2007, \aap, 471, 17

\bibitem[{{Athreya}(2009)}]{2009ApJ...696..885A}
{Athreya}, R. 2009, \apj, 696, 885

\bibitem[{{Axford} {et~al.}(1977){Axford}, {Leer}, \&
  {Skadron}}]{1977ICRC...11..132A}
{Axford}, W.~I., {Leer}, E., \& {Skadron}, G. 1977, in International Cosmic Ray
  Conference, Vol.~11, International Cosmic Ray Conference, 132--+

\bibitem[{{Baars} {et~al.}(1977){Baars}, {Genzel}, {Pauliny-Toth}, \&
  {Witzel}}]{1977A&A....61...99B}
{Baars}, J.~W.~M., {Genzel}, R., {Pauliny-Toth}, I.~I.~K., \& {Witzel}, A.
  1977, \aap, 61, 99

\bibitem[{{Bacchi} {et~al.}(2003){Bacchi}, {Feretti}, {Giovannini}, \&
  {Govoni}}]{2003A&A...400..465B}
{Bacchi}, M., {Feretti}, L., {Giovannini}, G., \& {Govoni}, F. 2003, \aap, 400,
  465

\bibitem[{{Bagchi} {et~al.}(2006){Bagchi}, {Durret}, {Neto}, \&
  {Paul}}]{2006Sci...314..791B}
{Bagchi}, J., {Durret}, F., {Neto}, G.~B.~L., \& {Paul}, S. 2006, Science, 314,
  791

\bibitem[{{Barrena} {et~al.}(2007){Barrena}, {Boschin}, {Girardi}, \&
  {Spolaor}}]{2007A&A...467...37B}
{Barrena}, R., {Boschin}, W., {Girardi}, M., \& {Spolaor}, M. 2007, \aap, 467,
  37

\bibitem[{{Becker} {et~al.}(1995){Becker}, {White}, \&
  {Helfand}}]{1995ApJ...450..559B}
{Becker}, R.~H., {White}, R.~L., \& {Helfand}, D.~J. 1995, \apj, 450, 559

\bibitem[{{Bell}(1978{\natexlab{a}})}]{1978MNRAS.182..147B}
{Bell}, A.~R. 1978{\natexlab{a}}, \mnras, 182, 147

\bibitem[{{Bell}(1978{\natexlab{b}})}]{1978MNRAS.182..443B}
{Bell}, A.~R. 1978{\natexlab{b}}, \mnras, 182, 443

\bibitem[{{Bertin} \& {Arnouts}(1996)}]{1996A&AS..117..393B}
{Bertin}, E. \& {Arnouts}, S. 1996, \aaps, 117, 393

\bibitem[{{Blandford} \& {Eichler}(1987)}]{1987PhR...154....1B}
{Blandford}, R. \& {Eichler}, D. 1987, \physrep, 154, 1

\bibitem[{{Blandford} \& {Ostriker}(1978)}]{1978ApJ...221L..29B}
{Blandford}, R.~D. \& {Ostriker}, J.~P. 1978, \apjl, 221, L29

\bibitem[{{Blasi} \& {Colafrancesco}(1999)}]{1999APh....12..169B}
{Blasi}, P. \& {Colafrancesco}, S. 1999, Astroparticle Physics, 12, 169

\bibitem[{{B{\"o}hringer} {et~al.}(2004){B{\"o}hringer}, {Schuecker}, {Guzzo},
  {Collins}, {Voges}, {Cruddace}, {Ortiz-Gil}, {Chincarini}, {De Grandi},
  {Edge}, {MacGillivray}, {Neumann}, {Schindler}, \&
  {Shaver}}]{2004A&A...425..367B}
{B{\"o}hringer}, H., {Schuecker}, P., {Guzzo}, L., {et~al.} 2004, \aap, 425,
  367

\bibitem[{{B{\"o}hringer} {et~al.}(2000){B{\"o}hringer}, {Voges}, {Huchra},
  {McLean}, {Giacconi}, {Rosati}, {Burg}, {Mader}, {Schuecker}, {Simi{\c c}},
  {Komossa}, {Reiprich}, {Retzlaff}, \& {Tr{\"u}mper}}]{2000ApJS..129..435B}
{B{\"o}hringer}, H., {Voges}, W., {Huchra}, J.~P., {et~al.} 2000, \apjs, 129,
  435

\bibitem[{{Bonafede} {et~al.}(2009{\natexlab{a}}){Bonafede}, {Feretti},
  {Giovannini}, {Govoni}, {Murgia}, {Taylor}, {Ebeling}, {Allen}, {Gentile}, \&
  {Pihlstr{\"o}m}}]{2009A&A...503..707B}
{Bonafede}, A., {Feretti}, L., {Giovannini}, G., {et~al.} 2009{\natexlab{a}},
  \aap, 503, 707

\bibitem[{{Bonafede} {et~al.}(2009{\natexlab{b}}){Bonafede}, {Giovannini},
  {Feretti}, {Govoni}, \& {Murgia}}]{2009A&A...494..429B}
{Bonafede}, A., {Giovannini}, G., {Feretti}, L., {Govoni}, F., \& {Murgia}, M.
  2009{\natexlab{b}}, \aap, 494, 429

\bibitem[{{Bonafede} {et~al.}(2011){Bonafede}, {Govoni}, {Feretti}, {Murgia},
  {Giovannini}, \& {Br{\"u}ggen}}]{2011A&A...530A..24B}
{Bonafede}, A., {Govoni}, F., {Feretti}, L., {et~al.} 2011, \aap, 530, A24+

\bibitem[{{Briggs}(1995)}]{briggs_phd}
{Briggs}, D.~S. 1995, PhD thesis, New Mexico Institute of Mining Technology,
  Socorro, New Mexico, USA

\bibitem[{{Brown} \& {Rudnick}(2011)}]{2011MNRAS.412....2B}
{Brown}, S. \& {Rudnick}, L. 2011, \mnras, 412, 2

\bibitem[{{Brunetti} {et~al.}(2009){Brunetti}, {Cassano}, {Dolag}, \&
  {Setti}}]{2009A&A...507..661B}
{Brunetti}, G., {Cassano}, R., {Dolag}, K., \& {Setti}, G. 2009, \aap, 507, 661

\bibitem[{{Brunetti} {et~al.}(2008){Brunetti}, {Giacintucci}, {Cassano},
  {Lane}, {Dallacasa}, {Venturi}, {Kassim}, {Setti}, {Cotton}, \&
  {Markevitch}}]{2008Natur.455..944B}
{Brunetti}, G., {Giacintucci}, S., {Cassano}, R., {et~al.} 2008, \nat, 455, 944

\bibitem[{{Brunetti} {et~al.}(2001){Brunetti}, {Setti}, {Feretti}, \&
  {Giovannini}}]{2001MNRAS.320..365B}
{Brunetti}, G., {Setti}, G., {Feretti}, L., \& {Giovannini}, G. 2001, \mnras,
  320, 365

\bibitem[{{Buote}(2001)}]{2001ApJ...553L..15B}
{Buote}, D.~A. 2001, \apjl, 553, L15

\bibitem[{{Cassano} \& {Brunetti}(2005)}]{2005MNRAS.357.1313C}
{Cassano}, R. \& {Brunetti}, G. 2005, \mnras, 357, 1313

\bibitem[{{Cassano} {et~al.}(2010{\natexlab{a}}){Cassano}, {Brunetti},
  {R{\"o}ttgering}, \& {Br{\"u}ggen}}]{2010A&A...509A..68C}
{Cassano}, R., {Brunetti}, G., {R{\"o}ttgering}, H.~J.~A., \& {Br{\"u}ggen}, M.
  2010{\natexlab{a}}, \aap, 509, A68+

\bibitem[{{Cassano} {et~al.}(2006){Cassano}, {Brunetti}, \&
  {Setti}}]{2006MNRAS.369.1577C}
{Cassano}, R., {Brunetti}, G., \& {Setti}, G. 2006, \mnras, 369, 1577

\bibitem[{{Cassano} {et~al.}(2007){Cassano}, {Brunetti}, {Setti}, {Govoni}, \&
  {Dolag}}]{2007MNRAS.378.1565C}
{Cassano}, R., {Brunetti}, G., {Setti}, G., {Govoni}, F., \& {Dolag}, K. 2007,
  \mnras, 378, 1565

\bibitem[{{Cassano} {et~al.}(2008){Cassano}, {Brunetti}, {Venturi}, {Setti},
  {Dallacasa}, {Giacintucci}, \& {Bardelli}}]{2008A&A...480..687C}
{Cassano}, R., {Brunetti}, G., {Venturi}, T., {et~al.} 2008, \aap, 480, 687

\bibitem[{{Cassano} {et~al.}(2010{\natexlab{b}}){Cassano}, {Ettori},
  {Giacintucci}, {Brunetti}, {Markevitch}, {Venturi}, \&
  {Gitti}}]{2010ApJ...721L..82C}
{Cassano}, R., {Ettori}, S., {Giacintucci}, S., {et~al.} 2010{\natexlab{b}},
  \apjl, 721, L82

\bibitem[{{Cavagnolo} {et~al.}(2009){Cavagnolo}, {Donahue}, {Voit}, \&
  {Sun}}]{2009ApJS..182...12C}
{Cavagnolo}, K.~W., {Donahue}, M., {Voit}, G.~M., \& {Sun}, M. 2009, \apjs,
  182, 12

\bibitem[{{Clarke} \& {Ensslin}(2006)}]{2006AJ....131.2900C}
{Clarke}, T.~E. \& {Ensslin}, T.~A. 2006, \aj, 131, 2900

\bibitem[{{Clowe} {et~al.}(2006){Clowe}, {Brada{\v c}}, {Gonzalez},
  {Markevitch}, {Randall}, {Jones}, \& {Zaritsky}}]{2006ApJ...648L.109C}
{Clowe}, D., {Brada{\v c}}, M., {Gonzalez}, A.~H., {et~al.} 2006, \apjl, 648,
  L109

\bibitem[{{Condon} {et~al.}(1998){Condon}, {Cotton}, {Greisen}, {Yin},
  {Perley}, {Taylor}, \& {Broderick}}]{1998AJ....115.1693C}
{Condon}, J.~J., {Cotton}, W.~D., {Greisen}, E.~W., {et~al.} 1998, \aj, 115,
  1693

\bibitem[{{Cornwell} \& {Perley}(1992)}]{1992A&A...261..353C}
{Cornwell}, T.~J. \& {Perley}, R.~A. 1992, \aap, 261, 353

\bibitem[{{Cotton}(2008)}]{2008PASP..120..439C}
{Cotton}, W.~D. 2008, \pasp, 120, 439

\bibitem[{{Crawford} {et~al.}(1995){Crawford}, {Edge}, {Fabian}, {Allen},
  {Bohringer}, {Ebeling}, {McMahon}, \& {Voges}}]{1995MNRAS.274...75C}
{Crawford}, C.~S., {Edge}, A.~C., {Fabian}, A.~C., {et~al.} 1995, \mnras, 274,
  75

\bibitem[{{Dennison}(1980)}]{1980ApJ...239L..93D}
{Dennison}, B. 1980, \apjl, 239, L93

\bibitem[{{Dolag} \& {En{\ss}lin}(2000)}]{2000A&A...362..151D}
{Dolag}, K. \& {En{\ss}lin}, T.~A. 2000, \aap, 362, 151

\bibitem[{{Donnert} {et~al.}(2010{\natexlab{a}}){Donnert}, {Dolag}, {Brunetti},
  {Cassano}, \& {Bonafede}}]{2010MNRAS.401...47D}
{Donnert}, J., {Dolag}, K., {Brunetti}, G., {Cassano}, R., \& {Bonafede}, A.
  2010{\natexlab{a}}, \mnras, 401, 47

\bibitem[{{Donnert} {et~al.}(2010{\natexlab{b}}){Donnert}, {Dolag}, {Cassano},
  \& {Brunetti}}]{2010MNRAS.407.1565D}
{Donnert}, J., {Dolag}, K., {Cassano}, R., \& {Brunetti}, G.
  2010{\natexlab{b}}, \mnras, 407, 1565

\bibitem[{{Drury}(1983)}]{1983RPPh...46..973D}
{Drury}, L.~O. 1983, Reports on Progress in Physics, 46, 973

\bibitem[{{Ebeling} {et~al.}(1998){Ebeling}, {Edge}, {Bohringer}, {Allen},
  {Crawford}, {Fabian}, {Voges}, \& {Huchra}}]{1998MNRAS.301..881E}
{Ebeling}, H., {Edge}, A.~C., {Bohringer}, H., {et~al.} 1998, \mnras, 301, 881

\bibitem[{{Ebeling} {et~al.}(2002){Ebeling}, {Mullis}, \&
  {Tully}}]{2002ApJ...580..774E}
{Ebeling}, H., {Mullis}, C.~R., \& {Tully}, R.~B. 2002, \apj, 580, 774

\bibitem[{{En{\ss}lin} {et~al.}(2011){En{\ss}lin}, {Pfrommer}, {Miniati}, \&
  {Subramanian}}]{2011A&A...527A..99E}
{En{\ss}lin}, T., {Pfrommer}, C., {Miniati}, F., \& {Subramanian}, K. 2011,
  \aap, 527, A99+

\bibitem[{{Ensslin} {et~al.}(1998){Ensslin}, {Biermann}, {Klein}, \&
  {Kohle}}]{1998A&A...332..395E}
{Ensslin}, T.~A., {Biermann}, P.~L., {Klein}, U., \& {Kohle}, S. 1998, \aap,
  332, 395

\bibitem[{{Feretti} {et~al.}(2006){Feretti}, {Bacchi}, {Slee}, {Giovannini},
  {Govoni}, {Andernach}, \& {Tsarevsky}}]{2006MNRAS.368..544F}
{Feretti}, L., {Bacchi}, M., {Slee}, O.~B., {et~al.} 2006, \mnras, 368, 544

\bibitem[{{Feretti} {et~al.}(2001){Feretti}, {Fusco-Femiano}, {Giovannini}, \&
  {Govoni}}]{2001A&A...373..106F}
{Feretti}, L., {Fusco-Femiano}, R., {Giovannini}, G., \& {Govoni}, F. 2001,
  \aap, 373, 106

\bibitem[{{Giacintucci} {et~al.}(2008){Giacintucci}, {Venturi}, {Macario},
  {Dallacasa}, {Brunetti}, {Markevitch}, {Cassano}, {Bardelli}, \&
  {Athreya}}]{2008A&A...486..347G}
{Giacintucci}, S., {Venturi}, T., {Macario}, G., {et~al.} 2008, \aap, 486, 347

\bibitem[{{Giovannini} {et~al.}(2010){Giovannini}, {Bonafede}, {Feretti},
  {Govoni}, \& {Murgia}}]{2010A&A...511L...5G}
{Giovannini}, G., {Bonafede}, A., {Feretti}, L., {Govoni}, F., \& {Murgia}, M.
  2010, \aap, 511, L5+

\bibitem[{{Giovannini} {et~al.}(2009){Giovannini}, {Bonafede}, {Feretti},
  {Govoni}, {Murgia}, {Ferrari}, \& {Monti}}]{2009A&A...507.1257G}
{Giovannini}, G., {Bonafede}, A., {Feretti}, L., {et~al.} 2009, \aap, 507, 1257

\bibitem[{{Giovannini} \& {Feretti}(2000)}]{2000NewA....5..335G}
{Giovannini}, G. \& {Feretti}, L. 2000, New Astronomy, 5, 335

\bibitem[{{Giovannini} {et~al.}(2011){Giovannini}, {Feretti}, {Girardi},
  {Govoni}, {Murgia}, {Vacca}, \& {Bagchi}}]{2011A&A...530L...5G}
{Giovannini}, G., {Feretti}, L., {Girardi}, M., {et~al.} 2011, \aap, 530, L5+

\bibitem[{{Giovannini} {et~al.}(1991){Giovannini}, {Feretti}, \&
  {Stanghellini}}]{1991A&A...252..528G}
{Giovannini}, G., {Feretti}, L., \& {Stanghellini}, C. 1991, \aap, 252, 528

\bibitem[{{Giovannini} {et~al.}(1999){Giovannini}, {Tordi}, \&
  {Feretti}}]{1999NewA....4..141G}
{Giovannini}, G., {Tordi}, M., \& {Feretti}, L. 1999, New Astronomy, 4, 141

\bibitem[{{Girardi} {et~al.}(2006){Girardi}, {Boschin}, \&
  {Barrena}}]{2006A&A...455...45G}
{Girardi}, M., {Boschin}, W., \& {Barrena}, R. 2006, \aap, 455, 45

\bibitem[{{Govoni} \& {Feretti}(2004)}]{2004IJMPD..13.1549G}
{Govoni}, F. \& {Feretti}, L. 2004, International Journal of Modern Physics D,
  13, 1549

\bibitem[{{Govoni} {et~al.}(2001){Govoni}, {Feretti}, {Giovannini},
  {B{\"o}hringer}, {Reiprich}, \& {Murgia}}]{2001A&A...376..803G}
{Govoni}, F., {Feretti}, L., {Giovannini}, G., {et~al.} 2001, \aap, 376, 803

\bibitem[{{Govoni} {et~al.}(2004){Govoni}, {Markevitch}, {Vikhlinin}, {van
  Speybroeck}, {Feretti}, \& {Giovannini}}]{2004ApJ...605..695G}
{Govoni}, F., {Markevitch}, M., {Vikhlinin}, A., {et~al.} 2004, \apj, 605, 695

\bibitem[{{Govoni} {et~al.}(2005){Govoni}, {Murgia}, {Feretti}, {Giovannini},
  {Dallacasa}, \& {Taylor}}]{2005A&A...430L...5G}
{Govoni}, F., {Murgia}, M., {Feretti}, L., {et~al.} 2005, \aap, 430, L5

\bibitem[{{Hales} {et~al.}(2007){Hales}, {Riley}, {Waldram}, {Warner}, \&
  {Baldwin}}]{2007MNRAS.382.1639H}
{Hales}, S.~E.~G., {Riley}, J.~M., {Waldram}, E.~M., {Warner}, P.~J., \&
  {Baldwin}, J.~E. 2007, \mnras, 382, 1639

\bibitem[{{Hoeft} \& {Br{\"u}ggen}(2007)}]{2007MNRAS.375...77H}
{Hoeft}, M. \& {Br{\"u}ggen}, M. 2007, \mnras, 375, 77

\bibitem[{{Jaffe}(1977)}]{1977ApJ...212....1J}
{Jaffe}, W.~J. 1977, \apj, 212, 1

\bibitem[{{Jeltema} \& {Profumo}(2011)}]{2011ApJ...728...53J}
{Jeltema}, T.~E. \& {Profumo}, S. 2011, \apj, 728, 53

\bibitem[{{Jones} \& {Ellison}(1991)}]{1991SSRv...58..259J}
{Jones}, F.~C. \& {Ellison}, D.~C. 1991, Space Science Reviews, 58, 259

\bibitem[{{Kang} \& {Ryu}(2011)}]{2011ApJ...734...18K}
{Kang}, H. \& {Ryu}, D. 2011, \apj, 734, 18

\bibitem[{{Katgert} {et~al.}(1996){Katgert}, {Mazure}, {Perea}, {den Hartog},
  {Moles}, {Le Fevre}, {Dubath}, {Focardi}, {Rhee}, {Jones}, {Escalera},
  {Biviano}, {Gerbal}, \& {Giuricin}}]{1996A&A...310....8K}
{Katgert}, P., {Mazure}, A., {Perea}, J., {et~al.} 1996, \aap, 310, 8

\bibitem[{{Kempner} \& {Sarazin}(2001)}]{2001ApJ...548..639K}
{Kempner}, J.~C. \& {Sarazin}, C.~L. 2001, \apj, 548, 639

\bibitem[{{Kempner} {et~al.}(2003){Kempner}, {Sarazin}, \&
  {Markevitch}}]{2003ApJ...593..291K}
{Kempner}, J.~C., {Sarazin}, C.~L., \& {Markevitch}, M. 2003, \apj, 593, 291

\bibitem[{{Keshet}(2010)}]{2010arXiv1011.0729K}
{Keshet}, U. 2010, ArXiv e-prints

\bibitem[{{Keshet} \& {Loeb}(2010)}]{2010ApJ...722..737K}
{Keshet}, U. \& {Loeb}, A. 2010, \apj, 722, 737

\bibitem[{{Kocevski} {et~al.}(2007){Kocevski}, {Ebeling}, {Mullis}, \&
  {Tully}}]{2007ApJ...662..224K}
{Kocevski}, D.~D., {Ebeling}, H., {Mullis}, C.~R., \& {Tully}, R.~B. 2007,
  \apj, 662, 224

\bibitem[{{Koester} {et~al.}(2007){Koester}, {McKay}, {Annis}, {Wechsler},
  {Evrard}, {Bleem}, {Becker}, {Johnston}, {Sheldon}, {Nichol}, {Miller},
  {Scranton}, {Bahcall}, {Barentine}, {Brewington}, {Brinkmann}, {Harvanek},
  {Kleinman}, {Krzesinski}, {Long}, {Nitta}, {Schneider}, {Sneddin}, {Voges},
  \& {York}}]{2007ApJ...660..239K}
{Koester}, B.~P., {McKay}, T.~A., {Annis}, J., {et~al.} 2007, \apj, 660, 239

\bibitem[{{Krymskii}(1977)}]{1977DoSSR.234R1306K}
{Krymskii}, G.~F. 1977, Akademiia Nauk SSSR Doklady, 234, 1306

\bibitem[{{Liang} {et~al.}(2000){Liang}, {Hunstead}, {Birkinshaw}, \&
  {Andreani}}]{2000ApJ...544..686L}
{Liang}, H., {Hunstead}, R.~W., {Birkinshaw}, M., \& {Andreani}, P. 2000, \apj,
  544, 686

\bibitem[{{Macario} {et~al.}(2011){Macario}, {Markevitch}, {Giacintucci},
  {Brunetti}, {Venturi}, \& {Murray}}]{2011ApJ...728...82M}
{Macario}, G., {Markevitch}, M., {Giacintucci}, S., {et~al.} 2011, \apj, 728,
  82

\bibitem[{{Macario} {et~al.}(2010){Macario}, {Venturi}, {Brunetti},
  {Dallacasa}, {Giacintucci}, {Cassano}, {Bardelli}, \&
  {Athreya}}]{2010A&A...517A..43M}
{Macario}, G., {Venturi}, T., {Brunetti}, G., {et~al.} 2010, \aap, 517, A43+

\bibitem[{{Malkov} \& {O'C Drury}(2001)}]{2001RPPh...64..429M}
{Malkov}, M.~A. \& {O'C Drury}, L. 2001, Reports on Progress in Physics, 64,
  429

\bibitem[{{Marini} {et~al.}(2004){Marini}, {Bardelli}, {Zucca}, {De Grandi},
  {Cappi}, {Ettori}, {Moscardini}, {Tormen}, \&
  {Diaferio}}]{2004MNRAS.353.1219M}
{Marini}, F., {Bardelli}, S., {Zucca}, E., {et~al.} 2004, \mnras, 353, 1219

\bibitem[{{Markevitch} {et~al.}(2005){Markevitch}, {Govoni}, {Brunetti}, \&
  {Jerius}}]{2005ApJ...627..733M}
{Markevitch}, M., {Govoni}, F., {Brunetti}, G., \& {Jerius}, D. 2005, \apj,
  627, 733

\bibitem[{{Maughan} {et~al.}(2008){Maughan}, {Jones}, {Forman}, \& {Van
  Speybroeck}}]{2008ApJS..174..117M}
{Maughan}, B.~J., {Jones}, C., {Forman}, W., \& {Van Speybroeck}, L. 2008,
  \apjs, 174, 117

\bibitem[{{Mazure} {et~al.}(1996){Mazure}, {Katgert}, {den Hartog}, {Biviano},
  {Dubath}, {Escalera}, {Focardi}, {Gerbal}, {Giuricin}, {Jones}, {Le Fevre},
  {Moles}, {Perea}, \& {Rhee}}]{1996A&A...310...31M}
{Mazure}, A., {Katgert}, P., {den Hartog}, R., {et~al.} 1996, \aap, 310, 31

\bibitem[{{Miniati} {et~al.}(2000){Miniati}, {Ryu}, {Kang}, {Jones}, {Cen}, \&
  {Ostriker}}]{2000ApJ...542..608M}
{Miniati}, F., {Ryu}, D., {Kang}, H., {et~al.} 2000, \apj, 542, 608

\bibitem[{{Noordam}(2004)}]{2004SPIE.5489..817N}
{Noordam}, J.~E. 2004, in Society of Photo-Optical Instrumentation Engineers
  (SPIE) Conference Series, Vol. 5489, Society of Photo-Optical Instrumentation
  Engineers (SPIE) Conference Series, ed. J.~M. {Oschmann}, Jr., 817--825

\bibitem[{{Nuza} {et~al.}(2011){Nuza}, {Hoeft}, {van Weeren}, {Gottl\"ober}, \&
  {Yepes}}]{nuza}
{Nuza}, S.~E., {Hoeft}, M., {van Weeren}, R.~J., {Gottl\"ober}, S., \& {Yepes},
  G. 2011, \mnras,~to be submitted

\bibitem[{{Oegerle} \& {Hill}(2001)}]{2001AJ....122.2858O}
{Oegerle}, W.~R. \& {Hill}, J.~M. 2001, \aj, 122, 2858

\bibitem[{{Paul} {et~al.}(2010){Paul}, {Iapichino}, {Miniati}, {Bagchi}, \&
  {Mannheim}}]{2010arXiv1001.1170P}
{Paul}, S., {Iapichino}, L., {Miniati}, F., {Bagchi}, J., \& {Mannheim}, K.
  2010, ArXiv e-prints

\bibitem[{{Perley}(1989)}]{1989ASPC....6..259P}
{Perley}, R.~A. 1989, in Astronomical Society of the Pacific Conference Series,
  Vol.~6, Synthesis Imaging in Radio Astronomy, ed. R.~A. {Perley}, F.~R.
  {Schwab}, \& A.~H. {Bridle}, 259--+

\bibitem[{{Perley} \& {Taylor}(1999)}]{perleyandtaylor}
{Perley}, R.~T. \& {Taylor}, G.~B. 1999, {VLA Calibrator Manual}, Tech. rep.,
  NRAO

\bibitem[{{Petrosian}(2001)}]{2001ApJ...557..560P}
{Petrosian}, V. 2001, \apj, 557, 560

\bibitem[{{Pizzo} \& {de Bruyn}(2009)}]{2009A&A...507..639P}
{Pizzo}, R.~F. \& {de Bruyn}, A.~G. 2009, \aap, 507, 639

\bibitem[{{Popesso} {et~al.}(2004){Popesso}, {B{\"o}hringer}, {Brinkmann},
  {Voges}, \& {York}}]{2004A&A...423..449P}
{Popesso}, P., {B{\"o}hringer}, H., {Brinkmann}, J., {Voges}, W., \& {York},
  D.~G. 2004, \aap, 423, 449

\bibitem[{{Reid} {et~al.}(1999){Reid}, {Hunstead}, {Lemonon}, \&
  {Pierre}}]{1999MNRAS.302..571R}
{Reid}, A.~D., {Hunstead}, R.~W., {Lemonon}, L., \& {Pierre}, M.~M. 1999,
  \mnras, 302, 571

\bibitem[{{Rengelink} {et~al.}(1997){Rengelink}, {Tang}, {de Bruyn}, {Miley},
  {Bremer}, {R\"ottgering}, \& {Bremer}}]{1997A&AS..124..259R}
{Rengelink}, R.~B., {Tang}, Y., {de Bruyn}, A.~G., {et~al.} 1997, \aaps, 124,
  259

\bibitem[{{Ricker} \& {Sarazin}(2001)}]{2001ApJ...561..621R}
{Ricker}, P.~M. \& {Sarazin}, C.~L. 2001, \apj, 561, 621

\bibitem[{{Rines} \& {Diaferio}(2006)}]{2006AJ....132.1275R}
{Rines}, K. \& {Diaferio}, A. 2006, \aj, 132, 1275

\bibitem[{{Roettiger} {et~al.}(1999){Roettiger}, {Burns}, \&
  {Stone}}]{1999ApJ...518..603R}
{Roettiger}, K., {Burns}, J.~O., \& {Stone}, J.~M. 1999, \apj, 518, 603

\bibitem[{{R\"ottgering} {et~al.}(1997){R\"ottgering}, {Wieringa}, {Hunstead},
  \& {Ekers}}]{1997MNRAS.290..577R}
{R\"ottgering}, H.~J.~A., {Wieringa}, M.~H., {Hunstead}, R.~W., \& {Ekers},
  R.~D. 1997, \mnras, 290, 577

\bibitem[{{Roy} {et~al.}(2010){Roy}, {Gupta}, {Pen}, {Peterson}, {Kudale}, \&
  {Kodilkar}}]{2010ExA....28...25R}
{Roy}, J., {Gupta}, Y., {Pen}, U., {et~al.} 2010, Experimental Astronomy, 28,
  25

\bibitem[{{Rudnick} \& {Lemmerman}(2009)}]{2009ApJ...697.1341R}
{Rudnick}, L. \& {Lemmerman}, J.~A. 2009, \apj, 697, 1341

\bibitem[{{Shan} {et~al.}(2010){Shan}, {Qin}, {Fort}, {Tao}, {Wu}, \&
  {Zhao}}]{2010MNRAS.406.1134S}
{Shan}, H., {Qin}, B., {Fort}, B., {et~al.} 2010, \mnras, 406, 1134

\bibitem[{{Skillman} {et~al.}(2011){Skillman}, {Hallman}, {O'Shea}, {Burns},
  {Smith}, \& {Turk}}]{2011ApJ...735...96S}
{Skillman}, S.~W., {Hallman}, E.~J., {O'Shea}, B.~W., {et~al.} 2011, \apj, 735,
  96

\bibitem[{{Slee} {et~al.}(2001){Slee}, {Roy}, {Murgia}, {Andernach}, \&
  {Ehle}}]{2001AJ....122.1172S}
{Slee}, O.~B., {Roy}, A.~L., {Murgia}, M., {Andernach}, H., \& {Ehle}, M. 2001,
  \aj, 122, 1172

\bibitem[{{Small} {et~al.}(1998){Small}, {Ma}, {Sargent}, \&
  {Hamilton}}]{1998ApJ...492...45S}
{Small}, T.~A., {Ma}, C., {Sargent}, W.~L.~W., \& {Hamilton}, D. 1998, \apj,
  492, 45

\bibitem[{{Struble} \& {Rood}(1999)}]{1999ApJS..125...35S}
{Struble}, M.~F. \& {Rood}, H.~J. 1999, \apjs, 125, 35

\bibitem[{{van Weeren} {et~al.}(2011{\natexlab{a}}){van Weeren}, {Hoeft},
  {R{\"o}ttgering}, {Br{\"u}ggen}, {Intema}, \& {van
  Velzen}}]{2011A&A...528A..38V}
{van Weeren}, R.~J., {Hoeft}, M., {R{\"o}ttgering}, H.~J.~A., {et~al.}
  2011{\natexlab{a}}, \aap, 528, A38+

\bibitem[{{van Weeren} {et~al.}(2009{\natexlab{a}}){van Weeren},
  {R{\"o}ttgering}, {Bagchi}, {Raychaudhury}, {Intema}, {Miniati},
  {En{\ss}lin}, {Markevitch}, \& {Erben}}]{2009A&A...506.1083V}
{van Weeren}, R.~J., {R{\"o}ttgering}, H.~J.~A., {Bagchi}, J., {et~al.}
  2009{\natexlab{a}}, \aap, 506, 1083

\bibitem[{{van Weeren} {et~al.}(2011{\natexlab{b}}){van Weeren},
  {R{\"o}ttgering}, \& {Br{\"u}ggen}}]{2011A&A...527A.114V}
{van Weeren}, R.~J., {R{\"o}ttgering}, H.~J.~A., \& {Br{\"u}ggen}, M.
  2011{\natexlab{b}}, \aap, 527, A114+

\bibitem[{{van Weeren} {et~al.}(2009{\natexlab{b}}){van Weeren},
  {R{\"o}ttgering}, {Br{\"u}ggen}, \& {Cohen}}]{2009A&A...508...75V}
{van Weeren}, R.~J., {R{\"o}ttgering}, H.~J.~A., {Br{\"u}ggen}, M., \& {Cohen},
  A. 2009{\natexlab{b}}, \aap, 508, 75

\bibitem[{{van Weeren} {et~al.}(2009{\natexlab{c}}){van Weeren},
  {R{\"o}ttgering}, {Br{\"u}ggen}, \& {Cohen}}]{2009A&A...505..991V}
{van Weeren}, R.~J., {R{\"o}ttgering}, H.~J.~A., {Br{\"u}ggen}, M., \& {Cohen},
  A. 2009{\natexlab{c}}, \aap, 505, 991

\bibitem[{{van Weeren} {et~al.}(2010){van Weeren}, {R{\"o}ttgering},
  {Br{\"u}ggen}, \& {Hoeft}}]{2010Sci...330..347V}
{van Weeren}, R.~J., {R{\"o}ttgering}, H.~J.~A., {Br{\"u}ggen}, M., \& {Hoeft},
  M. 2010, Science, 330, 347

\bibitem[{{Venturi} {et~al.}(2011){Venturi}, {Giacintucci}, {Dallacasa},
  {Brunetti}, {Cassano}, {Macario}, \& {Athreya}}]{2011MNRAS.tmpL.251V}
{Venturi}, T., {Giacintucci}, G., {Dallacasa}, D., {et~al.} 2011, \mnras, L251+

\bibitem[{{Venturi} {et~al.}(2007){Venturi}, {Giacintucci}, {Brunetti},
  {Cassano}, {Bardelli}, {Dallacasa}, \& {Setti}}]{2007A&A...463..937V}
{Venturi}, T., {Giacintucci}, S., {Brunetti}, G., {et~al.} 2007, \aap, 463, 937

\bibitem[{{Venturi} {et~al.}(2008){Venturi}, {Giacintucci}, {Dallacasa},
  {Cassano}, {Brunetti}, {Bardelli}, \& {Setti}}]{2008A&A...484..327V}
{Venturi}, T., {Giacintucci}, S., {Dallacasa}, D., {et~al.} 2008, \aap, 484,
  327

\bibitem[{{Voges} {et~al.}(2000){Voges}, {Aschenbach}, {Boller}, {Brauninger},
  {Briel}, {Burkert}, {Dennerl}, {Englhauser}, {Gruber}, {Haberl}, {Hartner},
  {Hasinger}, {Pfeffermann}, {Pietsch}, {Predehl}, {Schmitt}, {Trumper}, \&
  {Zimmermann}}]{2000IAUC.7432R...1V}
{Voges}, W., {Aschenbach}, B., {Boller}, T., {et~al.} 2000, \iaucirc, 7432, 1

\bibitem[{{Voges} {et~al.}(1999){Voges}, {Aschenbach}, {Boller},
  {Br{\"a}uninger}, {Briel}, {Burkert}, {Dennerl}, \& {et
  al.}}]{1999A&A...349..389V}
{Voges}, W., {Aschenbach}, B., {Boller}, T., {et~al.} 1999, \aap, 349, 389

\bibitem[{{Wardle} \& {Kronberg}(1974)}]{1974ApJ...194..249W}
{Wardle}, J.~F.~C. \& {Kronberg}, P.~P. 1974, \apj, 194, 249

\bibitem[{{White}(2000)}]{2000MNRAS.312..663W}
{White}, D.~A. 2000, \mnras, 312, 663

\end{thebibliography}

\clearpage
\end{document}